\documentclass{macros/elsarticle}

\usepackage[hide]{macros/ed}
\usepackage{url}
\usepackage{wrapfig}
\usepackage{amssymb,amsfonts,amsmath,amsthm}
\usepackage{tikz}
\usetikzlibrary{arrows}
\usetikzlibrary{mmt}
\usepackage{xkeyval}
\usepackage{listings}
\usepackage{macros/mmt_listings}
\usepackage{paralist} 
\let\hdashline\hline
\usepackage{mathdots} 

\usepackage[bookmarks,bookmarksopen,bookmarksnumbered,linkcolor=blue,citecolor=green,urlcolor=red,colorlinks,breaklinks]{hyperref}

\usepackage{macros/basics}
\usepackage{reltable}
\usepackage{macros/ded}
\usepackage{macros/mmt_new}
\usepackage[nobook]{macros/theorems} 
\usepackage{macros/variant}
\newcommand{\modexp}{\false} 

\makeatletter
\renewenvironment{example}[1][]{\def\@test{#1}%
\par\medskip\noindent\refstepcounter{theorem}%
\hbox{\textsl{Example \arabic{theorem}}~\ifx\@test\@empty\else(#1)\fi}}
{\par\medskip}
\makeatother
\newcommand{\snippet}[1]{\texttt{#1}} 
\newcommand{\defemph}[1]{{\bf#1}}
\newcommand{\element}[1]{{\tt#1}}
\newcommand{\omdoc}{\sc{OMDoc}}
\newcommand{\openmath}{\sc{OpenMath}}
\newcommand{\mathml}{\sc{MathML}}
\newcommand{\tntbase}{\sc{TNTBase}}
\newcommand{\Ypsilon}{\Upsilon}
\begin{document}
\begin{frontmatter}

\title{A Scalable Module System}

\author[jub]{Florian Rabe}
\ead{f.rabe@jacobs-university.de}
\ead[url]{http://kwarc.info/frabe/}

\author[jub]{Michael Kohlhase}
\ead{m.kohlhase@jacobs-university.de}
\ead[url]{http://kwarc.info/kohlhase/}
\address[jub]{Jacobs University Bremen, Computer Science, Germany}

\begin{abstract}
  Symbolic and logic computation systems ranging from computer algebra systems to theorem
  provers are finding their way into science, technology, mathematics and engineering. But
  such systems rely on explicitly or implicitly represented mathematical knowledge that
  needs to be managed to use such systems effectively. 

  While mathematical knowledge management (MKM) ``\emph{in the small}'' is well-studied,
  scaling up to large, highly interconnected corpora remains difficult. We hold that in
  order to realize MKM ``\emph{in the large}'', we need representation languages and
  software architectures that are designed systematically with large-scale processing in
  mind.

  Therefore, we have designed and implemented the {\mmt} language -- a module system for
  mathematical theories. {\mmt} is designed as the simplest possible language that
  combines a module system, a foundationally uncommitted formal semantics, and
  web-scalable implementations.  Due to a careful choice of representational primitives,
  {\mmt} allows us to integrate existing representation languages for formal mathematical
  knowledge in a simple, scalable formalism. In particular, {\mmt} abstracts from the
  underlying mathematical and logical foundations so that it can serve as a standardized
  representation format for a formal digital library. Moreover, {\mmt} systematically
  separates logic-dependent and logic-independent concerns so that it can serve as an
  interface layer between computation systems and MKM systems.
\end{abstract}

\end{frontmatter}

\ednote{check spacing of figures}

\newpage
\setcounter{tocdepth}{4}
\tableofcontents
\newpage

\section{Introduction}\label{sec:mmt:intro}
   Mathematics is one of the oldest areas of human knowledge and provides science with
modeling tools and a knowledge representation regime based on rigorous language. However,
mathematical knowledge is far too vast to be understood by one person -- it has been
estimated that the total amount of published mathematics doubles every ten to fifteen
years~\cite{odlyzko}. Indeed, for example, Zentralblatt Math~\cite{zblmath} maintains a
database of 2.9 million reviews for articles from 3500 journals from 1868 to 2010.

The currently practiced way to organize mathematical knowledge is to have
humans build a cognitive representation of the contents in their minds and to communicate their results in natural -- i.e., informal -- language with
interspersed formulas. This process is well-suited for doing mathematics ``in
the small'' where human creativity is needed to create new mathematical insights.
But the sheer volume of mathematical knowledge precludes this approach from organizing
mathematics ``in the large'': Except for prestige projects such as the classification of finite simple groups~\cite{solomon:fsg},
collaboration in mathematics is largely small-scale.

But this leads to increasing specialization and missed opportunities for knowledge transfer, and the question of supporting the management and dissemination of mathematical knowledge in the large remains difficult.
This problem has been tackled in the field of \emph{mathematical knowledge management} (MKM), which uses explicitly annotated content as the basis for mathematical software services such as semantics-based searching and navigation.
MKM in the large has been pioneered in the field of \emph{formal methods in software engineering}, where a sound logical foundation and the incorruptibility of
computers are combined to verify computer systems. These computer-aided proofs rely on large amounts of formal knowledge about the programming language constructs and data structures, and the productivity of formal methods is restricted in practice by the effectivity of managing this knowledge.

We currently see five obstacles for large scale computerized MKM:
\begin{list}{}{\setlength{\leftmargin}{0.5em}\setlength{\itemindent}{2.5em}}
\item[\bf Informality] As computer programs still lack any real understanding of mathematics,
  human mathematicians must make structures in mathematical knowledge sufficiently
  explicit. This usually means that the knowledge has to be formalized, i.e., represented
  in a formal, logical system. While it is generally assumed that all mathematical
  knowledge can in principle be formalized, this is so expensive that it is seldom even attempted.
\item[\bf Logical Heterogeneity] One of the advantages of informal, but rigorous mathematics is that it does not
  force the choice of a formal system.  There are many formal systems, each optimized for
  expressing and reasoning about different aspects of mathematical knowledge. All attempts to find the ``mother of all logical systems''
  (and convince others to use it) have failed. Even though logics themselves can be made
  the objects of mathematical investigation and even of formalization (in logical
  frameworks), we do not have scalable methods for efficiently dealing with
  heterogeneous, i.e., multi-logic, presentations of mathematical knowledge.
\item[\bf Foundational Assumptions] Logical heterogeneity is not only a matter of optimization because different developments of mathematical knowledge make different foundational assumptions. For example, classical mathematics usually assumes some kind of set theory as a
  foundation and embraces a platonist philosophy. But there are different ones of
  differing expressivity, such as those with and those without the axiom of choice.
  Other mathematicians even reject the law of excluded middle or insist on constructive witnesses for existential
  theorems. Corresponding developments often take a more formalist stance and use type theoretic foundations.
\item[\bf Modularity] Modern developments of mathematical knowledge are highly
  modular. They take pains to identify minimal sets of assumptions so that results are
  applicable at the most general possible level. This modularity and the mathematical
  practice of ``framing'', i.e., of viewing objects of interest in terms of already
  understood structures, must be supported to even approach human capabilities of managing
  mathematical knowledge in computer systems.
\item[\bf Global Scale] Mathematical research and applications are
  distributed globally, and mathematical knowledge is highly interlinked by explicit and implicit
  references. Therefore, a computer-supported management
  system for mathematical knowledge must support global interlinking and framing as well as management
  algorithms that scale up to very large (global) data sets.
\end{list}

In this paper we contribute to a uniform solution of four of the five challenges: We give a globally scalable module system for mathematical theories ({\mmt}) that abstracts from and mediates between different logics and foundations.\footnote{We have already solved the integration of formal and informal mathematical knowledge in the {\omdoc} format, whose formal part is a predecessor of the work presented in this paper. We plan to integrate this solution with the much stronger formal basis of {\mmt} in the future.}
With this, we lay a conceptual and technical foundation for formal MKM in the large.

Because our solution draws intuitions from the fields of mathematics, formal methods, and knowledge management, we give a comprehensive overview over the relevant language features and introduce a terminology for them in Sect.~\ref{sec:mmt:aspects}. This gives
us a solid footing to describe the central design choices underlying {\mmt} in
Sect.~\ref{sec:mmt:overview}. Then we describe the formal syntax of {\mmt} in
Sect.~\ref{sec:mmt:syntax} and an inference system that defines the well-formed
expressions in Sect.~\ref{sec:mmt:validity}. In Sect.~\ref{sec:mmt:metaanalysis}, we
discuss the meta-theoretical properties of {\mmt}, which include a
flattening algorithm that defines the semantics of modular {\mmt}-expressions.
The semantics of {\mmt} is parametric in what we call foundations, and we look at particular
foundations in Sect.~\ref{sec:mmt:specific_foundations}.
Then we discuss the web scalability of {\mmt} and our implementations in Sect.~\ref{sec:mmtweb}
and~\ref{sec:mmtweb:implementation}.
Finally we give an extensive discussion of related representation languages in Sect.~\ref{sec:mmt:related} and conclude in Sect.~\ref{sec:mmt:conc}.

 \section{Features of Knowledge Representation Languages}\label{sec:mmt:aspects}
   In order to compare {\mmt} to other representation languages for mathematical knowledge,
we will first develop a classification vocabulary that will allow us to place {\mmt} in
the taxonomy of modular, machine-processable knowledge representation languages. We will also use this vocabulary in Sect.~\ref{sec:mmt:related} to compare {\mmt} to other module systems.

By a \defemph{module system}, we mean a formal language that provides constructs to express high-level design patterns such as namespaces, imports, parametricity, encapsulation, etc. Very often the modular features of a language can be separated from the non-modular ones. In that case, we call the fragment containing no modularity the \defemph{base language}. Typical base languages are logics, type theories, or programming languages. Base language and module system can be designed together or independently, and in the latter case the module system may be designed before or after the base language. We will sometimes use the phrases \defemph{modular expression} and \defemph{base expression} to distinguish expressions of the module system and the base language.

\subsection{Packages and Modules}

Module systems typically feature one or both of two main scoping devices. Unfortunately,
these overlap, and even when they are distinguished, there is no universal convention on
how to name them. We will use the names \emph{package} and \emph{module}.  Other names in
common use for package are ``library'', ``namespace'', and ``module''; the latter is the
one used in Modula~\cite{modula}, one of the first systems with this functionality.  Other
names used instead of \emph{module} are ``theory'', ``signature'', ``specification'',
``(type) class'', ``(module) type'', and ``locale''.

\defemph{Packages} provide scopes for the grouping of related toplevel declarations into
-- possibly nested -- components. The main purpose of packages is \defemph{namespace
  management}: Packages have names, and their named toplevel declarations are identified
by a qualified name: a pair of a package name and a declaration name. This facilitates
reuse and distribution of declarations over files and networks. Often the packaging
structure is transparent to the semantics of the language; in that case the semantics of
packages is that they identify and locate the available toplevel declarations.

When identifying these declarations, we distinguish \defemph{open} and
\defemph{closed} packaging. With open packages, all packages can refer all toplevel
declarations in all other packages via qualified names. With closed packages, import declarations are necessary as only explicitly imported declarations are accessible.
In both cases, import declarations are often used to make the imported declarations available without qualification.

When locating these declarations, we speak of \defemph{logical} package identifiers if package identifiers are different from physical locations as given by file systems, databases, and networks; otherwise, we speak of \defemph{physical} identifiers. With logical identifiers, the location of resources requires a resolution algorithm that maps logical identifiers to physical locations. This resolution can be relegated to an extra-linguistic \defemph{catalog}. Catalogs provide an abstraction layer that makes the distribution of resources over physical locations transparent to the language and avoids conflicts due to naming conventions of operating systems and storage solutions. Using URI-based package identifiers, logical identifiers can be made globally unique to support global interlinking.
\medskip

Typically, the declarations in a package are module declarations, and a package can be seen as a group of modules. But there are also languages featuring only packages or only modules. In the former case, every package can be considered to contain a single unnamed module; this is the case in many XML-related languages where the packages are called namespaces such as in XQuery~\cite{xquery}. In the latter case, all modules share the same namespace, which can be considered to form a single unnamed package; this is the case in SML where a configuration file is used to list the files over which the modules are distributed. We will call the latter \defemph{single-package} module systems.
\medskip

Like packages, \defemph{modules} are scoped groups of declarations. But contrary to packages, modules are opaque to the language semantics and are used to realize modular design patterns such as inheritance, instantiation, and hiding. For example, moving a declaration between packages has no semantic consequences except that references to the moved declaration must be updated. But a module has a meaning itself that will be affected if a declaration is removed or added. For example, a mathematical theory should be represented as a module because moving axioms between theories changes the semantics; a mathematical paper should be represented as a package because some parts may be relegated to other papers.

Typically, languages provide a number of different types of declarations that may occur in a module. The most typical declarations are sorts and types, constants and values, operations and functions, and predicates. These are usually named. Further examples of named or unnamed declarations are axioms, theorems, inference rules, abbreviations, or notations for parsing and printing. A named declaration within a module can often be identified as a triple of package name, module name, and declaration name.

\subsection{Inheritance}

In the simplest case, \defemph{inheritance} is a binary relation between modules, which is
usually seen as an inheritance graph whose nodes are the modules and whose edges
make up the inheritance relation. The individual edges are called \defemph{imports}: If $T$ inherits from $S$, then $T$ imports all knowledge items of $S$, which then become available in $T$. An important distinction is whether the individual imports are \defemph{named} or \defemph{unnamed}. In the former case, the name of the import is
available to refer to (\emph{i}) the imported module as a whole, or (\emph{ii}) the
imported knowledge items via qualified names.

Other names for named imports are ``structure'' and ``instance''. Other names for unnamed imports are ``mixin'', ``inclusion'', ``inheritance'', and ``definitional morphism/link''.

Inheritance leads to a \defemph{diamond situation} when the same module is imported in two
different ways. The language may \defemph{identify} multiple imports of the same module or
\defemph{distinguish} them. For named imports, the distinguish-semantics is natural because the multiply imported knowledge items can have different qualified names. But then \defemph{sharing} declarations are necessary to force the
identification of these items. The identify-semantics is more natural with unnamed
imports. Then \defemph{renaming} declarations are needed to force the distinction of
multiply imported knowledge items.

A related problem is the \defemph{import name clash}, which arises when unnamed imports
import from different modules which happen to contain knowledge items with the same local
name. In large-scale developments, this is a very typical situation, which can be difficult to
detect.  Here module systems may signal an \defemph{error}, the knowledge item imported first can be \defemph{shadowed} by the one imported later, the name of the module can be used to form a unique \defemph{qualified} name, or \defemph{overload/identify}-semantics can be used. In the latter case, overloading resolution is used to disambiguate a reference to a knowledge item; and knowledge items that cannot be distinguished in this way (e.g., because they have the same types) are identified.

A more complex form of inheritance is \defemph{instantiation}. It means that when
importing $S$ into $T$, some names declared in $S$ may be mapped to expressions of
$T$. This set of mappings can be seen as the passing of argument values over $T$ to parameters of $S$. If instantiations are possible, multiple imports of the same module with different instantiations should be distinguished. Therefore, the distinguish-semantics is more natural. But it is also possible to identify two imports iff they use the same instantiations.

Module systems differ as to what kind of mappings are allowed. Some systems only allow the
map of $S$-symbols to $T$-symbols. This has the advantage that it is easier to check
whether a map is well-typed. Other systems allow mapping symbols to composed
expressions. And systems with named imports, can permit the map of an import itself to a
realization (see below).

Another difference is which symbols or imports may be instantiated: We speak of a
\defemph{free} instantiation if arbitrary symbols or imports can be instantiated. Free
instantiations must \defemph{explicitly} associate some names of $S$ with expressions of
$T$. And we speak of \defemph{interfaced} instantiation if the declarations of $S$ are
divided into two blocks, and only the declarations in the first block --- the interface
--- are available for instantiations. Interfaced instantiations are often
\defemph{implicit}: The order of declarations in the interface of $S$ must correspond to
the order of provided $T$-expressions. Furthermore, instantiations may be
{\defemph{total}} or {\defemph{partial}}: Total instantiations provide expressions for all
symbols or imports in (the interface of) $S$. Finally, some systems restrict inheritance
to axioms; in such systems, imports must carry instantiations for all symbols; we speak of
\defemph{axiom-inheritance}.

A further distinction regards the relation between the imports and the other declarations. We speak of \defemph{separated} imports if all imports must be given at the beginning of the module; otherwise, we call them \defemph{interspersed} imports. Separated imports are conceptually easier, but less expressive: At the beginning of a module, less syntactic material is available to form expressions that can be used in instantiations.

More general forms of imports permit \defemph{hiding} and \defemph{filtering} of declarations. Both are similar syntactically but not semantically. When importing from $S$ to $T$, filtering a declaration of $S$ means to exclude that declaration from the import. Hiding is more complicated -- one way to think of it is that if a declaration is hidden, it is still imported but rendered inaccessible. In both cases, it is necessary to maintain a dependency relation between declarations: If a declarations is hidden or filtered, so must be all declarations that depend on it. 

Hiding can be quite difficult to formalize but has an elegant interpretation in the context of algebraic specification. There, it is used to represent the hiding of implementation details or auxiliary constants. For example, implementations of a specification $S$ must also implement the hidden functions of $S$, but are considered equal if they differ only in the implementation of hidden functions. More precisely, we speak of \defemph{simple} hiding.

\defemph{Complex} hiding arises if not only declarations, i.e., atomic expressions, can be hidden but composed expressions as well. Syntactically, a complex hiding from $S$ to $T$ can be seen as a morphism from $T$ to $S$ in a category of specifications. Then simple hiding is the special case where this morphism is an inclusion. Complex hiding has the appeal that instantiation and hiding become dual to each other.

\subsection{Realizations}\label{sec:mmt:modules-as-types}

Many module systems use a concept that we will call \defemph{realization}. Its treatment can vary substantially between systems, which makes it more difficult to describe abstractly. The common intuition is that we can often think of a module as a specification, an interface, or a behavioral description. Then the realizations are the objects that conform to such a specification. Further names used instead of ``realization'' are ``interpretation'', ``structure'', ``instance'', and ``(module) term/value/expression''. Very often it is fruitful to consider modules as types and apply the intuitions of type theory to them. Then a realization is a value that is typed by a module.

For example, in SML, the structures are the realizations of the signatures. In Java, the instances of concrete classes are the realizations of the abstract classes and the interfaces. In logic, the models are the realizations of the theories. In formal specification, the implementations are the realizations of the specifications.

More concretely, a realization of a module $S$ in terms of some context $C$ must provide values over $C$ for all symbols declared in the module $S$. Two special cases are of particular importance.

Firstly, if $C$ is the empty context -- or more precisely: the global environment implicitly determined by the base language -- we speak of \defemph{grounded} realizations. For example, in formal specification, the grounded realizations of $S$ are the programs implementing $S$; the implicit global environment is given by the built-in datatypes and values of the programming language. In logic, the grounded realizations are the models of $S$; the implicit global environment is given by the foundation of mathematics, e.g., set theory.

Secondly, if $C$ is another module $T$, we obtain the notions of ``views from $S$ to $T$'' and of ``functors from $T$ to $S$''. They are dual in the sense that a view from $S$ to $T$ is a functor from $T$ to $S$ and vice versa. But because they are often associated with very different intuitions, this duality is rarely explicated. We can also recover imports as a special case of realizations akin to views.

\defemph{Functors} are associated with the intuitions of type theory: If modules are seen as types and realizations as values, then functors are the module-level analogue of functions. If $r(x)$ is a realization of $T$ that is given in terms of a realization $x$ of $S$, then $\lambda$-abstraction yields a functor $\lambda x:S.r(x)$. For such a functor, \defemph{functor application} maps a realization of $S$ to a realization of $T$ by $\beta$-reduction. A module system is \defemph{higher-order} if functors may take other functors as arguments.

\defemph{Views} are associated with the intuitions of category theory: Many declarative languages can be naturally formulated as categories with modules as objects and views as morphisms. A view from $S$ to $T$ interprets all declarations of $S$ in terms of $T$, and this often yields a \defemph{homomorphic extension} that maps expressions over $S$ to expressions over $T$: All symbols in an $S$-expression are replaced with their $T$-definition provided by $r$. Other names used instead of ``view'' are ``signature/theory/specification morphism'' and ``postulated morphism/link''.

\defemph{Imports} are similar to views in that every import from $S$ to $T$ yields a realization of $S$ in terms of $T$. Using the intuitions of type theory, declaring a named import from $S$ can be seen as the declaration of a symbol of type $S$.
\medskip

The duality between views and functors is connected to a duality of two important translation functions. Consider a realization $r$ of $S$ in terms of $T$.
The \defemph{syntactic translation} maps $S$-expressions to $T$-expressions using the homomorphic extension of $r$. This is closely related to the intuition of $r$ as a view from $S$ to $T$.
The \defemph{semantic translation} maps realizations of $T$ to realizations of $S$ by functor application. This is closely related to the intuition of $r$ as a functor from $T$ to $S$.

For example, let $S$ be the theory of monoids and $T$ the theory of groups, and let $r$ realize every symbol of the language of monoids by its analogue in the language of groups.
Then the syntactic translation maps an expression in the language of monoids to the corresponding expression in the language of groups. And the semantic translation maps every group to itself seen as a monoid.

A language that features realizations may or may not provide concrete syntax for these two translations. A language that can talk about both translations may also state the duality between them as an adjunction between two functors in the sense of category theory.
\medskip

Finally, we have the notion of \defemph{subtyping} between modules. If every expression over $S$ is also an expression over $T$, then $S$ is a \defemph{syntactic subtype} of $T$. Dually, if every realization of $S$ is also a realization of $T$, then $S$ is a \defemph{semantic subtype} of $T$. If both subtyping relations are present in a language, then they are usually opposites of each other.

More concretely, $S$ is a syntactic subtype of $T$ iff there is a realization $r$ of $S$ in terms of $T$ whose syntactic and semantic translations are inclusions. Then we speak of \defemph{nominal subtyping} if $r$ is an import, and of \defemph{structural subtyping} if $r$ is a view.

\subsection{Semantics}

There are two ways to give a formal semantics of modular
expressions. We speak of a \defemph{model theoretical} semantics if models are used to interpret modules. This is typical in the
algebraic specification community. We speak of a \defemph{proof theoretical} semantics if
the semantics is given by typing judgments and inference rules.

Often for some or all modular expressions, there is an expression of the base language with the same semantics. In the type and proof theory community, this is often built-in: The semantics of a modular expression is defined by transforming it into a non-modular one; this is called \defemph{elaboration}. In contrast, in languages with a model theoretical semantics, it is a theorem about the semantics and often called \defemph{flattening}.

We say that a module system is \defemph{conservative} if every modular expression can be
flattened or elaborated into an expression of the base language. Language features that typically prevent conservativity are higher-order functors and hiding.

Similar to conservativity is the \defemph{internalization} of a module system. For certain
languages, it is usually possible to represent the module level judgment $s$ as a
realization of the module $S$ as a typing judgment of the base language. This is possible
if the base language features record types, in which all declarations that can occur in a
module may also occur as fields in a record. Then modules are records, realizations are
values, and functors are functions. However, such expressive record types are often not
present and can often only be added at great cost, e.g., an internalized module system for
simple type theory requires type polymorphism. Moreover, in languages where declarations
build on each other, dependent record types are needed.

\subsection{Genericity}

A \defemph{logical framework} is a formal representation system that provides an uncommitted set of primitives. Such a framework can be used as a meta-language to define other languages. We call a module system \defemph{generic} if it is not specific to a certain base language, but defined within a logical framework. A generic module system is parametrized by an arbitrary base language defined within the logical framework.

We distinguish further whether the logical framework is based on \defemph{set theory} or \defemph{type theory}. The former typically has a model theoretical, the latter a proof theoretical semantics. The choice of framework often implies a foundational commitment because the framework must make some assumptions about the base language.

For example, set/model theoretical module systems may assume the semantics of the base
language as an institution. An example is ASL based on the framework of
institutions~\cite{institutions,asl}. This implies a commitment to a certain axiomatic set
theory in which models and institutions are given. But for example, if the foundation
includes axioms for choice or large cardinals, the models of the same module differ.

Similarly, a type/proof theoretical module system may assume the semantics of the base
language as a system of judgments and inference rules. An example is the locale module
system based on the logical framework Isabelle~\cite{isabelle,isabelle_locales}. This
implies a commitment to a formal language in which judgments and inference rules are
described. But different logical frameworks permit the representation of different object
logics.

We use the term \defemph{foundation} to refer to the mathematical theory that formalizes
this implicit commitment: the axiomatic set theory in the former, and the logical
framework in the latter case. We call a module system \defemph{foundation-independent} if
it avoids such a commitment.  This can for instance be achieved by explicitly representing
the foundation itself as a module. Foundation-indepen\-dent module systems are not only
parametric in the base language but also in the foundation used to express the semantics
of the base language.

\subsection{Degree of Formality}

Mathematics has traditionally been written in natural language with interspersed
formulas. This is different from the fully formal style that is often used in computer-supported mathematics. Even though the focus of {\mmt} is on formal languages, it is worthwhile to discuss informal languages as well because many aspects of module systems are independent of the degree of formality.

\defemph{Formal} languages are based on a formal syntax with a precisely defined semantics. The syntax is based on a formal grammar that can be implemented so that computers can parse and understand it. A typical service that a computer can offer for a formal language is the \defemph{validation} of knowledge to guarantee correctness. Computers can also automatically generate knowledge, such as in automated theorem proving where the generated knowledge item is a proof. This category also includes controlled grammars of natural language that are used to give formal representations a more human-friendly appearance.

\defemph{Informal} languages do not have a formal syntax and are based on unrestricted natural language. While mathematicians use informal language rigorously to obtain an unambiguous semantics, this semantics can only be understood by humans but not by machines. Therefore, only shallow machine-processing services are available such as authoring, storing, and distributing papers and books.

But mathematicians frequently use formal objects within natural language. This has motivated the design of \defemph{semi-formal} representation languages that combine formal and informal representations and degrade gracefully when the latter is used. The automated type-setting provided by {\LaTeX} is a simple example; here the formal representation aspects include the structuring of text into, e.g., definitions, theorems, and formulas.

Note that in the example of {\LaTeX}, the formulas themselves are not formal in our sense: While formal symbols are used, the representation is still human-oriented, and machines can usually not determine the syntax tree of a formula from its {\LaTeX} representation. Such representations are called \defemph{presentation}-based and distinguished from \defemph{content}-based representations that make the syntax tree accessible to machines.

\subsection{Scalability}

For machine-processable representation languages, performance and language design are not
always orthogonal. We are specifically interested in language aspects that affect
scalability.

We call a module system \defemph{web standard-compliant} if it provides a concrete syntax
that uses XML~\cite{xml} for all language expressions and URIs~\cite{uri} for all
identifiers.  XML enables standardized document fragment access by technologies like
XPath~\cite{xpath} and document fragment aggregation by XQuery \cite{xquery}. Deployment
on web servers allows distributed storage and flexible access methods. URIs provide a
standardized and flexible language for logical identifiers. They support the unambiguous
identification of all meaningful components of modular theories and provide an abstraction
layer over physical locations. An \defemph{XML catalog} can translate URIs into their
physical locations represented as URLs.

A common feature in implementations of formal languages is a distinction between
\defemph{internal} and \defemph{external} syntax. The latter is more relaxed in order to ease reading and writing for humans, whereas the
latter is stricter and fully disambiguated to ease machine-processing. A
\defemph{reconstruction} algorithm is used to obtain the internal representation from the
external one. For programming languages, this is usually called
\defemph{compilation}. Typical steps of the reconstruction algorithm are parsing of infix
operators using precedences, disambiguation of overloaded symbol names, inference of
omitted types, and automated proof search to discharge incurred proof
obligations. Moreover, often the internal syntax is non-modular, and the reconstruction
includes the elaboration or flattening.

If different systems are to communicate mathematical knowledge, a complex reconstruction algorithm can be problematic. If internal syntax is communicated, human-oriented information is lost; when external syntax is communicated, the receiving system must implement the costly reconstruction. Therefore, we speak of \defemph{authoring-oriented} languages if the reconstruction algorithm is complex and of \defemph{interchange-oriented} languages if it is simple (or even the identity).

%
%

We speak of \defemph{incremental} processing if modular expressions can be processed
step-wise. We say that a language is \defemph{decomposable} if there is an algorithm that
decompose a modular declaration into a sequence of \defemph{atomic} declarations with an
acyclic dependency relation. We say that a language is \defemph{order-invariant} if the
semantics is independent of the order of declarations as as long as the order respects the
dependency relation. Any decomposable, order-invariant language permits streaming of
documents and optimized storage in databases.

The flattening (elaboration) operation is usually defined by induction on expressions and
leads to an exponential increase in size. We speak of \defemph{eager} flattening if every
induction step requires the recursive flattening of all sub-expressions. If we regard
flattening as the evaluation of a modular expression, this corresponds to call-by-value
evaluation. We speak of \defemph{lazy} flattening if a corresponding call-by-reference
evaluation is possible. In the latter case, the exponential blow-up may be avoided.


 \section{Central Features of MMT}\label{sec:mmt:overview}
     We will now discuss the central design goals that have guided the development of {\mmt} in
terms of the concepts introduced above. For other systems with different applications and
design choices see Sect.~\ref{sec:mmt:related}.

\paragraph*{A Generic Formal Module System} 

{\mmt} is a generic, formal module system for mathematical knowledge. It is designed to be
applicable to a large collection of declarative formal base languages, and all {\mmt}
notions are fully abstract in the choice of base language.

{\mmt} is designed to be applicable to all base languages based on
\defemph{theories}. Theories are modules in the sense of Sect.~\ref{sec:mmt:aspects}, in
the simplest case they are defined by a set of typed \defemph{symbols} (the signature) and
a set of \defemph{axioms} describing the properties of the symbols. A \defemph{signature
  morphism} $\sigma$ from a theory $S$ to a theory $T$ translates or interprets the
symbols of $S$ in $T$.

If we have entailment relations for the formulas of $S$ and $T$, a signature morphism is
particularly interesting if it translates all theorems of $S$ to theorems of $T$; this is
called a \defemph{theory morphism}. Using the \defemph{Curry-Howard representation},
{\mmt} drops the distinction between symbols and axioms and between signatures and
theories altogether, and only uses theories. Axioms are constants whose type is the
asserted proposition, and theorem are defined constants whose definiens is a proof.

The flat fragment of {\mmt} provides a generic syntax for theories and theory morphisms
(called \emph{views} in {\mmt}). A view from $\qS$ to $\qT$ is a list of
\defemph{assignments} $\maps{\qc}{\omega}$ where $\qc$ is an $\qS$-constant (axiom) and
$\omega$ is a $\qT$-term (proof). Such a list of assignments induces a
\defemph{homomorphic} translation of $\qS$-terms to $\qT$-terms by replacing every $\qc$
with the corresponding $\omega$. Such translations are often called \emph{structural},
\emph{recursive}, or \emph{compositional}.

Full {\mmt} adds the most general form of inheritance: interspersed named imports (called
\emph{structures} in {\mmt}) carrying free, explicit, and partial instantiations. In
particular, we choose named imports to avoid the problems caused by the diamond situation
and import name clashes, which occur frequently in large-scale developments.

{\mmt} has been designed in the tradition of the semi-formal {\omdoc} language, and an
extension of {\mmt} to cover informal knowledge is poised to culminate in a successor to
{\omdoc}. But in this paper, we will focus on the formal aspects only. We will nonetheless
discuss the relation to semi-formal languages below. To ensure machine-processing {\mmt}
uses a content-oriented representation building on {\openmath}~\cite{openmath} and
akin to {\omdoc}~\cite{omdoc}. We have designed and implemented an extension
of {\mmt} with notation definitions that transform {\mmt}-content representations into
presentation-oriented formats \cite{project:mmt}, but this will not be the focus of this
work.

\begin{figure}
\begin{center}
\begin{tikzpicture}
\node (Ob) at (-0.5,-2) {Object};
\node (Te) at (-1.5,-1) {Term};
\node (Mo) at (0.5,-1) {Morphism};
\node (C) at (-1.5,0) {$Con$};
\node (St) at (0.5,0) {$Str$};
\node (Sy) at (-0.5,1) {$Sym$};
\node (CA) at (2,0) {$ConAss$};
\node (StA) at (4,0) {$StrAss$};
\node (A) at (3,1) {$Ass$};
\node (T) at (-0.5,2) {$Thy$};
\node (L) at (0.5,2) {$Link$};
\node (V) at (3,2) {$View$};
\node (M) at (1.25,3) {$Mod$};
\node (D) at (1.25,4) {$Doc$};

\draw[dotted,-\arrowtip] (C) -- (Te);
\draw[dotted,-\arrowtip] (L) .. controls (1,1) and (1,0) .. (Mo);
\draw[dashed,-\arrowtip] (C) -- (Sy);
\draw[dashed,-\arrowtip] (St) -- (Sy);
\draw[dashed,-\arrowtip] (CA) -- (A);
\draw[dashed,-\arrowtip] (StA) -- (A);
\draw[dashed,-\arrowtip] (T) -- (M);
\draw[dashed,-\arrowtip] (V) -- (M);
\draw[dashed,-\arrowtip] (St) -- (L);
\draw[dashed,-\arrowtip] (V) -- (L);
\draw[dashed,-\arrowtip] (Te) -- (Ob);
\draw[dashed,-\arrowtip] (Mo) -- (Ob);
\draw[-\arrowtip] (Sy) -- (T);
\draw[-\arrowtip] (A) -- (L);
\draw[-\arrowtip] (M) -- (D);

\draw[dashed,-\arrowtip] (-5,3.5) -- (-4,3.5);
\node[right] at (-4,3.5) {subconcept of};
\draw[-\arrowtip] (-5,3) -- (-4,3);
\node[right] at (-4,3) {declared in};
\draw[dotted,-\arrowtip] (-5,2.5) -- (-4,2.5);
\node[right] at (-4,2.5) {used to form};
\end{tikzpicture}
\caption{{\mmt} Ontology}\label{fig:mmt:ontology}
\end{center}
\end{figure}
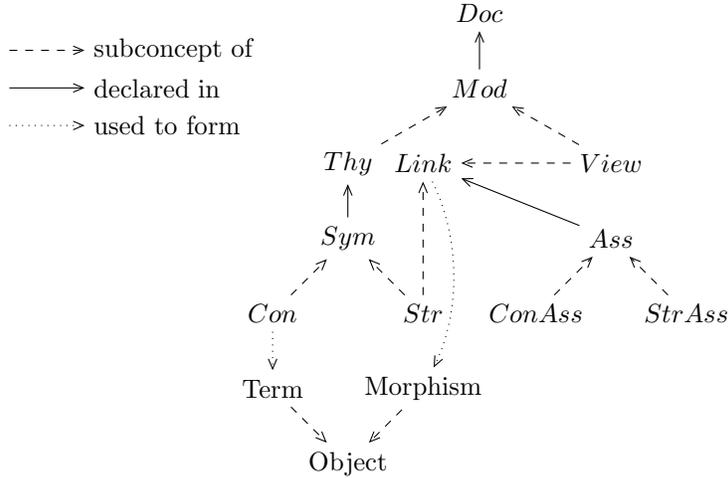

\paragraph*{A Simple Ontology}
A scalable module system must be both expressive and simple, which forms a difficult trade-off. Therefore, {\mmt} carefully picks only a few primitive language features: The ontology of {\mmt} language features is so simple that it can be visualized in a single graph, see Fig.~\ref{fig:mmt:ontology}. {\mmt} concepts are distinguished into four levels: the document, module, symbol, and object level.

Expressions at the document level are the \defemph{documents}, which act as
packages. {\mmt} systematically follows the intuition that documents are transparent to
the semantics. Therefore, scalable knowledge management services can be implemented easily
at the document level. Documents are open packages -- every document may refer to every
other document as long as the dependency relation is acyclic -- and the distribution of
modules into documents is transparent. Logical identifiers are used for all knowledge
items and are given as {\mmt} URIs, and the translation of URIs into URLs is relegated to
an extra-linguistic catalog; thus, {\mmt} documents provide namespace management and
abstract from physical locations.


Documents contain \defemph{modules}, and {\mmt} uses only two kinds of module
declarations: \defemph{theories} and \defemph{views}.
{\mmt} does not need other module declarations because both grounded realizations and functors can be represented as views.
Most declarative languages, can be stated naturally as a category. The objects are sets of declarations and are represented as {\mmt} theories. And the morphisms are translations between theories, which are represented as {\mmt} views.

More precisely, {\mmt} theories contain \defemph{symbol declarations}, and views contain \defemph{symbol
  assignments}. A view from theory $S$ to theory $T$ must realize all $S$-symbols in terms
of $T$-objects. Consequently, for every kind of symbol declaration, there is a
corresponding kind of objects.  {\mmt} uses only two kinds of symbol declarations:
\defemph{Constants} represent all declarations of the base language, and
\defemph{structures} represent inheritance between theories (see below). A constant
assignment provides a $T$-term for an $S$-constant, and structure assignments provide a
$T$-morphism for an $S$-structure.

Objects are complex expressions that represent mathematical expressions, formulas,
etc. {\mmt} only uses two kinds of objects: terms and morphisms. Constants occur as the
atomic terms, and structures and views as the atomic morphisms.  The grammar for
\defemph{terms} is motivated by the {\openmath} grammar \cite{openmath}. It uses generic
constructs for application and binding to form complex terms in a way that is general
enough to represent most mathematical languages. {\mmt} achieves this by relegating the
semantics of terms to a foundation (see below).

\defemph{Morphisms} from $S$ to $T$ are realizations of $S$ over $T$. We take the concept
of \defemph{links} from development graphs~\cite{devgraphs} to unify the two atomic
morphisms: Structures are morphisms induced by imports, views are morphisms declared (and
proved) explicitly. Complex morphisms are formed by composition. The representation of
realizations as morphisms has the advantage that {\mmt} can easily provide concrete syntax
for the two translations induced by a realization: The syntactic translation is given by
applying morphisms to terms, and the semantic translation by composition of
morphisms. Thus, {\mmt} can capture the semantics of realizations while being parametric
in the semantics of terms.

\paragraph*{A Simple Semantics using Theory Graphs}
The semantics of a collection of {\mmt} documents is given as a \defemph{theory graph},
which serves as a compact specification of a collection of mathematical theories and their
relations. The nodes of a theory graph are the theories; the edges are the links. Each
path in a theory graph yields a theory morphism. In particular, if a declarative language is given as a category whose components are represented as {\mmt} theories and morphisms, then diagrams in that category are represented as {\mmt} theory graphs. It
is a crucial observation that theory graphs are universal in the sense that they arise
naturally and in the same way in any declarative language. Using theory graphs, {\mmt} can
capture the semantics of modular theories generically.

\begin{figure}[hpt]
\begin{center}
\begin{tikzpicture}
\begin{scope}[inner sep=0pt]
\node[thy] (Mon) at (0,0) {$\mathll[c]{\cn{Monoid}\\ \color{gray}\cn{comp},\;\cn{unit}}$};
\node[thy] (CGr) at (0,3) {$\mathll[c]{\cn{CGroup}\\ \color{gray}\cn{mon},\;\cn{inv}}$};
\node[thy] (Ring) at (-3,3) {$\mathll{\cn{Ring}\\\color{gray}\cn{add}\\\color{gray}\cn{mult}}$};
\node[thy] (Int) at (3,3) {$\mathll[c]{\cn{integers}\\ \color{gray}0,+,-}$};
\node[draw,dashed,rounded corners] (v2) at (1,4.5) {
 $\mathll{
   \cn{v2} \\
   \color{gray}
   \begin{array}{l|l}
     \mathll{
             \maps{\cnpath{mon,comp}}{+} \\
             \maps{\cnpath{mon,unit}}{0}
     } &
     \maps{\cn{mon}}{\cn{v1}} \\
     \maps{\cn{inv}}{-} & \maps{\cn{inv}}{-}
   \end{array}
 }$
};
\end{scope}
\draw[struct](Mon) -- node[left,near end]{$\color{gray}\cn{mon}$} (CGr);
\draw[struct](CGr) -- node[above]{$\color{gray}\cn{add}$}(Ring);
\draw[struct](Mon) -- node[left]{$\color{gray}\cn{mult}$} (Ring);
\draw[view](Mon) -- node[right=1cm,draw,dashed,rounded corners]{
  $\mathll{
    \cn{v1}\\
    \color{gray} \maps{\cn{comp}}{+} \\
    \color{gray} \maps{\cn{unit}}{0}
  }$
} (Int);
\draw[view](CGr) -- node[above]{$\cn{v2}$}(Int);
\draw[struct] (1.5,.25) -- +(.8,0);
\draw[view] (1.5,-.25) -- +(.8,0);
\node[right] at (2.3,.25) {structure};
\node[right] at (2.3,-.25) {view};
\end{tikzpicture}
\caption{A Theory Graph for Elementary Algebra}\label{fig:mmt:syntax:algebra_example}
\end{center}
\end{figure}
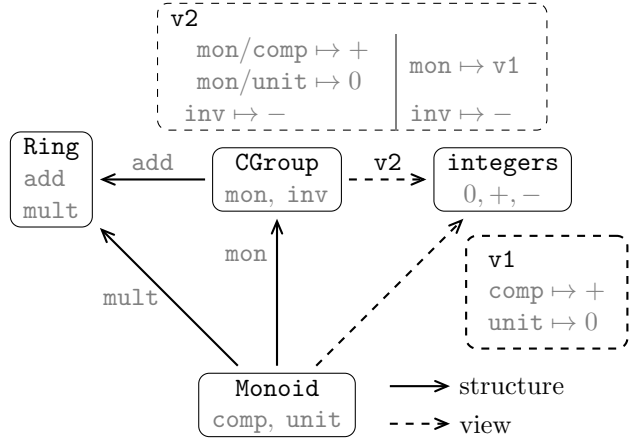

\begin{example}[Running Example: Elementary Algebra]\label{ex:mmt:algebra}
  For a simple example, consider the theory graph on the right with nodes for the theories
  of monoids, 
\begin{wrapfigure}{r}{3.7cm}\footnotesize
\vspace{-1em}
\fbox{\begin{tikzpicture}[xscale=.4,yscale=.4]
\node[thy] (mon) at (7,2.5) {$\cn{Monoid}$};
\node[thy] (cgr) at (7,5) {$\cn{CGroup}$};
\node[thy] (ring) at (1,5) {$\cn{Ring}$};
\draw[struct](mon) -- node[left] {$\cn{mon}$} (cgr);
\draw[struct](mon) -- node[below] {$\cn{mult}$} (ring);
\draw[struct](cgr) -- node[above] {$\cn{add}$}(ring);
\end{tikzpicture}}
\vspace{-1em}
\end{wrapfigure}
commutative groups, and rings, and three structures between them. The theory
  $\cn{Monoid}$ might declare symbols for composition and unit, and axioms
  for associativity and neutrality. The theory of commutative groups is an extension of
  the theory of monoids: it arises by adding symbols and axioms to
  $\cn{Monoid}$. Therefore, we only need to represent those added symbols and axioms in
  $\cn{CGroup}$ and add a structure $\cn{mon}$ importing from $\cn{Monoid}$.

  Fig.~\ref{fig:mmt:syntax:algebra_example} gives a more detailed view of the theory graph
  adding the symbols in the theory nodes, but eliding the axioms.
  $\cn{Ring}$ declares two structures for addition and multiplication, and the distinguish-semantics yields two different monoid operations for addition and multiplication.
  
  Our running example shows a slight complication in the case of first-order logic: We can declare a symbol for the first-order universe either in $\cn{Monoid}$ or in the theory $\cn{FOL}$, which we will introduce in Ex.~\ref{ex:mmt:elalg-meta}. Both choices are justified, and we will assume the latter for the sake of our example.
\end{example}

Structures in {\mmt} are always named and the distinguish-semantics is used in the case of diamonds. Qualified identifiers for the imported constants are formed by concatenating the structure name and the name of the imported symbols.
For example, the theory $\cn{Ring}$ from Fig.~\ref{fig:mmt:syntax:algebra_example} can access the symbols $\cnpath{add,mon,comp}$ (addition), $\cnpath{add,mon,unit}$ (zero), $\cnpath{add,inv}$ (additive inverse), $\cnpath{mult,comp}$ (multiplication), and $\cnpath{mult,unit}$ (one).

Both structures and views from $\qS$ to $\qT$ are defined by a list of assignments
$\sigma$ that assigns $\qT$-objects to $\qS$-symbols, and both induce theory morphisms
from $\qS$ to $\qT$ that map all $\qS$-objects to $\qT$-objects. This can be utilized to
obtain the identify-semantics: sharing declarations are special cases of
assignments in structures.

\begin{example}[Sharing via Instantiation]
Consider Ex.~\ref{ex:mmt:algebra} but with the change that we declare a symbol $\cn{univ}$ in $\cn{Monoid}$ for the first-order universe. Then $\cn{univ}$ must be shared between the two imports from $\cn{Monoid}$ to $\cn{Ring}$.
We obtain an asymmetric sharing declaration by first declaring the import $\cn{add}$ and then adding the assignment $\maps{\cn{univ}}{\cnpath{add,mon,univ}}$ to the structure $\cn{mult}$ in order to
identify the two copies of the universe.
Alternatively, we can give a symmetric sharing declaration by declaring $\cn{univ}$ in $\cn{Ring}$ as well and adding the assignment $\maps{\cn{univ}}{\cn{univ}}$ to both structures.

We will see that whole structures can be shared in the same way.
\end{example}

While a view relates two fixed theories without changing either one, structures from $\qS$
to $\qT$ occur within $\qT$ and change $\qT$ by including a copy of $\qS$. Thus,
structures induce theory morphisms by definition, and views correspond to representation
theorems.

\begin{example}[Views (continued from Ex.~\ref{ex:mmt:algebra})] \label{ex:mmt:algebra-integers}
  The node on the right side of the graph in Fig.~\ref{fig:mmt:syntax:algebra_example}
  represents a theory for the integers declaring the constants $0$, $+$, and $-$. The fact
  that the integers are a monoid is represented by the view $\cn{v1}$. It is a theory
  morphism that explicitly gives the interpretations of all symbols: $\maps{\cn{comp}}{+}$ and
  $\maps{\cn{unit}}{0}$. If we did not omit axioms, this view would also have to interpret
  all the axioms of $\cn{Monoid}$ as proof terms.

  The view $\cn{v2}$ is particularly interesting because there are two ways to represent
  the fact that the integers are a commutative group. In the first variant, all constants
  of $\cn{CGroup}$ are interpreted separately: $\cn{inv}$ as $-$ and the two imported
  constants $\cnpath{mon,comp}$ and $\cnpath{mon,unit}$ as $+$ and $0$, respectively. In
  the second variant $\cn{v2}$ is constructed modularly by importing the existing view
  $\cn{v1}$: The {\mmt} structure assignment $\maps{\cn{mon}}{\cn{v1}}$ maps all symbol
  imported by $\cn{mon}$ according to $\cn{v1}$. The intuition behind a structure
  assignment is that it makes the right triangle commute: $\cn{v2}$ is defined such that
  $\cn{v2}\,\circ\,\cn{mon}=\cn{v1}$. Clearly, both variants lead to
  the same theory morphism; the second one is conceptually more complex but eliminates
  redundancy because it is structured.
\end{example}

\paragraph*{Partial Morphisms}
The assignments defining a structure may be (and typically are) partial whereas a view
should be total. In order to treat structures and views uniformly, we admit
\defemph{partial views} as well. This is not only possible, but in fact desirable. A
typical scenario when working with views is that some of the specific assignments making
up the view constitute proof obligations and must be found by costly
procedures. Therefore, it is reasonable to represent partial views, namely views where
some proof obligations have already been discharged whereas others remain open.

\begin{example}[Partial Morphisms (continued from Ex.~\ref{ex:mmt:algebra-integers})]
\label{ex:mmt:algebra-partial}
  Consider for instance the situation in Fig.~\ref{fig:mmt:syntax:algebra_example} but
  this time taking axioms into account. Recall that under the Curry-Howard correspondence,
  axioms are just symbols whose types is given by the asserted formula. So we would have additional constants $\cn{assoc}$ and $\cn{neut}$ for
  associativity and the properties of the neutral element in $\cn{Monoid}$, the constants
  $\cn{inv\_ax}$ and $\cn{comm}$ for the properties of the inverse element and commutativity in
  $\cn{CGroup}$, and finally the constant $\cn{dist}$ for distributivity in $\cn{Ring}$.

  Thus, the views $\cn{v1}$ and $\cn{v2}$ are clearly partial views, and the missing
  assignments for $\cn{assoc}$ and $\cn{neut}$ in $\cn{v1}$ and for $\cn{inv\_ax}$ and
  $\cn{comm}$ in $\cn{v2}$ are proof obligations that need to be discharged by proving the
  translated axioms in theory $\cn{integers}$. If these proof terms are known, they can be added to the views as assignments to the respective (axiom) constants. In this situation, the structured view $\cn{v2}$ shows its strength: It imports the constant assignments from $\cn{v1}$ that discharge proof obligations so that these proofs do not have to be repeated.
\end{example}

Partial morphisms also arise when representations are inherently partial. For example, we can give a one-sided inverse to the structure $\cn{mon}$ in Fig.~\ref{fig:mmt:syntax:algebra_example} by mapping $\cnpath{mon,comp}$ and $\cnpath{mon,unit}$ to $\cn{comp}$ and $\cn{unit}$.

{\mmt} introduces \defemph{filtering} to obtain a semantics for partial morphisms: All
constants for which a view does not provide an assignment are implicitly filtered, i.e.,
are mapped to a special term $\hid$. If a link $\ql$ from $\qS$ to $\qT$ filters a $\qS$-constant that has a definiens, this is harmless because the filtered constant can be replaced with its definiens. But if undefined constants are filtered, {\mmt} enforces the strictness of filtering: All terms depending on a filtered constant, are also filtered. In that case, we speak of filtered terms, which are also represented by $\hid$.

\paragraph*{A Foundation-Independent Semantics}
Mathematical knowledge is described using very different foundations. Most of them can be
grouped into set theory and type theory. Within each group there are numerous variants,
e.g., Zermelo-Fraenkel~\cite{zermelo,fraenkel} or G\"odel-Bernays set
theory~\cite{goedelsettheory,bernays}, or set theories with or without the axiom of
choice. Therefore, scalability across semantic domains requires a foundation-independent
representation language. It is a unique feature of {\mmt} to provide such a high level of
genericity and still be able to give a rigorous semantics in terms of theory graphs and a
foundation-independent \defemph{flattening theorem}.

The semantics of {\mmt} is given proof theoretically by flattening in order to avoid a
commitment to a particular model theory. This also makes {\mmt} conservative over the base
language so that we can combine {\mmt} with arbitrary base languages without affecting
their semantics. Therefore, we have to exclude non-conservative language features, but we
have shown in \cite{rabe:combining:10,HR:folsound:10} that despite the proof theoretical
semantics of {\mmt}, model theoretical module systems can be represented in
{\mmt}. Moreover, we have given an extension of {\mmt} with hiding in
\cite{CHKMR:hiding:11}.

Foundation-independence is achieved by representing all logics, logical frameworks, and
the foundational languages themselves simply as theories. For example, an {\mmt} theory
graph based on ZFC set theory starts with a theory that declares the symbols of ZFC such
as $\in$ and $\subset$. Moreover, {\mmt} does not prescribe a set of well-typed
terms. Instead, {\mmt} uses generic term formation operators, and any term may occur as
the type of any other
term. 

We recover this loss of precision by formalizing the notion of \emph{meta-languages},
which pervades mathematical discourse. Let us write $M/T$ to express that we work in the
object language $T$ using the meta-language $M$. For example, most of mathematics is
carried out in $\cn{FOL}/\cn{ZFC}$, i.e., first-order logic is the meta-language, in which
set theory is defined. $\cn{FOL}$ itself might be defined in a logical framework such as
$LF$~\cite{lf}, and within $ZFC$, we can define the language of natural numbers, which
yields $\cn{LF}/\cn{FOL}/\cn{ZFC}/\cn{Nat}$. In {\mmt}, all of these languages are
represented as theories.  In many ways $M/T$ behaves like an import from $M$ to $T$, but
using only an import would fail to describe the meta-relationship. Therefore, {\mmt} uses
a binary \defemph{meta-theory} relation between theories.

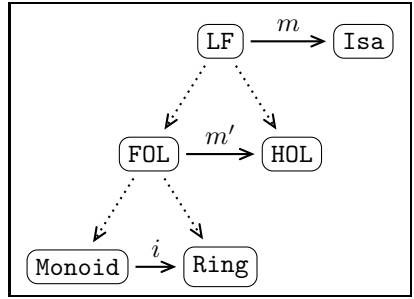
\begin{wrapfigure}r{5.2truecm}
\fbox{
\begin{tikzpicture}[xscale=.95]
\node[thy] (A) at (0,3)  {$\cn{LF}$};
\node[thy] (A') at (2,3)  {$\cn{Isa}$};
\node[thy] (C) at (-1,1.5)   {$\cn{FOL}$};
\node[thy] (C') at (1,1.5) {$\cn{HOL}$};
\node[thy] (E) at (-2,0) {$\cn{Monoid}$};
\node[thy] (E') at (0,0)  {$\cn{Ring}$};
\draw[meta](A) -- (C);
\draw[meta](A) -- (C');
\draw[meta](C) -- (E);
\draw[meta](C) -- (E');
\draw[struct](A) --node[above] {$m$} (A');
\draw[struct](C) --node[above] {$m'$} (C');
\draw[struct](E) --node[above] {$i$} (E');
\end{tikzpicture}
}
\caption{Meta-Theories}\label{fig:mmt:intro:metatheory}\vspace*{-1em}
\end{wrapfigure}
In the example in Fig~\ref{fig:mmt:intro:metatheory} and generally in this paper, the
meta-theory relation is visualized using dotted inclusion morphisms. The theory $\cn{FOL}$
for first-order logic is the meta-theory for $\cn{Monoid}$ and $\cn{Ring}$. And the theory
$\cn{LF}$ for the logical framework LF is the meta-theory of $\cn{FOL}$ and the theory
$\cn{HOL}$ for higher-order logic. Note how the meta-theory can indicate both to humans
and to machines how $T$ is to be interpreted. For example, interpretations of
$\cn{Monoid}$ are always stated relative to a fixed interpretation of $\cn{FOL}$.

The importance of meta-theories $M/T$ in {\mmt} is that $M$ defines the semantics of
$T$. More precisely, a \defemph{foundational theory} declares all primitive concepts and
axioms of the foundational language and occurs as the upper-most meta-theory -- like
$\cn{LF}$ and $\cn{Isabelle}$ in the example in Fig~\ref{fig:mmt:intro:metatheory}. The
semantics of the foundational theory is called the \defemph{foundation}; it is given
externally and assumed by {\mmt}, and it induces the semantics of all other theories.
Formally, {\mmt} assumes that the foundation for the foundational theory $M$ defines
typing and equality judgments for arbitrary theories $T$ with (possibly indirect)
meta-theory $M$.

The choice of typing and equality is motivated by their universal importance in the formal
languages of mathematics and computer science. Here we should clarify that, from an {\mmt}
perspective, languages like untyped set theory are in fact typed languages, if only
coarsely-typed: For example, typical formalizations of set theory at least distinguish
types for sets, propositions, and proofs, and a concise definition of axiom schemes
naturally leads to a notion of function types.

\begin{example}[Meta-Theories (continued from Ex.~\ref{ex:mmt:algebra-partial})]
\label{ex:mmt:elalg-meta}
\begin{wrapfigure}r{6.3cm}\vspace*{-1em}
\fbox{\begin{tikzpicture}[xscale=1.2,yscale=1.5]
  \node[thy] (folsyn) at (0,1) {$\cn{FOL}$};
  \node[thy] (zfc) at (3,1) {$\cn{ZFC}$};
  \node[thy] (ring) at (-1,0) {$\cn{Ring}$};
  \node[thy] (monoid) at (1,-1) {$\cn{Monoid}$};
  \node[thy] (cgroup) at (1,0) {$\cn{CGroup}$};
  \node[thy] (int) at (3,0) {$\cn{Integer}$};
  
  \draw[struct] (monoid) -- node[right]{$\cn{mon}$} (cgroup);
  \draw[struct] (monoid) -- node[left]{$\cn{mult}$} (ring);
  \draw[view] (monoid) -- node[right]{$\cn{v1}$} (int);
  \draw[struct] (cgroup) -- node[above]{$\cn{add}$} (ring);
  \draw[view] (cgroup) -- node[above]{$\cn{v2}$} (int);
  \draw[meta] (folsyn) -- (monoid);
  \draw[meta] (folsyn) -- (ring);
  \draw[meta] (folsyn) -- (cgroup);
  \draw[meta] (zfc) -- (int);
  \draw[view] (folsyn) -- node[above]{$\cn{FOLSem}$} (zfc);
 \end{tikzpicture}}\vspace*{-1em}
\end{wrapfigure}
We can add meta-theories by adding a theory $\cn{FOL}$ for first-order logic, which occurs
as the meta-theory of monoids, groups, and rings. In particular, $\cn{FOL}$ declares symbols
for the first-order universe and the connectives and quantifiers.
We use a theory for $\cn{ZFC}$ as the meta-theory of the integers. In that case the views $\cn{v1}$ and $\cn{v2}$ are only meaningful relative to an interpretation of first-order logic in set theory. In {\mmt},
this interpretation is given as a view $\cn{FOLSem}$ from $\cn{FOL}$ to $\cn{ZFC}$ which is
attached to $\cn{v1}$ and $\cn{v2}$ as a meta-morphism.
$\cn{FOLSem}$ represents the inductive interpretation function that defines the semantics of first-order logic in set theory.
\end{example}

\paragraph*{Little Logics and Little Foundations}
The little theories methodology~\cite{littletheories} strives to state every mathematical
theorem in the theory with the smallest possible set of axioms in order to maximize
theorem reuse. Using the \defemph{foundations-as-theories} approach of, we can extend it
to the \defemph{little logics} and \defemph{little foundations} methodology.

{\mmt} provides a uniform module system for theories, logics, and foundational
languages. Thus, we can use structures to represent inheritance at the level of logical
foundations and views to represent formal translations between them. For example, the
morphisms $m$ and $m'$ in Fig.~\ref{fig:mmt:intro:metatheory} indicate possible
translations on the levels of logical frameworks and logics, respectively. Therefore, just
like in the little theories approach, we can prove meta-logical results in the simplest
logic or foundation that is expressive enough and then use views to move results between
foundations.

\begin{example}[Proof and Model Theory of First-Order Logic]\label{ex:mmt:fosem}
In~\cite{HR:folsound:10}, we formalize the syntax, proof theory, and model theory and
prove the soundness of first-order logic in $\cn{MMT}/\cn{LF}$. We use the theory graph given in the commutative diagram on the right. We represent the syntax -- i.e., the connectives and quantifiers -- in the theory $\cn{FOLSyn}$. (This theory was called $\cn{FOL}$ in Ex.~\ref{ex:mmt:elalg-meta}.) For the proof theory, the theory $\cn{FOLPf}$ imports $\cn{FOLSyn}$ and adds constants for the rules of a calculus for first-order logic encoded via the Curry-Howard correspondence.

For the definition of the model theory in $\cn{FOLMod}$, we should use set theory as the meta-theory.
However, doing proofs in set theory as needed for the soundness proof is tedious. Therefore, we use higher-order logic $\cn{HOL}$ as the meta-theory of $\cn{FOLMod}$; it is expressive enough to carry out the soundness proof but permits typed reasoning.
The syntax is interpreted in the model theory by a view $\cn{FOLSem1}$.

Then the view $\cn{refine}$ from $\cn{HOL}$ to $\cn{ZFC}$ proves that $\cn{ZFC}$ is a refinement of $\cn{HOL}$. We reuse $\cn{refine}$ to give a morphism $\cn{FOLSem2}$ that interprets $\cn{FOLMod}$ in $\cn{ZFC}$. The composition of $\cn{FOLSem1}$ and $\cn{FOLSem2}$ yields the view $\cn{FOLSem}$ from Ex.~\ref{ex:mmt:elalg-meta}.

Finally the soundness proof -- which shows that all proof terms over $\cn{FOLPf}$ induce valid statements over $\cn{FOLMod}$ -- is represented as the view $\cn{sound}$. $\cn{sound}$ is given as a structured view: It imports the view $\cn{FOLSem1}$ using the structure assignment $\maps{\cn{syn}}{\cn{FOLSem1}}$.

\begin{wrapfigure}{r}{6.1cm}\vspace*{-1em}
\fbox{\begin{tikzpicture}[yscale=1.2]
  \node[thy] (folsyn) at (-1,3) {$\cn{FOLSyn}$};
  \node[thy] (folpf) at (1,4) {$\cn{FOLPf}$};
  \node[thy] (folpf2) at (3.5,4) {$\cn{FOLPf}$};
  \node[thy] (foldesc) at (3.5,2.5) {$\cn{FOLd}$};
  \node[thy] (folmod) at (1,2) {$\cn{FOLMod}$};
  \node[thy] (hol) at (1,1) {$\cn{HOL}$};
  \node[thy] (zfc) at (3.5,1) {$\cn{ZFC}$};
  \draw[meta] (foldesc) -- (zfc);
  \draw[struct] (folpf2) -- node [left] {$\cn{fol}$} (foldesc);
  \draw[struct] (folsyn) -- node [left] {$\cn{syn}$} (folpf);
  \draw[view] (folsyn) -- node [left] {$\cn{FOLSem1}$} (folmod);
  \draw[view] (folpf) -- node [right] {$\cn{sound}$} (folmod);
  \draw[meta] (hol) -- (folmod);
  \draw[view] (hol) -- node [below] {$\cn{refine}$} (zfc);
  \draw[view] (folmod) -- node [very near start,right=.2cm] {$\cn{FOLSem2}$} (zfc);
\end{tikzpicture}}
\vspace*{-1em}
\end{wrapfigure}

To establish the views into $\cn{ZFC}$, it must have proof rules of its own. We use a variant of first-order logic as the meta-theory of $\cn{ZFC}$, namely $\cn{FOLd}$. It arises by importing $\cn{FOLPf}$ and then adding a description operator $\iota$.
This yields two morphisms from $\cn{FOLPf}$ to $\cn{ZFC}$: one via the import $\cn{fol}$ and the meta-theory relation, and one as the composition of $\cn{sound}$ and $\cn{FOLSem2}$.
These morphisms are not equal: If $\cn{o}$ is the type of formulas in $\cn{FOLsyn}$, then the former morphism maps $\cn{o}$ to the type of propositions of $\cn{ZFC}$; but the latter maps $\cn{o}$ to the set of boolean truth values.
Note that to express this difference in our commutative diagram, we must use two copies of the node $\cn{FOLPf}$.

Moreover, in \cite{HR:folsound:10}, all theories have $\cn{LF}$ as the ultimate meta-theory, which we omitted from the diagram on the right.
In addition, \cite{HR:folsound:10} gives all theories and views using our little logics approach, e.g., using separate theories for each connective. Thus, we can reuse these fragments to define other logics as we do in \cite{project:latin}.
\end{example}

\paragraph*{Built-in Web-Scalability}
Most module systems in mathematics and computer science are designed with the implicit assumption that all theories of a graph are retrieved from a single file system or server and are processed by loading them into the working memory of a single process. These assumptions are becoming increasingly unrealistic in the face of the growing size of both mathematical knowledge and formalized mathematical knowledge. Moreover, this mathematical knowledge is represented in different formal languages, which are processed with different implementations.

{\mmt} is designed as a representation language that scales well to large inter-linked document collections that are processed with a wide variety of systems across networks and implementation languages. Therefore, {\mmt} offers integration support through web standards-compliance, incremental processing of large theory graphs, and an interchange-oriented fully disambiguated external syntax.

Scalable transport of {\mmt} documents must be mediated by standardized protocols and
formats. While the use of XML as concrete syntax is essentially orthogonal to the language
design, the use of \defemph{URIs as identifiers} is not because it imposes subtle
constraints that can be hard to meet a posteriori. In {\mmt}, all constants, including
imported ones, that are available in a theory have canonical URIs. {\mmt} uses tripartite
URIs $\triple{doc}{mod}{sym}$ formed from a document URI $doc$, a module name $mod$, and a
qualified symbol name $sym$. For example, if the theory graph from
Fig.~\ref{fig:mmt:intro:metatheory} is given in a document with URI
\lstinline|http://cds.omdoc.org/mmt/paper/example|, then the constant $\cn{unit}$ imported
from $\cn{Monoid}$ into $\cn{CGroup}$ has the URI
\lstinline|http://cds.omdoc.org/mmt/paper/example?CGroup?mon/unit|.

Note that theories are containers for declarations, and relations between theories define the declarations that are available
in a given theory. Therefore, if every available constant has a canonical identifier, the syntax of identifiers is inherently connected to the possible relations between theories. Consequently, and maybe surprisingly, defining the canonical identifiers is almost as difficult as defining the semantics of the whole language.

All {\mmt} definitions and algorithms are designed with incremental processing in mind. In particular, {\mmt} is decomposable and order-invariant. For example, the declaration $T=\{s_1:\tau_1,\;s_2:\tau_2\}$ of a theory $T$ with two typed symbols yields the atomic declarations $T=\{\}$, $T?s_1:\tau$, and $T?s_2:\tau_2$. Documents, views, and structures are decomposed accordingly. This ``unnesting'' of declarations is possible because every declaration has a canonical URI so that declarations can be taken out of context for transport and storage and re-assembled later. 

The understanding of structures and their induced declarations is crucial to achieve web-scalability. Languages with imports and instantiations tend to be much more complex than flat ones making them harder to specify and implement. Therefore, the semantics of modularity must often remain opaque to generic knowledge management services, an undesirable situation. Because {\mmt} has a simple and foundation-independent flattening semantics, modularity can be made transparent whenever a system is unable to process it.

Moreover, the flattening of {\mmt} is lazy: Every structure declaration can be eliminated individually without recursively flattening the imported theory. Thus, systems gain the flexibility to flatten {\mmt} documents partially and on demand.

 \section{Syntax}\label{sec:mmt:syntax}
   We will now develop the abstract syntax of {\mmt}, our formal module system that realizes the features described in the last section. We introduce the syntax in Sect.~\ref{sec:mmt:syntax:main}. Then we use {\mmt} to give a precise definition of the concept of ``realizations'' in Sect.~\ref{sec:mmt:foundations}. In Sect~\ref{sec:mmt:syntax:lookup} and~\ref{sec:mmt:syntax:normalize}, we introduce auxiliary functions for lookup and normalization that are used to talk about {\mmt} theory graphs.


   \subsection{{\mmt} Theory Graphs}\label{sec:mmt:syntax:main}
     The {\mmt} syntax for theory graphs distinguishes the {\defemph{module}},
{\defemph{symbol}}, and {\defemph{object}} level. We defer the description of the document
level to Sect.~\ref{sec:mmtweb} because documents are by construction transparent to the
semantics.

\subsubsection{Grammar}
\newcommand{\flatsyn}[1]{{\color{gray}#1}}
\begin{figure}[ht]
\begin{center}
\begin{tabular}{|llcl|}\hline
Theory graph  & $\TG$ & $\bnfas$ & $\cdot \bnfalt \TG,\;\mmtTth \bnfalt \TG,\;View$ \\
Theory        & $\mmtTth$& $\bnfas$ & $\thdeclm{\qT}{[\qM]}{\theta} \iv{\modexp}{\bnfalt \thdef{\qT}{\Theta}}$ \\
View          & $View$    &  $\bnfas$ & $\vwdeclm{\ql}{\qS}{\qT}{[\mu]}{\sigma} \bnfalt \vwdef{\ql}{\qS}{\qT}{\mu}$ \\
\hdashline
Theory body & $\theta$   & $\bnfas$ & $\cdot \bnfalt \theta,\;Con \bnfalt \flatsyn{\theta,\;Str}$\\
Constant    & $Con$      & $\bnfas$ & $\symdd{\qc}{\omega}{\omega} \bnfalt \symdd{\qc}{\omega}{}
                            \bnfalt    \symdd{\qc}{}{\omega} \bnfalt \symdd{\qc}{}{}$ \\
Structure   & $Str$      & $\bnfas$ & $\flatsyn{\impddm{\qi}{\qS}{[\mu]}{\sigma}} \bnfalt \flatsyn{\dimpdd{\qi}{\qS}{\mu}}$ \\
Link body   & $\sigma$   & $\bnfas$ & $\cdot \bnfalt \sigma,\;ConAss \bnfalt \flatsyn{\sigma,\;StrAss}$ \\
Ass. to constant  & $ConAss$ & $\bnfas$ & $\maps{\qc}{\omega}$ \\
Ass. to structure & $StrAss$ & $\bnfas$ & $\flatsyn{\maps{\qi}{\mu}}$ \\
\hdashline
Variable context & $\Ypsilon$ & $\bnfas$ & $\emptycon \bnfalt \Ypsilon,\;\yps[:\omega][=\omega]$ \\
Term   & $\omega$ & $\bnfas$
       & $\hid
  \bnfalt \spath{\qT}{\qc}
  \bnfalt \yps
  \bnfalt \flatsyn{\ma{\omega}{\mu}}
  \bnfalt \oma{\omega,\omega^+}$
  \\ &&& $
  \bnfalt \ombind{\omega}{\Ypsilon}{\omega}
  $\\
\iv{\modexp}{Theory & $\Theta$ & $\bnfas$ 
       & $\qT \bnfalt \poii{\Theta}{\Theta} \bnfalt \poi{\mu}{\Theta}{\Theta}$\\
}
Morphism & $\mu$ & $\bnfas$ 
       & $\mmtident{\qT}
  \bnfalt \ql
  \bnfalt \ö{\mu}{\mu}
\iv{\modexp}{  \bnfalt \poiim{\mu}{\mu} \bnfalt \poim{\mu}{\mu}{\mu} \bnfalt \poiw{\mu}{\Theta}}
$\\
\hline
Document identifier    & $g$        & $\bnfas$ & URI, no query, no fragment\\
Module identifier      & $\qS,\qT,\qM,\ql$  & $\bnfas$ & $\mpath{g}{\qI}$\\
Symbol identifier      &                    &  & $\mpath{\qT}{\qI}$\\
Local identifier       & $\qc,\qi,\qI$      & $\bnfas$ & $\uI [\localpathsep\uI]^+$\\
Names                  & $\uI,\yps$         & $\bnfas$ & pchar$^+$ \\
  & URI, pchar & & see RFC 3986~\cite{uri} \\
\hline
\end{tabular}
\caption{The Grammar for {\mmt} Expressions}\label{fig:mmt:grammar}
\end{center}
\end{figure}


The {\mmt} grammar is given in Fig.~\ref{fig:mmt:grammar} where $^+$, $|$, and $[-]$
denote non-empty repetition, alternative, and optional parts, respectively. Note that
several non-terminal symbols correspond directly to concepts of the {\mmt} ontology given
in Sect.~\ref{sec:mmt:overview}.  In order to state the flattening theorem below, we also
introduce the flat {\mmt} syntax; it arises by removing the productions given in gray.  We
will call a theory graph, module, or definiens {\defemph{flat}}, iff it can be expressed
in the flat {\mmt} syntax.

The meta-variables we will use are given in Fig.~\ref{fig:mmt:metavars}. References to named {\mmt} knowledge items are Latin letters, {\mmt} objects and lists of knowledge items are Greek letters. We will occasionally use $\_$ as an unnamed meta-variable for irrelevant values. 

In the following we describe the syntax of {\mmt} and its intended semantics in a bottom-up manner, i.e., identifiers, object level, symbol level, and module level. Alternatively, the following subsections can be read in top-down order. 

\begin{figure}[ht]
\begin{center}
\begin{tabular}{|l|l|l|}\hline
Level        & Declaration                       & Expression    \\\hline
Module       & theory $\qT,\qS,\qR,\qM$          & theory graph $\TG$ (set of modules)\\
             & link $\ql$                        & \\\hdashline
Symbol       & constant $\qc$                    & theory body $\theta$ (set of symbols)\\
						 & structure $\qh,\qi$               & link body $\sigma$ (set of assignments) \\\hdashline
Object       & variable $\yps$                   & term $\omega$ \\
             &                                   & morphism $\mu$ \\    
\hline
\end{tabular}
\caption{Meta-Variables}\label{fig:mmt:metavars}
\end{center}
\end{figure}

\subsubsection{Identifiers}

All {\mmt} identifiers are URIs and the productions for URIs given in RFC 3986~\cite{uri} are part of the {\mmt} grammar. We distinguish identifiers of documents, modules, and symbols.

Document identifiers $g$ are URIs without queries or fragments (The query and fragment components of a URI are those starting with the special characters ? and \#, respectively.).

Module identifiers are formed by pairing a document identifier $g$ with a local module identifier $\qI$ valid in that document. We use $\globalpathsep$ as a
separating character. Similarly, symbol identifiers $\mpath{\qT}{\qc}$ arise by pairing a
theory identifier with an local identifier valid in that theory.

Local identifiers may be qualified and are thus lists of names separated by $\localpathsep$. Finally, names are non-empty strings of pchars. pchar is defined in RFC 3986 and produces any Unicode character where certain reserved characters must be \%-encoded; reserved characters are ?/\#[]\@\% and all characters generally illegal in URIs.

\begin{example}[Continued from Ex.~\ref{ex:mmt:algebra}]\label{ex:mmt:algebra:consts}
We assume that the {\mmt} theory graph for the running example is located in a document with some URI $e$. Then the {\mmt} URIs of theories and views are for example $\mpath{e}{\cn{Ring}}$ and $\mpath{e}{\cn{v1}}$. The {\mmt} URIs of the constants available in the theory $\mpath{e}{\cn{Ring}}$ are
\begin{itemize}
 \item $\triple{e}{\cn{Ring}}{\cnpath{add,mon,comp}}$,
 \item $\triple{e}{\cn{Ring}}{\cnpath{add,mon,unit}}$,
 \item $\triple{e}{\cn{Ring}}{\cnpath{add,inv}}$,
 \item $\triple{e}{\cn{Ring}}{\cnpath{mult,comp}}$,
 \item $\triple{e}{\cn{Ring}}{\cnpath{mult,unit}}$.
\end{itemize}
The identifiers of structures are special because they may be considered both as symbol level and as module level knowledge items. This is reflected in {\mmt} by giving structures two identifiers. Consider the structure that imports $\cn{Monoid}$ into $\cn{CGroup}$: If we want to emphasize its nature as a declaration within $\cn{CGroup}$, we use the symbol identifier $\triple{e}{\cn{CGroup}}{\cn{mon}}$; if we want to emphasize its nature as a morphism, we use the module identifier $\mpath{e}{\cnpath{CGroup,mon}}$. Consequently, the non-terminal symbol $\ql$ for links may refer both to a view and to a structure (as expected).
\end{example}

\subsubsection{The Object Level}

Following the OpenMath approach, {\mmt} objects are distinguished into terms and
morphisms. \defemph{Terms} $\omega$ are formed from:
\begin{itemize}
\item {\defemph{constants}} $\spath{\qT}{\qc}$ referring to constant $\qc$ declared in theory $\qT$,
\item {\defemph{variables}} $\yps$ declared in an enclosing binder,
\item {\defemph{applications}} $\oma{\omega,\omega_1,\ldots,\omega_n}$ of $\omega$ to arguments $\omega_i$,
\item {\defemph{bindings}} $\ombind{\omega_1}{\Ypsilon}{\omega_2}$ by a binder $\omega_1$ of a list of variables $\Ypsilon$ with body $\omega_2$,
\ednote{MK: I like using $\beta$ better than outer brackets.
I reverted to $\beta$. Actually, I prefer (binder [context] scope) because that's how it looks with Twelf HOAS. But square brackets are already used in the grammar.}
\item {\defemph{morphism applications}} $\ma{\omega}{\mu}$ of $\mu$ to $\omega$,
\item a {\defemph{special term}} $\hid$ for filtered terms (see below).
\end{itemize}

Variable contexts are lists of variable declarations. Parallel to constant declarations, variables carry an optional type and an optional definiens. The scope of a bound variable consists of the types and definitions of the succeeding variable declarations and the body of the binder.

For every occurrence of a term, there is a \defemph{home theory} against which the term is checked. For occurrences in constant declarations, this is the containing theory. For occurrences in assignments, this is the codomain of the containing link.
We call a term $t$ that is well formed in a theory $\qT$ a
\defemph{term over} $\qT$. Terms over $\qT$ may use $\spath{\qT}{\qc}$ to refer to a previously declared
$\qT$-constant $\qc$. And if $\qi$ is a previously declared structure instantiating $\qS$, and $\qc$ is a constant
declared in $\qS$, then $\qT$ may use $\spath{\qT}{\qi,\qc}$ to refer to the copy of $\qc$
induced by $\qi$. Note that {\mmt} assumes that the declarations occur in an order that respects their dependencies; we will see later that the precise order chosen does not matter.
{\mmt} does not impose a specific typing relation between terms. In
particular, well-formed terms may be untyped or may have multiple types.

\begin{example}[Continued from Ex.~\ref{ex:mmt:algebra:consts}]\label{ex:mmt:algebra:axioms}
  The running example only contains constants. Complex terms arise when types and axioms
  are covered. For example, the type of the inverse in a commutative group is
  $\oma{\arr,\iota,\iota}$. Here $\arr$ represents the function type constructor and
  $\iota$ the carrier set. These two constants are not declared in the example. Instead,
  we will add them in Ex.~\ref{ex:mmt:synfull} by giving $\cn{CGroup}$ a meta-theory, in
  which these symbols are declared. A more complicated term is the axiom for
  left-neutrality of the unit:
\[\omega_e \;:= \;
\ombind{\forall}
       {\ombvar{\yps}{\iota}{}}
       {
          \oma{=,
             \oma{\triple{e}{\cn{Monoid}}{\cn{comp}}, \triple{e}{\cn{Monoid}}{\cn{unit}}, x},
             x
          }
       }
.\] 
Here $\forall$ and $=$ are further constants that are inherited from the meta-theory.
\end{example}

\defemph{Morphisms} are built up from links and compositions. If $\qi$ is a structure
declared in $\qT$ that imports from $\qS$, then $\lpath{\qT}{\qi}$ is a link from $\qS$ to $\qT$. Similarly, every view $\qm$ from $\qS$
to $\qT$ is a link. Composition is written $\ö{\mu}{\mu'}$ where $\mu$ is applied before
$\mu'$, i.e., composition is in diagrammatic order. The identity morphism of the theory
$\qT$ is written $\mmtident{\qT}$. A morphism application $\ma{\omega}{\mu}$ takes a term
$\omega$ over $\qS$ and a morphism $\mu$ from $\qS$ to $\qT$, and returns a term over
$\qT$.

Just like a structure declared in $\qT$ is both a symbol of $\qT$ and a link into $\qT$, a
morphism from $\qS$ to $\qT$ can be regarded as a composed object over $\qT$. To stress
this often fruitful perspective, we also call the codomain of a morphism its \defemph{home
  theory}, and the domain its \defemph{type}. Then morphism composition $\ö{\mu'}{\mu}$
can be regarded as the application of $\mu$ to $\mu'$: It takes a morphism $\mu'$ with
home theory $\qS$ and type $\qR$ and returns a morphism with home theory $\qT$ of the same
type.

\begin{example}[Continued from Ex~\ref{ex:mmt:algebra:axioms}]\label{ex:mmt:algebra:morphism}
In the running example, an example morphism is
 \[\mu_e\;:=\; \ö{\mpath{e}{\cnpath{CGroup,mon}}}{\mpath{e}{\cn{v2}}}.\]
 It has domain $\mpath{e}{\cn{Monoid}}$ and codomain $\mpath{e}{\cn{integers}}$. The intended semantics of the term $\ma{\omega_e}{\mu_e}$ is that it yields the result of applying $\mu_e$ to $\omega_e$, i.e.,
 \[\ombind{\forall}
          {\ombvar{\yps}{\iota}{}}
          {\oma{=,\oma{+,0,x},x}}.\]
Here, we assume $\mu_e$ has no effect on those constants that are inherited from the meta-theory. We will make that more precise below by using the identity as a meta-morphism.
\end{example}

We define a straightforward abbreviation for the application of morphisms to whole contexts:

\begin{definition}
We define $\ma{\Yps}{\mu}$ by
\[\ma{\cdot}{\mu}\;\; :=\;\; \cdot \tb \mand\tb
  \ma{\big(\Yps,\ombvar{\yps}{\tau}{\delta}\big)}{\mu} \;\;:= \;\;
  \ma{\Yps}{\mu},\;\ombvar{\yps}{\ma{\tau}{\mu}}{\ma{\delta}{\mu}}
\]
Here we assume $\ma{\undef}{\mu}=\undef$ to avoid case distinctions.
\end{definition}

The analogy between terms and morphisms is summarized in Fig.~\ref{fig:mmt:objects}.
\begin{figure}[ht]
\begin{center}\footnotesize
\begin{tabular}{|l|llll|}\hline
          & Atomic object  & Complex object   & Type           & Checked relative to\\\hline
Terms     & constant       & term             & term           & home theory\\
Morphisms & link           & morphism         & domain         & codomain\\ 
\hline
\end{tabular}
\caption{The Object Level}\label{fig:mmt:objects}
\end{center}
\end{figure}

\subsubsection{The Symbol Level}

We distinguish four symbol level concepts as given in Fig.~\ref{fig:mmt:statementlevel}: constants and structures, and assignments to them.

\begin{figure}[ht]
\begin{center}
\begin{tabular}{|l|ll|}\hline
          & Declaration   & Assignment  \\\hline
Terms     & of a constant $Con$  & to a constant $\maps{\qc}{\omega}$ \\
Morphisms & of a structure $Str$ & to a structure $\maps{\qi}{\mu}$ \\ 
\hline
\end{tabular}
\caption{The Symbol Level}\label{fig:mmt:statementlevel}
\end{center}
\end{figure}

A {\defemph{constant declaration}} of the form $\symdd{\qc}{\tau}{\delta}$ declares a
constant $\qc$ of type $\tau$ with definition $\delta$. Both the type and the definition
are optional yielding four kinds of constant declarations. If both are given, then
$\delta$ must have type $\tau$. In order to unify these four kinds, we will sometimes
write $\undef$ for an omitted type or definition.

Recall that via the Curry-Howard representation, a theorem can be declared as a constant with the asserted proposition as the type and the proof as the definiens. Similarly, (derived) inference rules are declared as (defined) constants.

A {\defemph{structure declaration}} of the form $\impddm{\qi}{\qS}{[\mu]}{\sigma}$ in a
theory $\qT$ declares a structure $\qi$ instantiating the theory $\qS$ defined by
assignments $\sigma$. Such structures can have an optional meta-morphism $\mu$ (see
below).  Alternatively, structures may be introduced as an abbreviation for an existing
morphism: $\dimpdd{\qi}{\qS}{\mu}$.  While the domain of a structure is given explicitly
(in the style of a type), the codomain is the theory in which the structure is
declared. Consequently, if $\dimpdd{\qi}{\qS}{\mu}$ is declared in $\qT$, $\mu$ must be a
morphism from $S$ to $\qT$.

Just like symbols are the constituents of theory bodies, assignments are the constituents
of link bodies. Let $\ql$ be a link from $\qS$ to $\qT$. A \defemph{assignment to a
  constant} of the form $\maps{\qc}{\omega}$ in the body of $\ql$ expresses that $\ql$ maps
the constant $\qc$ of $\qS$ to the term $\omega$ over $\qT$. Assignments of the form
$\maps{\qc}{\hid}$ are special: They express that the constant $\qc$ is
\defemph{filtered}, i.e., $\ql$ is undefined for $\qc$.

If $\qi$ is a structure declared in $\qS$ and $\mu$ a morphism over (i.e., into) $\qT$,
then an \defemph{assignment to a structure} of the form $\maps{\qi}{\mu}$ expresses that
$\ql$ maps $\qi$ to $\mu$. This means that the triangle $\ö{\lpath{\qS}{\qi}}{\;\ql}=\mu$
commutes.

Both kinds of assignments must type-check to ensure that typing is preserved by theory
morphisms. In the case of constants, this means that the term $\omega$ must type-check
against $\ma{\tau}{\ql}$ where $\tau$ is the type of $\qc$ declared in $\qS$. In the case
of structures, it means that $\mu$ must be a morphism from $\qR$ to $\qT$ where $\qR$ is
the type, i.e., the domain, of $\qi$.

\paragraph*{Induced Symbols}
Intuitively, the semantics of a structure $\qi$ with domain $\qS$ declared in $\qT$ is that all symbols of $\qS$ are copied into $\qT$. For example, if $\qS$ contains a constant $\qc$, then an induced constant $\lapp{\qi}{\qc}$ is available in $\qT$. In other words, $\localpathsep$ is used as the operator that dereferences structures.

Similarly, every assignment to a structure induces assignments to constants. Continuing the above example, if a link with domain $\qT$ contains an assignment to $\qi$, this induces assignments to the induced constants $\lapp{\qi}{\qc}$. Furthermore, assignments may be \defemph{deep} in the following sense: If $\qc$ is a constant of $\qS$, a link with domain $\qT$ may also contain assignments to the induced constant $\lapp{\qi}{\qc}$. Of course, this can lead to clashes if a link contains assignments for both $\qi$ and $\lapp{\qi}{\qc}$; links with such clashes will not be well-formed.

\begin{example}[Continued from Ex.~\ref{ex:mmt:algebra:morphism}]\label{ex:mmt:algebra:dec}
The symbol declarations in the theory $\cn{CGroup}$ are written formally like this:
\[\impdd{\cn{mon}}{\mpath{e}{\cn{Monoid}}}{} \tb\mand\tb \symdd{\cn{inv}}{\oma{\arr,\iota,\iota}}{}.\]
The former induces the constants $\triple{e}{\cn{CGroup}}{\cnpath{mon,comp}}$ and $\triple{e}{\cn{CGroup}}{\cnpath{mon,unit}}$.
$\cn{Ring}$ contains only the two structures \[\impdd{\cn{add}}{\mpath{e}{\cn{CGroup}}}{} \tb\mand\tb \impdd{\cn{mult}}{\mpath{e}{\cn{Monoid}}}{}.\]

Instead of inheriting a symbol $\iota$ for the first-order universe from the meta-theory, we can declare a symbol $\cn{univ}$ in $\cn{Monoid}$. Then $\cn{Ring}$ would inherit two instances of $\cn{univ}$, which must be shared. $\cn{Ring}$ would contain the two structures
\[\mathll{\impdd{\cn{add}}{\mpath{e}{\cn{CGroup}}}{} \nl
 \impdd{\cn{mult}}{\mpath{e}{\cn{Monoid}}}{\maps{\cnpath{mon,univ}}{\triple{e}{\cn{Ring}}{\cnpath{add,mon,univ}}}}}\]

Using an assignment to a structure, the assignments of the view $\cn{v2}$ look like this:
\[\maps{\cn{inv}}{\triple{e}{\cn{integers}}{-}} \tb\mand\tb \maps{\cn{mon}}{\mpath{e}{\cn{v1}}}.\]
The latter induces assignments for the induced constants $\triple{e}{\cn{CGroup}}{\cnpath{mon,comp}}$ as well as $\triple{e}{\cn{CGroup}}{\cnpath{mon,unit}}$. For example, $\triple{e}{\cn{CGroup}}{\cnpath{mon,comp}}$ is mapped to $\ma{\triple{e}{\cn{Monoid}}{\cn{comp}}}{\mpath{e}{\cn{v1}}}$.

The alternative formulation of the view $\cn{v2}$ arises if two deep assignments to the induced constants are used instead of the assignment to the structure $\cn{mon}$:
\[\maps{\cnpath{mon,comp}}{\triple{e}{\cn{integers}}{+}} \tb\mand\tb \maps{\cnpath{mon,unit}}{\triple{e}{\cn{integers}}{0}}\]
\end{example}

\subsubsection{The Module Level}

On the module level a {\defemph{theory declaration}} of the form
$\thdeclm{\qT}{[\qM]}{\theta}$ declares a theory $\qT$ defined by a list of symbol
declarations $\theta$, which we call the \defemph{body} of $\qT$. Theories have an optional
meta-theory $\qM$.  A {\defemph{View declarations}} of the form
$\vwdeclm{\qm}{\qS}{\qT}{[\mu]}{\sigma}$ declares a view $\qm$ from $\qS$ to $\qT$ defined
by a list of assignments $\sigma$ and by an optional meta-morphism $\mu$. Just like
structures, views may also be defined by an existing morphism:
$\vwdef{\qm}{\qS}{\qT}{\mu}$.

\paragraph*{Meta-Theories}
Above, we have already mentioned that theories may have meta-theories and that links may have meta-morphisms. Meta-theories provide a second dimension in the theory graph. If $\qM$ is the meta-theory of $\qT$, then $\qT$ may use all symbols of $\qM$. $\qM$ provides the syntactic material that $\qT$ can use to define the semantics of its symbols.

Because a theory $\qS$ with meta-theory $\qM$ implicitly imports all symbols of $\qM$, a
link from $\qS$ to $\qT$ must provide assignments for these symbols as well. This is the
role of the meta-morphism: Every link from $\qS$ to $\qT$ must provide a meta-morphism
from $\qM$ to $\qT$ (or any meta-theory of $\qT$).

\begin{example}[Continued from Ex.~\ref{ex:mmt:algebra:dec}]\label{ex:mmt:synfull}
%
We can now combine the situations from Ex.~\ref{ex:mmt:elalg-meta} and~\ref{ex:mmt:fosem} in one big {\mmt} theory graph.
In a document with URI $m$, we declare an {\mmt} theory for the logical framework as
\[\thdecl{\mpath{m}{\cn{LF}}}{\symdd{\cn{type}}{}{},\;\symdd{\cn{\arr}}{}{},\;\ldots}\]
where we only list the constants that are relevant for our running example: $\cn{type}$
represents the kind of types, and $\cn{\arr}$ is the function type constructor.

We declare a theory for first-order logic in a document with URI $f$ like this:
\[\thdeclm{\mpath{f}{\cn{FOLSyn}}}
    {\mpath{m}{\cn{LF}}}{\begin{array}{l}
     \symdd{\cn{\iota}}{\triple{m}{\cn{LF}}{\cn{type}}}{},\;
     \symdd{\cn{o}}{\triple{m}{\cn{LF}}{\cn{type}}}{},\;\\
     \symdd{\cn{equal}}{
        \oma{\triple{m}{\cn{LF}}{\cn{\arr}},
             \triple{}{}{\cn{\iota}},
             \triple{}{}{\cn{\iota}},
             \triple{}{}{\cn{o}}
        }
     }{},\; \ldots \end{array}}\]
   Here we already use relative identifiers (see Sect.~\ref{sec:mmtweb:URIencoding}) in
   order to keep the notation readable: Every identifier of the form $\triple{}{}{\qc}$ is
   relative to the enclosing theory: For example, $\triple{}{}{\cn{\iota}}$ resolves to
   $\triple{f}{\cn{FOLSyn}}{\cn{\iota}}$.  Again we restrict ourselves to a few constant
   declarations: The types $\cn{\iota}$ and $\cn{o}$ represent terms and formulas, and the
   equality operation takes two terms and returns a formula.

   Then the theories $\cn{Monoid}$, $\cn{CGroup}$, and $\cn{Ring}$ are declared using
   $\mpath{f}{\cn{FOLSyn}}$ as their meta-theory. For example, the declaration of the theory $\cn{CGroup}$ finally
   looks like this:
\[\thdeclm{\mpath{e}{\cn{CGroup}}}
                 {\mpath{f}{\cn{FOLSyn}}}
                 {\begin{array}{l}
                     \impddm{\cn{mon}}{\mpath{e}{\cn{Monoid}}}{\mmtident{\mpath{f}{\cn{FOLSyn}}}}{},\\
                     \symdd{\cn{inv}}{
                       \oma{\triple{m}{\cn{LF}}{\cn{\arr}},
                         \triple{f}{\cn{FOLSyn}}{\cn{\iota}},
                         \triple{f}{\cn{FOLSyn}}{\cn{\iota}}}
                     }{}
\end{array}}\]
Here the structure $\cn{mon}$ must have a meta-morphism translating from the meta-theory
of $\cn{Monoid}$ to the current theory, and that is simply the identity morphism of
$\mpath{f}{\cn{FOLSyn}}$ because $\cn{Monoid}$ and $\mpath{e}{\cn{CGroup}}$ have the same
meta-theory.  If the meta-theory of $\cn{integers}$ is $\cn{ZFC}$, then the meta-morphism of $\cn{v1}$ and $\cn{v2}$ is $\cn{FOLSem}$.
\end{example}


   \subsection{Realizations}\label{sec:mmt:foundations}
     {\mmt} permits an elegant and precise formulation of a general theory of realizations, which formalizes the intuitions 
we introduced in Sect.~\ref{sec:mmt:modules-as-types}. The central idea is to formalize the implicit global environment
as an {\mmt} theory $D$. In particular, {\mmt} naturally provides concrete syntax for both the syntactic and the semantic translations associated with views and functors.

We will consider examples from programming languages and logic. In the former
case, we use SML and $D$ is a theory for the global environment of SML. In the latter
case, $D$ is a theory for ZFC set theory.

\begin{definition}[Grounded Realizations]
An {\mmt}-theory with meta-theory $D$ is called a ``$D$-theory''. Then a \defemph{grounded realization} of the $D$-theory $\qS$ is a morphism from $\qS$ to $D$ that is the identity on $D$.
\end{definition}

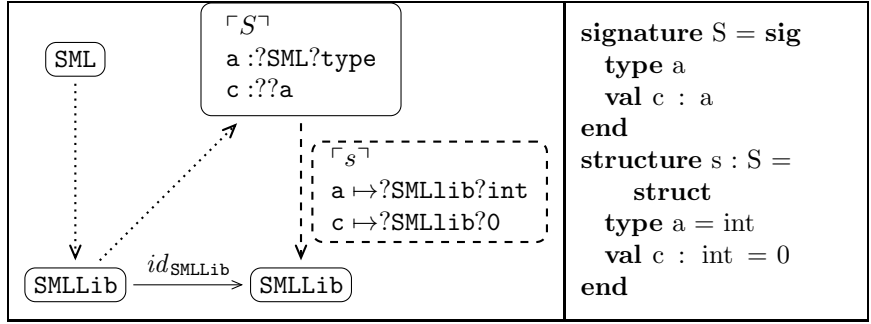
\begin{figure}[ht]\centering
  \begin{tabular}{|l|l|}\hline
\begin{tikzpicture}[scale=3]
      \node[thy] (sml) at (0,1) {$\cn{SML}$};
      \node[thy] (smllib) at (0,0) {$\cn{SMLLib}$};
      \node[thy] (smllib2) at (1,0) {$\cn{SMLLib}$};
      \node[thy] (s) at (1,1) {\begin{tabular}{l}
          $\encode{S}$\\ $\cn{a} : ?\cn{SML}?\cn{type}$\\$\cn{c} :  ??\cn{a}$
        \end{tabular}};
      \draw[meta] (sml) -- (smllib);
      \draw[meta] (smllib) -- (s);
      \draw[morph] (smllib) --node[above]{$\mmtident{\cn{SMLLib}}$} (smllib2);
      \draw[view] (s) -- 
      node [right,draw,rounded corners,inner sep=.1em,outer sep = 1ex]{\begin{tabular}{l}
          $\encode{s}$\\
          $\maps{\cn{a}}{?\cn{SMLlib}?\cn{int}}$\\
          $\maps{\cn{c}}{?\cn{SMLlib}?\cn{0}}$\\
        \end{tabular}}
     (smllib2);
    \end{tikzpicture}
 & 
  \begin{minipage}{3.6cm}\vspace*{-3cm}
\begin{lstlisting}[language=ML,basicstyle=\normalsize,aboveskip=-1em]
signature S = sig
  type a
  val c : a
end
structure s : S = struct
  type a = int
  val c : int = 0
end
\end{lstlisting}
  \end{minipage}
  \\\hline
\end{tabular}
\caption{Implementations in SML}\label{fig:mmt:SML}
\end{figure}

\begin{example}[Realizations in SML]\label{ex:mmt:SML}
The theory $\cn{SML}$ contains declarations for all primitives of the simple type theory underlying SML, such as $\verb|->|$, $\cn{fn}$, and $\cn{type}$. These constants are untyped and undefined. $\cn{SML}$ is used as the meta-theory of the theory $D=\cn{SMLLib}$, which extends $\cn{SML}$ with typed constants for all declarations of the SML basis library~\cite{sml_library}.

Now we represent SML signatures $S$ as $\cn{SMLLib}$-specifications and SML structures $s$ realizing $S$ as grounded realizations of $S$. For example, consider the simple SML signature $S$ and the structure $s$ realizing it given on the right of Fig.~\ref{fig:mmt:SML}. Its representation in {\mmt} is given by the commutative theory graph on the left side of the same figure.
$\encode{S}$ contains one {\mmt} constant declaration for every declaration in $S$. These constants have a type according to $S$ but no definiens. The view $\encode{s}$ maps every declaration of $\encode{S}$ to its value given by $s$.

More generally, SML structures $s$ may also contain declarations that do not correspond to declarations present in $S$. In that case, an auxiliary theory $\qT$ with meta-theory $\cn{SMLLib}$ is used that contains one declaration for every declaration in $s$. Then the view $\encode{s}$ arises as the partial view from $\qT$ to $\cn{SMLLib}$.
\end{example}

In Ex.~\ref{ex:mmt:SML}, a single meta-theory $\cn{SMLLib}$ is used because both SML signatures and SML structures may use the SML basis library. A common alternative is that the specification and the implementation language are separated into two different languages. We encounter this, for example, in logic where specifications (i.e., theories) are written using only the syntax of the logic whereas implementations (i.e., models) are given in terms of the semantic domain -- in our example ZFC set theory.

\begin{example}[Realizations in Logic]\label{ex:mmt:FOL}
\begin{wrapfigure}{r}{6cm}
\fbox{
\begin{tikzpicture}[xscale=2.6,yscale=1.5]
  \node[thy] (fol) at (0,1) {$\cn{FOL}$};
  \node[thy] (zfc) at (1,1) {$\cn{ZFC}$};
  \node[thy] (mon) at (0,0) {$\cn{Monoid}$};
  \node[thy] (monmod) at (1,0) {$\cn{MonoidMod}$};
  \draw[meta] (fol) -- (mon);
  \draw[meta] (zfc) -- (monmod);
  \draw[view] (fol) -- node [above]{$FOLSem$} (zfc);
  \draw[view] (mon) -- node [above=5]{$MonoidSem$}  (monmod);
  \node[thy] (zfc2) at (1.7,1) {$\cn{ZFC}$};
  \draw[morph] (zfc) -- node [above]{$\mmtident{\cn{ZFC}}$}  (zfc2);
  \draw[view] (monmod) -- node [right]{$\encode{M}$}  (zfc2);
\end{tikzpicture}
}
\vspace*{-1em}
\end{wrapfigure}

The theory graph on the right continues Ex.~\ref{ex:mmt:elalg-meta}. The theory of monoids
is represented as a $\cn{FOL}$-theory $\cn{Monoid}$.  To represent models as grounded
realizations, we need a $\cn{ZFC}$-theory $\cn{MonoidMod}$. This theory arises as the pushout of $\cn{Monoid}$ along $\cn{FOLSem}$ over $\cn{FOL}$. In {\mmt}, this pushout can be expressed easily:
  \[\thdeclm{\cn{MonoidMod}}{\cn{ZFC}}{\impddm{\cn{mon}}{\cn{Monoid}}{\cn{FOLSem}}{}}\]
Thus, $\cn{MonoidMod}$ declares the same local symbols as $\cn{Monoid}$ but translated along $\cn{FOLSem}$.

Then we can represent models $M$, i.e., monoids, as grounded realizations $\encode{M}$ of $\cn{MonoidMod}$.
Indeed, a monoid $M$ provides one value for every declaration of $\cn{Monoid}$, just like an {\mmt}-morphism.

So far, we have declared the universe in $\cn{FOL}$, which means that we actually need a family of views $\cn{FOLSem}(U)$, each of which interprets the universe as the set $U$. If we declare the universe in $\cn{Monoid}$ (rather than in $\cn{FOL}$), this example becomes more intuitive. Then $\cn{Monoid}$ and thus also $\cn{MonoidMod}$ have constant declarations for the names $\cn{univ}$, $\cn{comp}$, $\cn{unit}$, $\cn{assoc}$, and $\cn{neut}$.
Consequently, a monoid $M=(U,\circ,e)$ is encoded as the view $\encode{M}$ that contains assignments $\maps{\cn{univ}}{\encode{U}}$, $\maps{\cn{comp}}{\encode{\circ}}$, $\maps{\cn{unit}}{\encode{e}}$, $\maps{\cn{assoc}}{P}$, and $\maps{\cn{neut}}{Q}$. Here $\encode{U}$, $\encode{\circ}$, and $\encode{e}$ are the {\mmt}-terms over $\cn{ZFC}$ that represent the objects $U$, $\circ$, and $e$. Moreover, $P$ and $Q$ are the terms representing the necessary proofs that show that $M$ is indeed a monoid.
\end{example}

As we will see in Sect.~\ref{sec:mmt:validity}, {\mmt} guarantees that $\encode{s}$ and $\encode{M}$ preserve typing. Thus, the properties of implementing a specification and modeling a theory are captured naturally by the properties of {\mmt} theory morphisms.

\begin{definition}[Functors]
Given two $D$-theories $\qS$ and $\qT$, a \defemph{functor} from $\qS$ to $\qT$ is a morphism from $\qT$ to $\qS$ that is the identity on $D$.
Given such a functor $f$ and a grounded realization $r$ of $\qS$, the \defemph{functor application} is defined as the grounded realization $\ö{f}{r}$.
\end{definition}

\begin{wrapfigure}{r}{4cm}
\vspace*{-1em}
\fbox{
\begin{tikzpicture}[scale=1]
\node[thy] (D) at (0,4)  {$D$};
\node[thy] (B) at (0,2.5)  {$B$};
\node[thy] (S) at (-1,1) {$\qS$};
\node[thy] (T) at (1,1)   {$\qT$};
\draw[meta](D) edge[bend right=45] (S);
\draw[meta](D) edge [bend left=45] (T);
\draw[meta](D) -- (B);
\draw[struct](S) --node[near start,right] {$\cnpath{B,i}$} (B);
\draw[view](T) --node[near start,left] {$o$} (B);
\draw[morph](S) --node[left] {$r$} (D);
\draw[morph](T) --node[right] {$\ö{f}{r}$} (D);
\draw[morph](T) --node[below] {$f$} (S);
\end{tikzpicture}
}
\vspace*{-1em}
\end{wrapfigure}

It is often convenient to give such a functor as a triple $(B,i,o)$ as in the diagram on
the right. Here the body of the theory $B$ consists of a structure declaration
$\impddm{i}{\qS}{\mmtident{D}}{}$ followed by arbitrary constant declarations all of which
have a definiens. The intuition is that $B$ imports its input theory $\qS$ and then
implements the intended output theory $\qT$; the view $o$ determines how $\qT$ is
implemented by $B$.

Let $i^{-1}$ denote the view from $B$ to $\qS$, which inverts $i$, i.e., it maps every constant induced by the structure $i$ to the corresponding constant of $\qS$. Because all local constant declarations of $B$ have a definiens, $i^{-1}$ is total. Then we obtain the intended functor as the composition $f=\ö{o}{i^{-1}}$. Given a grounded realization $r$ of $\qS$, functor application is simply composition.

Note that we are flexible whether the intelligence of the functor is given in $B$ or in $o$. $B$ may contain defined constants for all declarations of $\qT$ already so that $o$ is just an inclusion. The opposite extreme arises if $B$ contains no declaration besides $i$ and the assignments in $o$ give the body of the functor.

Sometimes it is not desirable to use the view $i^{-1}$ because applying $i^{-1}$ to a $B$-term involves expanding all the definitions of $B$. In that case, we can use structure assignments to represent functor application.
Consider a $D$-theory $C$, which has access to a realization $r$ of $\qS$, i.e., $r$ is a morphism from $\qS$ to $C$. We wish to apply the functor given by $(B,i,o)$ to $r$ in order to obtain a realization of $\qT$, i.e., a morphism from $\qT$ to $C$. We can do that by using the following structure declaration in $C$
\[\impddm{\cn{apply}}{B}{\mmtident{D}}{\maps{i}{r}}\]
Now the composed morphism $\ö{o}{\cn{apply}}$ is the result of applying $(B,i,o)$ to $r$.

\begin{example}[SML (continued from Ex.~\ref{ex:mmt:SML})]\label{ex:mmt:SML:functors}
An SML functor
\begin{lstlisting}[frame=single,language=ML,basicstyle=\normalsize]
functor f(struct i : $S$) : $T$ = struct $\Sigma$ end
\end{lstlisting}
can be represented directly as a triple $(B,i,o)$ where $B$ is the theory \[\thdeclm{B}{\mmtident{\cn{SMLLib}}}{\impddm{i}{\encode{S}}{\mmtident{\cn{SMLLib}}}{},\;\encode{\Sigma}}\]
and the view $o$ from $\encode{T}$ to $B$ is an inclusion.

Functors with multiple arguments can be represented by first declaring an auxiliary theory that collects all the arguments of the functor.
\end{example}

\begin{example}[Logic (continued from Ex.~\ref{ex:mmt:FOL})]\label{ex:mmt:FOL:functors}
\begin{wrapfigure}{r}{5.8cm}
\vspace*{1em}
\fbox{
\begin{tikzpicture}[yscale=.8,xscale=1.1]
\node[thy] (D) at (0,4)  {$ZFC$};
\node[thy] (B) at (0,2.5)  {$\mpath{}{UnitGroup}$};
\node[thy] (S) at (-1,0.5) {$\mpath{}{\cn{MonoidMod}}$};
\node[thy] (T) at (1,0.5)   {$\mpath{}{\cn{GroupMod}}$};
\draw[meta](D) edge [bend right=45] (S);
\draw[meta](D) edge [bend left=45] (T);
\draw[meta](D) -- (B);
\draw[struct](S) --node[left] {$\mpath{}{\cnpath{UnitGroup,mon}}$} (B);
\draw[view](T) --node[right] {$o$} (B);
\end{tikzpicture}
}
\vspace*{-1em}
\end{wrapfigure}

Consider the functor that maps a monoid $M=(U,\circ,e)$ to its group of units (whose universe is the set $\{u\in U|\exists v\in U.u\circ v=v\circ u=e\}$). We represent it as a triple $(\mpath{}{\cn{UnitGroup}},mon,o)$ as in the diagram on the right. We assume that all involved modules are declared in the same document so that we can use relative identifiers (see Sect.~\ref{sec:mmtweb:URIencoding}) and declare
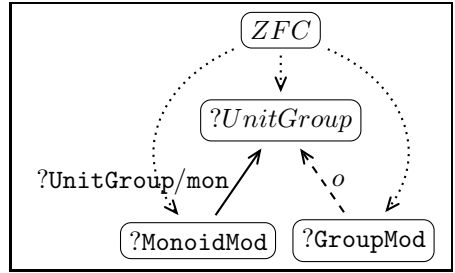
\[\thdeclm{\cn{UnitGroup}}{\mpath{}{\cn{ZFC}}}{\impddm{i}{\mpath{}{\cn{MonoidMod}}}{\mpath{}{\cn{FOLSem}}}{}}\]
\[\vwdeclm{\cn{o}}{\mpath{}{\cn{GroupMod}}}{\mpath{}{\cn{UnitGroup}}}{\mpath{}{\cn{FOLSem}}}{\sigma}\]
Here $\sigma$ contains the assignments that realize a group in terms of a set theory and an assumed monoid $\cn{mon}$. For example, $\sigma$ contains an assignment
\[\maps{\cn{univ}}{\oma{C, \triple{}{\cn{UnitGroup}}{\cnpath{i,univ}}, I}}\]
where we assume that $C$ is defined in $\cn{ZFC}$ such that $\oma{C,s,p}$ represents the
set $\{x\in s\,|\,p(x)\}$, and we use $I$ to represent the property of having an inverse
element.
\end{example}

We can strengthen the above representations considerably by using an additional meta-theory: A foundational theory for a logical framework that occurs as the meta-theory of $D$. For example, we can use $\cn{LF}$ as the meta-theory of $\cn{SML}$ and $\cn{ZFC}$. Then the constants occurring in $\cn{SML}$ and $\cn{ZFC}$ can be typed using the type theory of LF.

If the semantics of $\cn{LF}$ is given in terms of typing and equality judgments, then {\mmt} induces a precise semantics of realizations and functors that adequately represents that of, for example, SML and first-order logic. More generally, the type preservation of {\mmt} morphism formalizes the ``conforms-to'' relation between a specification and an implementation or between a model and a theory.

We follow this approach systematically in \cite{HR:folsound:10} as indicated in
Ex.~\ref{ex:mmt:elalg-meta}. In \cite{IR:foundations:10}, we show how to formalize other
foundations of mathematics. A corresponding representation of the semantics of $\cn{SML}$
in LF can be found in \cite{ml_in_twelf}.


   \subsection{Valid Declarations}\label{sec:mmt:syntax:lookup}
      In the following we define the valid declarations of a theory graph, which arise by adding all induced symbols and assignments. This corresponds to the flattening semantics of structures that eliminates structures and transforms {\mmt} theory graphs into flat ones.

\newcommand{\inp}[1]{{\color{red}#1}}
\newcommand{\outp}[1]{{\color{blue}#1}}

\begin{figure}[ht]
\begin{center}
\begin{tabular}{|l@{\tb}p{7cm}|}\hline
  Judgment & Intuition: in theory graph $\TG$ \ldots\\\hline
$\elabthy{\TG}{\inp{\qT}}{\outp{\theta}}$
   & $\qT$ is a theory in $\qT$ with body $\theta$.
   \\
$\elablink{\TG}{\inp{\ql}}{\outp{\qS}}{\outp{\qT}}{\outp{B}}$
   & $\ql$ is a link from $\qS$ to $\qT$ with definiens $B$.
   \\
$\elabsym{\TG}{\inp{\qT}}{\inp{\qc}}{\outp{\tau}}{\outp{\delta}}$
   & $\symdd{\qc}{\tau}{\delta}$ is an induced constant of $\qT$.
   \\
$\elabass{\TG}{\inp{\ql}}{\inp{\qc}}{\outp{\delta}}$
   & $\maps{\qc}{\delta}$ is an induced constant assignment of $\ql$.
   \\
$\imports{\qM}{\qT}$
   & $\qM$ is the meta-theory of $\qT$.
   \\
$\imports{\mu}{\ql}$
   & $\mu$ is the meta-morphism of $\ql$.
   \\
\hline
\end{tabular}
\caption{Judgments for Valid Declarations}\label{fig:mmt:elabjudgments}
\end{center}
\end{figure}

The judgments for valid declarations are given in Fig.~\ref{fig:mmt:elabjudgments}. All of
them are parametrized by a theory graph $\TG$. The first four judgments are functional in
the sense that they take identifiers as input (red) and return declarations (blue) as
output. The mutually recursive definitions of all judgments are given below.

\paragraph*{Valid Modules}
Firstly, the judgments $\elabthy{\TG}{\qT}{\theta}$ and $\elablink{\TG}{\ql}{\qS}{\qT}{B}$ define the structure of the {\mmt} theory graph, i.e., the valid module level identifiers. Here $B$ is of the form $\{\sigma\}$ or $\mu$ according to whether $\ql$ is defined by a link body or a morphism. Moreover, we write $\imports{\qM}{\qT}$ and $\imports{\mu}{\ql}$ to give the meta-theory and meta-morphism of a theory $\qT$ or a link $\ql$. These judgments are somewhat trivial because they hold iff a meta-theory or meta-morphism is provided explicitly in the syntax of the theory graph.

The first five rules in Fig.~\ref{fig:mmt:elablink} are straightforward: They simply cover the declaration of a theory, and the two 
possible ways each to declare a view or a structure. We use square brackets to denote the optional meta-theories or meta-morphisms, and we give the cases for $\imports{\qM}{\qT}$ and $\imports{\mu}{\ql}$ as second conclusions of a rule.

The only non-trivial rule is $\ELindstr$, which covers the case of induced structures:
$\lpath{\qT}{\qi,\qh}$ identifies the structure induced when a structure declaration $\qi$
instantiates $\qS$ and $\qS$ itself has a structure $\qh$. The induced structure is
defined to be equal to the composition of the two structures, which formalizes the
intended semantics of induced structures.\ednote{MK: a diagram would be helpful here}


\begin{fignd}{mmt:elablink}{Valid Modules}
\ianc{\thdeclm{\qT}{[\qM]}{\theta} \;\minn\; \TG}
     {\elabthy{\TG}{\qT}{\theta}\tb[\imports{\qM}{\qT}]}
     {\ELthy}
\\
\ianc{\vwdef{\ql}{\qS}{\qT}{\mu} \;\minn\; \TG}
     {\elablink{\TG}{\ql}{\qS}{\qT}{\mu}}
     {\ELviewdef}
\tb\tb
\ianc{\vwdeclm{\ql}{\qS}{\qT}{[\mu]}{\sigma} \;\minn\; \TG}
     {\elablink{\TG}{\ql}{\qS}{\qT}{\{\sigma\}}\tb[\imports{\mu}{\ql}]}
     {\ELview}
\tb\\
\ibnc{\elabthy{\TG}{\qT}{\theta}}
     {\dimpdd{\qi}{\qS}{\mu} \;\minn\;\theta}
     {\elablink{\TG}{\lpath{\qT}{\qi}}{\qS}{\qT}{\mu}}
     {\ELstrdef}
\\
\ibnc{\elabthy{\TG}{\qT}{\theta}}
     {\impddm{\qi}{\qS}{[\mu]}{\sigma} \;\minn\;\theta}
     {\elablink{\TG}{\lpath{\qT}{\qi}}{\qS}{\qT}{\{\sigma\}}\tb[\imports{\mu}{\lpath{\qT}{\qi}}]}
     {\ELstr}
\\
\icnc{\elabthy{\TG}{\qT}{\theta}}
     {\dimpdd{\qi}{\qS}{\_} \;\minn\;\theta}
     {\elablink{\TG}{\lpath{\qS}{\qh}}{\qR}{\qS}{\_}}
     {\elablink{\TG}{\lpath{\qT}{\qi,\qh}}{\qR}{\qT}{\ö{\lpath{\qS}{\qh}\;}{\lpath{\qT}{\qi}}}}
     {\ELindstr}
\end{fignd}

\paragraph*{Valid Symbols}
For every theory or link of $\TG$, we define the symbol level identifiers valid in it.
If $\elabthy{\TG}{\qT}{\_}$, we write $\elabsym{\TG}{\qT}{\qc}{\tau}{\delta}$ if $\symdd{\qc}{\tau}{\delta}$ is a valid constant declaration of $\qT$. To avoid case distinctions, we write $\bot$ for $\tau$ or $\delta$ if they are omitted.
If $\elablink{\TG}{\ql}{\_}{\_}{\_}$, we write $\elabass{\TG}{\ql}{\qc}{\delta}$ if $\maps{\qc}{\delta}$ is a valid assignment of $\qT$. 

The induced constants of a theory are defined by the rules in Fig.~\ref{fig:mmt:elabsym}. The rule $con$ simply handles explicit constant declarations. The remaining rules handle induced constants that arise by translating a declaration $\symdd{\qc}{\omt}{\omd}$ along a structure $\lpath{\qT}{\qi}$. In all cases, the type of the induced constant is determined by translating $\omt$ along $\lpath{\qT}{\qi}$. To avoid case distinctions, we assume $\ma{\undef}{\lpath{\qT}{\qi}}=\undef$, i.e., untyped constants induce untyped constants.

But three cases are distinguished to determine the definiens of the induced constant. Firstly, rule $\ELindcondef$ applies if the constant $\qc$ already has a definiens $\omd\neq\undef$. Then the induced constant has the translation of $\omd$ along $\lpath{\qT}{\qi}$ as its definiens. Otherwise, there are two further cases depending on the assignment provided by the structure $\lpath{\qT}{\qi}$ (see the respective rules in Fig.~\ref{fig:mmt:elabass}). If $\lpath{\qT}{\qi}$ provides the default assignment $\spath{\qT}{\lapp{\qi}{\qc}}$ the induced constant has no definiens (rule $\ELindconelse$). If $\lpath{\qT}{\qi}$ provides an explicit definiens $\omd$, it becomes the definiens of the induced constant (rule $\ELindconass$).

\begin{fignd}{mmt:elabsym}{Valid Constants}
\ibnc{\elabthy{\TG}{\qT}{\theta}}
     {\symdd{\qc}{\omt}{\omd}\;\minn\;\theta}
     {\elabsym{\TG}{\qT}{\qc}{\omt}{\omd}}
     {\ELcon}
\\
\icnc{\elablink{\TG}{\lpath{\qT}{\qi}}{\qS}{\qT}{\_}}
     {\elabsym{\TG}{\qS}{\qc}{\omt}{\omd}}
     {\omd\neq\undef}
     {\elabsym{\TG}{\qT}{\lapp{\qi}{\qc}}{\ma{\omt}{\lpath{\qT}{\qi}}}{\ma{\omd}{\lpath{\qT}{\qi}}}}
     {\ELindcondef}
\\
\icnc{\elablink{\TG}{\lpath{\qT}{\qi}}{\qS}{\qT}{\_}}
     {\elabsym{\TG}{\qS}{\qc}{\omt}{\undef}}
     {\elabass{\TG}{\lpath{\qT}{\qi}}{\qc}{\spath{\qT}{\lapp{\qi}{\qc}}}}
     {\elabsym{\TG}{\qT}{\lapp{\qi}{\qc}}{\ma{\omt}{\lpath{\qT}{\qi}}}{\undef}}
     {\ELindconelse}
\\
\icnc{\elablink{\TG}{\lpath{\qT}{\qi}}{\qS}{\qT}{\_}}
     {\elabsym{\TG}{\qS}{\qc}{\omt}{\undef}}
     {\elabass{\TG}{\lpath{\qT}{\qi}}{\qc}{\omd}}
     {\elabsym{\TG}{\qT}{\lapp{\qi}{\qc}}{\ma{\omt}{\lpath{\qT}{\qi}}}{\omd}}
     {\ELindconass}
\end{fignd}

The induced assignments of a link $\ql$ are defined by the rules in Fig.~\ref{fig:mmt:elabass}. The rule $\ELassdeflink$ defines the assignments of a link that is defined as $\mu$: Every undefined constant is translated along $\mu$.

For links that are defined by a list of assignments, four cases must be distinguished. Firstly, the rule $\ELass$ applies if there is an explicit assignment $\maps{\qc}{\omg}$ in $\ql$. Secondly, rule $\ELindass$ creates induced assignments to $\lapp{\qi}{\qc}$, which arise if there is an assignment of a morphism $\mu$ to the structure $\qh$ in a link $\ql$. Since $\lapp{\qh}{\qc}$ identifies the constant $\qc$ imported along $\qh$, the induced assignment arises by translating $\qc$ along $\mu$.

Finally, it is possible that neither rule $\ELass$ nor rule $\ELindass$ applies to $\qc$
-- namely if the body of $\ql$ contains neither an explicit nor an induced assignment for $\qc$. We
abbreviate that by ``$\qc\mtext{not\;covered\;by}\sigma$''. In that case the rules
$\ELdefassstr$ and $\ELdefassview$ define default assignments depending on whether $\ql$
is a structure or a view. If $\ql$ is a structure, $\qc$ is mapped to the induced constant
$\spath{\qT}{\qi,\qc}$ in rule $\ELdefassstr$. If $\ql$ is a view, $\qc$ is filtered via
rule $\ELdefassview$.

\begin{fignd}{mmt:elabass}{Valid Assignments}
\ibnc{\elablink{\TG}{\ql}{\qS}{\qT}{\mu}}
     {\elabsym{\TG}{\qS}{\qc}{\_}{\undef}}
     {\elabass{\TG}{\ql}{\qc}{\mab{\spath{\qS}{\qc}}{\mu}}}
     {\ELassdeflink}
\\
\ibnc{\elablink{\TG}{\ql}{\qS}{\qT}{\{\sigma\}}}
     {\maps{\qc}{\omg}\;\minn\;\sigma}
     {\elabass{\TG}{\ql}{\qc}{\omg}}
     {\ELass}
\\
\icnc{\elablink{\TG}{\ql}{\qS}{\qT}{\{\sigma\}}}
     {\myatop
      {\elablink{\TG}{\lpath{\qS}{\qh}}{\qR}{\qS}{\_}}
      {\elabsym{\TG}{\qS}{\lapp{\qh}{\qc}}{\_}{\undef}}
     }{\maps{\qh}{\mu}\;\minn\sigma}
     {\elabass{\TG}{\ql}{\lapp{\qh}{\qc}}{\mab{\spath{\qR}{\qc}}{\mu}}}
     {\ELindass}
\\
\icnc{\elablink{\TG}{\ql}{\qS}{\qT}{\{\sigma\}}}
     {\elabsym{\TG}{\qS}{\qc}{\_}{\undef}}
     {\myatop{\ql\mtext{structure}}{\qc\mtext{not\;covered\;by}\sigma}}
     {\elabass{\TG}{\ql}{\qc}{\spath{\qT}{\qi,\qc}}}
     {\ELdefassstr}
\\
\icnc{\elablink{\TG}{\ql}{\qS}{\qT}{\{\sigma\}}}
     {\elabsym{\TG}{\qS}{\qc}{\_}{\undef}}
     {\myatop{\ql\mtext{view}}{\qc\mtext{not\;covered\;by}\sigma}}
     {\elabass{\TG}{\ql}{\qc}{\hid}}
     {\ELdefassview}
\end{fignd}

\paragraph*{Clash-Freeness}
It is easy to prove that if $\elabsym{\TG}{\qS}{\qc}{\_}{\undef}$ and $\elablink{\TG}{\ql}{\qS}{\qT}{\_}$, then always $\elabass{\TG}{\ql}{\qc}{\omd}$ for some $\omd$, but $\omd$ is not necessarily unique. More generally, the elaboration judgments do not necessarily define functions from qualified identifiers to induced declarations. For example, a theory graph might declare the same module name twice or a theory might declare the same symbol name twice. To exclude theory graphs with such name clashes, we use the following definition:

\begin{definition}\label{def:mmt:clashfree}
A theory graph $\TG$ is called \emph{clash-free} if all of the following hold:
\begin{itemize}
	\item $\TG$ contains no two module declarations for the names $i$ and $j$ such that $i=j$ or such that $j$ is of the form $\lapp{i}{j'}$ and the body of $i$ contains a declaration for the name $j'$.
	\item There is no module in $\TG$ whose body contains two declarations for the names $i$ and $j$ such that $i=j$ or $j$ is of the form $\lapp{i}{j'}$.
\end{itemize}
\end{definition}

This definition is a bit complicated because it covers theory graphs and theories that explicitly declare qualified identifiers such as in a constant declaration $\symdd{\ompath{\qi,\qc}}{\tau}{\delta}$. In most languages, such declarations are forbidden. But such declarations are introduced when flattening the theory graph, and we want the flat theory graph to be well-formed as well. It is natural to solve this problem by assuming that the flattening algorithm can always generate fresh names for the induced constants. However, such a non-canonical choice of identifiers prevents interoperability.

Therefore, {\mmt} permits declarations that introduce qualified identifiers. This is in fact quite natural because deep assignments in links introduce assignments to qualified identifiers already. The definition of clash-freeness handles both theories and links uniformly: Theories may not explicitly declare both a structure $\qi$ and a constant $\lapp{\qi}{\qc}$, and links may not provide both an assignment for a structure $\qi$ and a deep assignment for an induced constant $\lapp{\qi}{\qc}$.

More precisely, we have:

\begin{lemma}\label{lem:mmt:clashfree}
If a theory graph $\TG$ is clash-free, then the judgments of Fig.~\ref{fig:mmt:elabjudgments} are well-defined functions where the red parameters are input and the blue ones output.
\end{lemma}
\begin{proof}
This follows by a simple induction over the derivations of the elaboration judgments.
\end{proof}

\begin{example}[Continued from Ex.~\ref{ex:mmt:algebra}]\label{ex:mmt:lookup}
  In our running example, we have the theory
  $\elabthy{\TG}{\mpath{e}{\cn{CGroup}}}{\ldots}$ and the structure
  $\elablink{\TG}{\mpath{e}{\cnpath{CGroup,mon}}}{\mpath{e}{\cn{Monoid}}}{\mpath{e}{\cn{CGroup}}}{\{\}}$. This
  structure has the induced assignment
  $\elabass{\TG}{\mpath{e}{\cnpath{CGroup,mon}}}{\cn{comp}}{\triple{e}{\cn{CGroup}}{\cn{mon},\cn{comp}}}$
  according to rule $\ELdefassstr$. And we have the induced constant
 \[\elabsym{\TG}{\mpath{e}{\cn{CGroup}}}{\lapp{\cn{mon}}{\cn{comp}}}
   {\ma{\oma{\triple{m}{\cn{LF}}{\cn{\arr}},
             \triple{f}{\cn{FOL}}{\cn{\iota}},\triple{f}{\cn{FOL}}{\cn{\iota}}, \triple{f}{\cn{FOL}}{\cn{\iota}}}}
       {\mpath{e}{\cnpath{CGroup,mon}}}}
   {\undef}\]
according to rule $\ELindconelse$.
\end{example}



   \subsection{Normal Terms}\label{sec:mmt:syntax:normalize}
     Because {\mmt} is foundation-independent, the equality relation on terms is transparent to {\mmt}. However, some concepts of {\mmt} influence the equality between terms. In particular, the result of a morphism application $\ma{\omega}{\mu}$ can be computed by homomorphically replacing all constants in $\omega$ with their assignments under $\mu$. In the sequel, we define this equality relation to the extent that it is imposed by {\mmt}.

We define a \defemph{normal form} $\rewr{\omega}$ for {\mmt} terms $\omega$. Normalization eliminates all morphism applications, expands all definitions, and enforces the strictness of filtering. The latter means that a term with a filtered subterm is also filtered. Technically, $\rewr{\omega}$ is relative to a fixed theory graph, but we will suppress that in the notation.

\begin{figure}[hbt]\centering
\begin{eqnarray*}
\rewr{\hid} & \hid \\
\rewr{x}      &:=& x \\
\rewr{\spath{\qT}{\qc}} 
       &:=& \cas{
           \rewr{\delta}      \mifc \elabsym{\TG}{\qT}{\qc}{\_}{\delta}\;  {\rm and}\; \delta\neq\bot \\
           \spath{\qT}{\qc}   \mothw
         } \\
\rewr{\oma{\omega_1,\ldots,\omega_n}} 
       &:=& \cas{
           \oma{\rewr{\omega_1},\ldots,\rewr{\omega_n}}
             \mifc \rewr{\omega_i}\neq\hid\mforall i  \\
           \hid
             \mothw
         } \\
\rewr{\ombind{\omega_0}{\Yps}{\omega_1}}
       &:=& \cas{
           \ombind{\rewr{\omega}}{\rewr{\Yps}}{\rewr{\omega_1}}
             \mifc \rewr{\omega_i}\neq\hid\mforall i, \rewr{\Yps}\neq\hid  \\
           \hid
             \mothw
         } \\
\rewr{\ma{\omega}{\mmtident{\qT}}}    &:=& \rewr{\omega} \\
\rewr{\ma{\omega}{\ö{\mu}{\mu'}}}    &:=& \rewr{\mab{\ma{\omega}{\mu}}{\mu'}} \\[.3cm]
\rewr{\ma{\hid}{\ql}}                 &:=& \hid \\
\rewr{\ma{\yps}{\ql}}                 &:=& \yps \\
\rewr{\ma{\oma{\omega_1,\ldots,\omega_n}}{\ql}} &:=&
     \rewr{\oma{\ma{\omega_1}{\ql},\ldots,\ma{\omega_n}{\ql}}} \\
\rewr{\ma{\ombind{\omega_0}{\Yps}{\omega_1}}{\ql}} &:=&
     \rewr{\ombind{\ma{\omega_0}{\ql}}{\ma{\Yps}{\ql}}{\ma{\omega_1}{\ql}}} \\
\rewr{\mab{\ma{\omega}{\mu}}{\ql}}    &:=& \rewr{\ma{\rewr{\ma{\omega}{\mu}}}{\ql}} \\     
\rewr{\mab{\spath{D}{\qc}}{\ql}} &:=&
  \cas{
    \rewr{\ma{\delta}{\ql}}                                      \mifc \delta\neq\undef \\
    \rewr{\mab{\spath{D}{\qc}}{\mu}}                             \mifc \delta=\undef,\;D\neq\qS,\;\imports{\mu}{\ql} \\
    \rewr{\delta'}                                               \mifc \delta=\undef,\;D=\qS,\;
                                                                       \elabass{\TG}{\ql}{\qc}{\delta'}
  }\\[1cm]
\rewr{\cdot} &:=& \cdot \tb\\
\rewr{\Yps,\ombvar{\yps}{\tau}{\delta}}
      &:=& \cas{
           \rewr{\Yps},\ombvar{\yps}{\rewr{\tau}}{\rewr{\delta}}
              \mifc \rewr{\Yps}\neq\hid,\;\rewr{\tau}\neq\hid\;{\rm and}\; \rewr{\delta}\neq\hid \\
           \hid
              \mothw
          }
\end{eqnarray*}
\caption{Normalization}\label{fig:normalize}
\end{figure}

$\rewr{\omega}$ is defined by structural induction using sub-inductions for the case of
morphism application. The definition is given in Fig.~\ref{fig:normalize}. There, we also define $\rewr{\Yps}$, the straightforward extension of normalization to contexts; as before, we assume $\ma{\undef}{\ql}=\undef$ to avoid case distinctions.

Among the cases in Fig.~\ref{fig:normalize}, the case $\mab{\spath{D}{\qc}}{\ql}$ is the most interesting. First of all, we assume $\elablink{\TG}{\ql}{\qS}{\qT}{\_}$ and $\elabsym{\TG}{D}{\qc}{\_}{\delta}$. Note that we permit the case $D\neq\qS$: Below we will see that in well-formed theory graphs $D$ must be $\qS$ or a possibly indirect meta-theory of $\qS$. The definition distinguishes three subcases.
If $\spath{D}{\qc}$ has a definiens $\delta\neq\undef$, it is expanded before applying $\ql$ (first subcase) -- firstly because $\ql$ should not have to give assignments for defined constants, and secondly because $\ql$ might filter the name $\qc$. Otherwise, if $D\neq\qS$, then $\ql$ must have a meta-morphism $\imports{\mu}{\ql}$, which is applied to $\spath{D}{\qc}$ (second case). Finally, if $D=\qS$, then $\ql$ must provide an assignment $\elabass{\TG}{\ql}{\qc}{\delta'}$ (third subcase).
\smallskip

We use a functional notation $\rewr{\omega}$ for the normal form. But technically, if the underlying theory graph is arbitrary, the normal form does not always exist uniquely, e.g., if the theory graph is not clash-free. We will show in Sect.~\ref{sec:mmt:validity} that $\rewr{\omega}$ exists uniquely if the underlying theory graph $\TG$ is well-formed, which justifies our notation.

\begin{example}[Continued from Ex.~\ref{ex:mmt:algebra}]\label{ex:mmt:normalize}\strut\\
Consider the type of the constant $\triple{e}{\cn{CGroup}}{\cnpath{mon,comp}}$ from Ex.~\ref{ex:mmt:lookup}. After two normalization steps, we obtain
\[\mathll{
   \rewr{\ma{\oma{\triple{m}{\cn{LF}}{\cn{\arr}},
            \triple{f}{\cn{FOL}}{\cn{\iota}},\triple{f}{\cn{FOL}}{\cn{\iota}},
            \triple{f}{\cn{FOL}}{\cn{\iota}}}}
       {\mpath{e}{\cnpath{CGroup,mon}}}
    }\;=\nl
   \oma{\rewr{\ma{\triple{m}{\cn{LF}}{\cn{\arr}}}{\mpath{e}{\cnpath{CGroup,mon}}}},
             \rewr{\ma{\triple{f}{\cn{FOL}}{\cn{\iota}}}{\mpath{e}{\cnpath{CGroup,mon}}}},
             \rewr{\ma{\triple{f}{\cn{FOL}}{\cn{\iota}}}{\mpath{e}{\cnpath{CGroup,mon}}}},
             \rewr{\ma{\triple{f}{\cn{FOL}}{\cn{\iota}}}{\mpath{e}{\cnpath{CGroup,mon}}}}
       
    }
}\]
And using $\imports{\mmtident{\mpath{f}{\cn{FOL}}}}{\mpath{e}{\cnpath{CGroup,mon}}}$ $\elabsym{\TG}{\mpath{f}{\cn{FOL}}}{\iota}{\_}{\undef}$, we obtain further
\[\rewr{\ma{\triple{f}{\cn{FOL}}{\cn{\iota}}}{\mpath{e}{\cnpath{CGroup,mon}}}}\;=\;
  \rewr{\ma{\triple{f}{\cn{FOL}}{\cn{\iota}}}{\mmtident{\mpath{f}{\cn{FOL}}}}}\;=\;
  \rewr{\triple{f}{\cn{FOL}}{\cn{\iota}}} \;=\;
  \triple{f}{\cn{FOL}}{\cn{\iota}}\]
so that the result of normalization is $\oma{\triple{m}{\cn{LF}}{\cn{\arr}},
            \triple{f}{\cn{FOL}}{\cn{\iota}},\triple{f}{\cn{FOL}}{\cn{\iota}},
            \triple{f}{\cn{FOL}}{\cn{\iota}}}$ as expected.
\end{example}



 \section{Well-formed Expressions}\label{sec:mmt:validity}
   In this section we define the well-formed {\mmt} expressions (also called valid
expressions). Only those are meaningful. First we define a set of judgments in
Sect.~\ref{sec:mmt:judgments}, and we give a set of inference rules for them in
Sect.~\ref{sec:mmt:rules:structural}-\ref{sec:mmt:rules:term}. Because {\mmt} is generic,
both the judgments and the rules are parametric in a foundation, which we define in
Sect.~\ref{sec:mmt:foundation}.



   \subsection{Judgments}\label{sec:mmt:judgments}
      The judgments for {\mmt} are given in Fig.~\ref{fig:mmt:judgments}. They are relative to a fixed foundation, which we omit from the notation.

\begin{figure}[ht]
\begin{center}
\begin{tabular}{|l@{\tb}l|}\hline
  Judgment & Intuition\\\hline
$\library{\TG}$ 
   & ${\TG}$ is a well-formed theory graph.
   \\
$\oterm[\Yps]{\TG}{\qT}{\omega}$
   & $\omega$ is structurally well-formed over $\TG$, $\qT$, and $\Yps$.
   \\
$\otermtype[\Yps]{\TG}{\qT}{\omega}{\omega'}$
   & $\omega$ is well-typed with type $\omega'$ over $\TG$, $\qT$, and $\Yps$.
   \\
$\otermequal[\Yps]{\TG}{\qT}{\omega}{\omega'}$ 
   & $\omega$ and $\omega'$ are equal over $\TG$, $\qT$, and $\Yps$.
   \\
$\omorphism{\TG}{\mu}{\qS}{\qT}$
   & $\mu$ is a well-typed morphism from $\qS$ to $\qT$.
   \\
$\omorphequal{\TG}{\mu}{\mu'}{\qS}{\qT}$
   & $\mu$ and $\mu'$ are equal as morphisms from $\qS$ to $\qT$.
   \\
\hline
\end{tabular}
\caption{Typing Judgments}\label{fig:mmt:judgments}
\end{center}
\end{figure}

For the \defemph{structural levels}, the inference system uses a single judgment $\library{\TG}$ for well-formed theory graphs. The inference rules will define how well-formed theory graphs can be extended incrementally. There are three kinds of extensions of a theory graph $\TG$:
\begin{itemize}
	\item add a module at the end of $\TG$ -- see the rules in Fig.~\ref{fig:mmt:modules},
	\item add a symbol at the end of the last module of $\TG$ (which must be a theory) -- see the rules in Fig.~\ref{fig:mmt:symbols},
	\item add an assignment to the last link of $\TG$ (which may be a view if $\TG$ ends in that view, or a structure if $\TG$ ends in a theory which ends in that structure) -- see the rules in Fig.~\ref{fig:mmt:assignments}.
\end{itemize}
When theories or links are added, their body is empty initially and populated incrementally by adding symbols and assignments, respectively. This has the effect that there is exactly one inference rule for every theory, view, symbol, or assignment, i.e., for every URI-bearing knowledge item.

For the \defemph{object level}, we use judgments for terms and for
morphisms. $\otermtype[\Yps]{\TG}{\qT}{\omega}{\omega'}$ and
$\otermequal[\Yps]{\TG}{\qT}{\omega}{\omega'}$ express \defemph{typing and equality of
  terms} in context $\Yps$ and theory $\qT$. These judgments are not defined
generically by {\mmt}; instead, they are defined by the foundation (see Sect.~\ref{sec:mmt:foundation}).
{\mmt} only provides the judgment $\oterm[\Yps]{\TG}{\qT}{\omega}$, for structurally well-formed terms; this is the strongest necessary condition for the well-formedness of $\omega$ that does not depend on the foundation. In all three judgments, we omit $\Yps$ when it is empty.

As before, we will occasionally write $\undef$ when the optional type or definition of a constant or variable is not present. For that case, it is convenient to extend the equality and typing judgment to $\undef$. We write $\otermtype[\Yps]{\TG}{\qT}{\omega}{\undef}$ to express that $\omega$ is a well-formed untyped value, and $\otermtype[\Yps]{\TG}{\qT}{\undef}{\omega}$ to express that $\omega$ is a well-formed type, i.e., a term that may occur on the right hand side of $:$. Moreover, we assume that $\otermequal{\TG}{\qT}{\undef}{\undef}$.

Contrary to the judgments for terms, all judgments for \defemph{typing and equality of morphisms} are defined foundation-independently by {\mmt}.
$\omorphism{\TG}{\mu}{\qS}{\qT}$ expresses that $\mu$ is a well-formed morphism from $\qS$
to $\qT$. Similarly, $\omorphequal{\TG}{\mu}{\mu'}{\qS}{\qT}$ expresses
equality.
This notation emphasizes the category theoretic intuition of morphisms with domain and codomain. If, instead, we prefer the type theoretic intuition of realizations as typed objects, we can use the notation $\ostructure{\TG}{\qT}{\mu}{\qS}$ (speak: $\mu$ is a well-typed realization of $\qS$ over $\qT$).

   \subsection{Foundations}\label{sec:mmt:foundation}
     Intuitively, foundations attach a semantics to the constants occurring in the
foundational theories. For the purposes of {\mmt}, this is achieved as follows:
\begin{definition}
A \defemph{foundation} is a definition of the judgments
$\otermtype[\Yps]{\TG}{\qT}{\omega}{\omega'}$ and
$\otermequal[\Yps]{\TG}{\qT}{\omega}{\omega'}$.
In order to avoid case distinctions, we require foundations to define these judgments also for the cases where $\omega$ or $\omega'$ are $\undef$.
\end{definition}

In theoretical accounts, foundations can be given, for example, as an inference system or a decision procedure, or via a denotational semantics. In the {\mmt} implementation, foundations are realized as oracles that are provided by plugins.

In fact, inspecting the rules of {\mmt} will show that {\mmt} only needs the special case of these judgments where $\Yps$ is the empty context. But it is useful to require the general case to permit future extensions of {\mmt}; moreover, for most foundations, the use of an arbitrary context makes the definitions easier.

While the details of the foundation are transparent to {\mmt}, it is useful to impose a regularity condition on foundations that captures some intuitions of typing and equality. First we need an auxiliary definition for the declaration-wise equality of contexts:

\begin{definition}
For two contexts $\Yps^j=\big(\ombvar{\yps}{\tau^j_1}{\delta^j_1},\ldots,\ombvar{\yps}{\tau^j_n}{\delta^j_n}\big)$ for $j=1,2$, we write $\otermequal[\Yps^0]{\TG}{\qT}{\Yps^1}{\Yps^2}$ iff for all $i=1,\ldots,n$
    \[\otermequal[\Yps^0,\Yps^1_{i-1}]{\TG}{\qT}{\tau^1_i}{\tau^2_i} \tb\mand\tb
      \otermequal[\Yps^0,\Yps^1_{i-1}]{\TG}{\qT}{\delta^1_i}{\delta^2_i}
    \]
where $\Yps^1_i:=\ombvar{\yps}{\tau^1_1}{\delta^1_1},\ldots,\ombvar{\yps}{\tau^1_i}{\delta^1_i}$.
Recall that we assume $\otermequal{\TG}{\qT}{\undef}{\undef}$ to avoid case distinctions.
\end{definition}

\begin{definition}[Regular Foundation]\label{def:foundation:regular}
A foundation is called \defemph{regular} if it satisfies the following conditions where $\TG$, $\qT$, and all terms are arbitrary:
\begin{enumerate}
	\item\label{def:regular:normalize}
	 The equality judgment respects normalization:
	  \[\otermequal{\TG}{\qT}{\omega}{\rewr{\omega}}\]
	\item\label{def:regular:congruence} The equality relation induced by
          $\otermequal[\Yps]{\TG}{\qT}{\omega}{\omega'}$ is an equivalence relation for
          every $\Yps$ and satisfies the following congruence laws (where $i$ runs over
          the respective applicable indices):
	  \[\mathll{
	    \otermequal[\Yps]{\TG}{\qT}{\omega_i}{\omega'_i}  \mimplies 
	       \otermequal[\Yps]{\TG}{\qT}{\oma{\omega_0,\ldots,\omega_n}}{\oma{\omega'_0,\ldots,\omega'_n}} \nl
	    \otermequal[\Yps_0]{\TG}{\qT}{\omega_i}{\omega'_i} \mand \otermequal[\Yps_0]{\TG}{\qT}{\Yps}{\Yps'} \mimplies \\
	        \otermequal[\Yps_0]{\TG}{\qT}
	           {\ombind{\omega_0}{\Yps}{\omega_1}}{\ombind{\omega'_0}{\Yps'}{\omega'_1}} \nl
	    \otermequal[\Yps]{\TG}{\qT}{\omega_i}{\omega'_i} \mand \otermtype[\Yps]{\TG}{\qT}{\omega_1}{\omega_2}  \mimplies 
	       \otermtype[\Yps]{\TG}{\qT}{\omega'_1}{\omega'_2} \\
	    \otermequal{\TG}{\qT}{\Yps}{\Yps'} \mand \otermequal[\Yps]{\TG}{\qT}{\omega_1}{\omega_2}  \mimplies 
	       \otermequal[\Yps']{\TG}{\qT}{\omega_1}{\omega_2} \\
	    \otermequal{\TG}{\qT}{\Yps}{\Yps'} \mand \otermtype[\Yps]{\TG}{\qT}{\omega_1}{\omega_2}  \mimplies 
	       \otermtype[\Yps']{\TG}{\qT}{\omega_1}{\omega_2}
	  }\]
	  
	  Note that we do not impose a congruence law for morphism application at this point.
	 
	\item\label{def:regular:morphism} Foundations preserve typing and equality along
          flat morphisms.  To state this precisely, assume flat theories $\qS$ and
          $\qT$. Moreover assume a mapping $f$ of constant identifiers to terms such that:
          Whenever $D=\qS$ or $D$ is a possibly indirect meta-theory of $\qS$ and
          $\symdd{c}{\tau}{\delta}$ is declared in $D$, then
          $\otermtype{\TG}{\qT}{f(\spath{D}{c})}{f(\tau)}$ and
          $\otermequal{\TG}{\qT}{f(\spath{D}{c})}{f(\delta)}$.
	
	Then we require that for two flat terms $\oterm{\TG}{\qS}{\omega_i}$
	 \[\mathll{
	 \otermtype{\TG}{\qS}{\omega_1}{\omega_2}  \mimplies \otermtype{\TG}{\qT}{f(\omega_1)}{f(\omega_2)} \nl
	 \otermequal{\TG}{\qS}{\omega_1}{\omega_2} \mimplies \otermequal{\TG}{\qT}{f(\omega_1)}{f(\omega_2)}
	 }\]
	 where $f(\omega)$ arises by replacing every constant $\mpath{D}{c}$ in $\omega$ with $f(\mpath{D}{c})$.
\end{enumerate}
\end{definition}

Regular foundations are uniquely determined by their action on flat terms so that the module system is transparent to the foundation:
\begin{lemma}\label{def:foundation:normalize}
For every regular foundation and arbitrary $\TG$, $\qT$, $\omega$, $\omega'$:
	  \[\mathll{
	    \otermequal{\TG}{\qT}{\omega}{\omega'} \miff     \otermequal{\TG}{\qT}{\rewr{\omega}}{\rewr{\omega'}} \nl
	    \otermtype{\TG}{\qT}{\omega}{\omega'}  \miff     \otermtype{\TG}{\qT}{\rewr{\omega}}{\rewr{\omega'}} \nl
	  }\]
\end{lemma}
\begin{proof}
The first equivalence follows easily using property (\ref{def:regular:normalize}), symmetry, and transitivity.
The second equivalence follows easily using  property (\ref{def:regular:normalize}), symmetry, and the last of the congruence properties.
\end{proof}

Note that the typing and equality judgments are only assumed for the foundational
theories. For all other theories, the typing and equality judgments are inherited from the
respective meta-theory. For example, a foundation for $\cn{SML}$ must specify the typing
and normalization relations of SML expressions. And a foundation for $\cn{ZFC}$ must
specify the well-formedness and provability of propositions, both of which we consider as
special cases of typing.

It is no coincidence that exactly these two judgments form the interface between {\mmt}
and the foundation: They are closely connected to the syntax of {\mmt} constant
declarations, which may carry types and definitions. If types can be declared for the
constants of a language, then the typing relation should be extended to all complex
expressions. This is necessary, for example, to check that theory morphisms preserve
types. Similarly, if the constants may carry definitions, an equality relation for complex
expressions becomes necessary. Vice versa, {\mmt} constant declarations provide the
foundation with the base cases for the definitions of typing and equality.



   \subsection{Inference Rules for the Structural Levels}\label{sec:mmt:rules:structural}
     When defining well-formed theory graphs, we assume for simplicity that all theory graphs
are clash-free. It is straightforward to extend all inference rules with additional
newness hypotheses for identifiers such that eventually $\library{\TG}$ implies that $\TG$
is clash-free. But we omit this here to simplify the notation.

\begin{wrapfigure}r{2.4cm}\vspace*{-1em}
  \fbox{\begin{tikzpicture}[xscale=1.5,yscale=1.2]
    \node[thy] (S) at (0,0) {$\qS$};
    \node[thy] (T) at (1,0) {$\qT$};
    \node[thy] (M) at (0,1) {$\qM$};
    \node[thy] (Mp) at (1,1) {$\qM'$};
    \draw[struct] (S) -- (T);
    \draw[view] (M) -- node [above] {$\mu$}  (Mp);
    \draw[meta] (M) -- (S);
    \draw[meta] (Mp) -- (T);
  \end{tikzpicture}}\vspace*{-1em}
\end{wrapfigure}
The rules in Fig.~\ref{fig:mmt:modules} define theory graphs as lists of modules. The rule
$\TGemptygraph$ starts with an empty theory graph, and the rules $\TGemptytheory$ and
$\TGemptyview$ add modules with empty bodies (that will be filled incrementally). Rule
$\TGemptyview$ unifies the cases whether $\qS$ has a meta-theory or not by using square
brackets for optional parts; whether $\qT$ has a meta-theory, is irrelevant. Finally
$\TGemptyviewmapall$ adds a view defined by a morphism.  

In rule $\TGemptyview$, one might intuitively expect the assumption
$\big[\imports{\qM}{\qS}\quad\imports{\qM'}{\qT}\quad\omorphism{\TG}{\mu}{\qM}{\qM'}\big]$
which is the situation depicted in the diagram on the right. That case is subsumed by
$\TGemptyview$ as we will see in the rules for morphisms below.

\begin{fignd}{mmt:modules}{Adding Modules}
\ianc{}
     {\library{\cdot}}
     {\TGemptygraph}
\tb\tb
\ibnc{\library{\TG}}
     {[\elabthy{\TG}{\qM}{\_}]}
     {\library{\TG,\;\thdeclm{\qT}{[\qM]}{\emptytheo}}}
     {\TGemptytheory}
\\
\ibnc{\library{\TG}}
     {\omorphism{\TG}{\mu}{\qS}{\qT}}
     {\library{\TG,\;\vwdef{\qm}{\qS}{\qT}{\mu}}}
     {\TGemptyviewmapall}
\\
\ienc{\library{\TG}}
     {\elabthy{\TG}{\qS}{\_}}
     {\elabthy{\TG}{\qT}{\_}}
     {\big[\imports{\qM}{\qS}}
     {\omorphism{\TG}{\mu}{\qM}{\qT}\big]}
     {\library{\TG,\;\vwdeclm{\qm}{\qS}{\qT}{[\mu]}{\emptyview}}}
     {\TGemptyview}
\end{fignd}

The rules in Fig.~\ref{fig:mmt:symbols} add symbols to theories. There are three cases
corresponding to the three kinds of symbols: constants, structures defined by a morphisms,
and structures defined by a list of assignments. The rule $\TGsymbol$ says that constant
declarations $\symdd{\qc}{\tau}{\delta}$ can be added if $\delta$ has type $\tau$. Recall
that this includes the cases where $\tau=\undef$ or $\delta=\undef$. 

Note also that $\oterm{\TG}{\qT}{\omd}$ and $\oterm{\TG}{\qT}{\omt}$ are necessary even
though we require $\otermtype{\TG}{\qT}{\omd}{\omt}$: Indeed in most type systems, the
latter would entail the former two, but in {\mmt} the typing judgment is given by the
foundation as an oracle, so we cannot be sure.

The rules $\TGemptyimport$ and $\TGemptyimportmapall$ are completely analogous to the rules $\TGemptyview$ and $\TGemptyviewmapall$. Again square brackets are used in $\TGemptyimport$ to unify the two cases where $\qS$ has a meta-theory or not. In all three rules it is irrelevant whether $\qT$ has a meta-theory or not; we indicate that by giving this optional meta-theory in gray.

\newcommand{\opt}[1]{{\color{gray}#1}}
\begin{fignd}{mmt:symbols}{Adding Symbols ($\TGv$ abbreviates$TG,\;\thdeclm{\qT}{\opt{\qM}}{\theta}$)}
\idnc{\library{\TG,\;\thdeclm{\qT}{\opt{\qM}}{\theta}}}
     {\oterm{\TGv}{\qT}{\omd}}
     {\oterm{\TGv}{\qT}{\omt}}
     {\otermtype{\TGv}{\qT}{\omd}{\omt}}
     {\library{\TG,\;\thdeclm{\qT}{\opt{\qM}}{\theta,\;\symdd{\qc}{\omt}{\omd}}}}
     {\TGsymbol}
\\
\ibnc{\library{\TG,\;\thdeclm{\qT}{\opt{\qM}}{\theta}}}
     {\omorphism{\TGv}{\mu}{\qS}{\qT}}
     {\library{\TG,\;\thdeclm{\qT}{\opt{\qM}}{\theta,\;\dimpdd{\qi}{\qS}{\mu}}}}
     {\TGemptyimportmapall}
\\
\idnc{\library{\TG,\;\thdeclm{\qT}{\opt{\qM}}{\theta}}}
     {\elabthy{\TG}{\qS}{\_}}
     {\big[\imports{\qM'}{\qS}}
     {\omorphism{\TGv}{\mu}{\qM'}{\qT}\big]}
     {\library{\TG,\;\thdeclm{\qT}{\opt{\qM}}{\theta,\;\impddm{\qi}{\qS}{[\mu]}{\emptyimp}}}}
     {\TGemptyimport}
\end{fignd}

The addition of assignments to a link $\ql$ is more complicated because assignments can be added to views or structures. {\mmt} treats both cases in the same way, which we want to stress by unifying the rules. Therefore, let $\lastlink{\TG}{\ql}{\qS}{\qT}$ denote that $\ql$ is a link occurring at the end of $\TG$, i.e., either
\begin{itemize}
\item $\ql$ refers to a view and $\TG=\ldots,\;\vwdeclm{\ql}{\qS}{\qT}{[\mu]}{\sigma}$ or
\item $\ql=\lpath{\qT}{\qi}$ refers to a structure and
  $\TG=\ldots,\;\thdeclm{\qT}{[\qM]}{\ldots,\;\impddm{\qi}{\qS}{[\mu]}{\sigma}}$,
\end{itemize}
and in that case let $\extend{\TG}{Ass}$ be the theory graph arising from $\TG$ by replacing $\sigma$ with $\sigma,Ass$.

Then the rules in Fig.~\ref{fig:mmt:assignments} add assignments to a link. $\TGinstsymb$ adds an assignment $\maps{\qc}{\omd}$ for an undefined constant $\qc$ of $\qS$. Such assignments are well-typed if $\omd$ is typed by the translation of the type of $\qc$ along $\ql$. Again we assume $\ma{\undef}{\ql}:=\undef$ to avoid case distinctions.

Rule $\TGinstsymb$ includes the case of $\omd=\hid$, i.e., undefined constants can be filtered by mapping them to $\hid$. For defined constants, filtering is the only possible assignment; this is covered by Rule $\TGhidesymb$.

\begin{fignd}{mmt:assignments}{Adding Assignments}
\ienc{\library{\TG}}
     {\lastlink{\TG}{\ql}{\qS}{\qT}}
     {\elabsym{\TG}{\qS}{\qc}{\omt}{\undef}}
     {\oterm{\TG}{\qT}{\omd}}
     {\otermtype{\TG}{\qT}{\omd}{\ma{\omt}{\ql}}}
     {\library{\extend{\TG}{\maps{\qc}{\omd}}}}
     {\TGinstsymb}
\\
\idnc{\library{\TG}}
     {\lastlink{\TG}{\ql}{\qS}{\qT}}
     {\elabsym{\TG}{\qS}{\qc}{\_}{\omd}}
     {\omd\neq\undef}
     {\library{\extend{\TG}{\maps{\qc}{\hid}}}}
     {\TGhidesymb}
\\
\ianc{\myatopp{
       \library{\TG} \tb
       \lastlink{\TG}{\ql}{\qS}{\qT} \tb
       \elablink{\TG}{\lpath{\qS}{\qh}}{\qR}{\qS}{\_} \tb
       \omorphism{\TG}{\mu}{\qR}{\qT}
     }{
       [\imports{\qM}{\qR}\tb
       \imports{\mu'}{\lpath{\qS}{\qh}}\tb
       \omorphequal{\TG}{\mu}{\ö{\mu'}{\ql}}{\qM}{\qT}]
     }{
       \otermequal{\TG}{\qT}{\ma{\delta}{\ql}}{\mab{\spath{\qR}{\qc}}{\mu}}\;\;\mbox{whenever}\;\; \elabsym{\TG}{\qS}{\lapp{\qh}{\qc}}{\_}{\delta},\;\delta\neq\undef
     }}
     {\library{\extend{\TG}{\maps{\qh}{\mu}}}}
     {\TGinstimp}
\end{fignd}

\begin{wrapfigure}{r}{5.2cm}
\begin{center}\vspace*{-2em}
\fbox{
\begin{tikzpicture}
\node[thy] (R) at (0,0) {$R$};
\node[thy] (S) at (2,0) {$S$};
\node (T) at (4,0) {$T$};
\draw[struct](R) -- node[below] {$\lpath{\qS}{\qh}$} (S);
\draw[morph](S) -- node[below] {$\ql$} (T);
\draw[morph](R) .. node[above] {$\mu$} controls (2,.5) .. (T);
\end{tikzpicture}
}
\end{center}\vspace*{-1em}
\end{wrapfigure}

The rule $\TGinstimp$ is similar to $\TGinstsymb$ except that adding assignments for structures is a bit more complicated. The first three hypotheses correspond to the rule $\TGinstimp$. The guiding intuition for the remaining hypotheses is that an assignment $\maps{\qh}{\mu}$ for a structure $\qh$ in $\qS$ should make the diagram on the right commute. From this intuition, we can immediately derive the typing requirements that $\mu$ must be a well-typed morphism from $\qR$ to $\qT$.

However, this is not sufficient yet to make the diagram commute. In general, the link $\ql$ already contains some assignments and possibly a meta-morphism so that the semantics of the composition $\ö{\lpath{\qS}{\qh}}{\ql}$ is already partially determined. Therefore, $\mu$ must agree with $\ö{\lpath{\qS}{\qh}}{\ql}$ whenever the latter is already determined.

This is easy for a possible meta-morphism of $\omorphism{\TG}{\mu'}{\qM}{\qS}$ of $\lpath{\qS}{\qh}$. The composition of $\mu'$ and $\ql$ must agree with the restriction of $\mu$ to $\qM$. Additionally, for all constants $\lapp{\qh}{\qc}$ of $\qS$ that have a definiens $\omd$, the translation of $\omd$ along $\ql$ must be equal to the translation of $\spath{\qR}{\qc}$ along $\mu$.

The rule $\TGinstimp$ is in fact inefficient because it requires to flatten $\qS$, i.e., to compute all induced constants $\lapp{\qh}{\qc}$ of $\qS$. But it is important for scalability to avoid this whenever possible. Therefore, we give some admissible rules of inference in Sect.~\ref{sec:mmt:modlevel} that use module-level reasoning to avoid flattening.


   \subsection{Inference Rules for Morphisms}\label{sec:mmt:rules:morph}
     Fig.~\ref{fig:mmt:morphism} gives the typing rules for morphisms. The rule $\TMmor$ handles links. $\TMid$ and $\TMcomp$ give identity and composition of morphisms. Meta-theories behave like inclusions with regard to composition of morphisms: The rules $\TMcovariant$ and $\TMcontravariant$ give the usual co- and contravariance rules.

Finally, we define the equality of morphisms in rule $\TMequal$. We use an extensional
equality that identifies two morphisms if they map the same argument to equal terms. This
is to equivalent to the special case where the morphisms agree for all undefined
constants.
If the domain has a meta-theory $\qM$, the meta-morphisms must be equal as well. This is checked recursively by requiring $\omorphequal{\TG}{\mu}{\mu'}{\qM}{\qT}$.
\ednote{MK: we should have a diagram for the meta-morphism case of $\TMequal$}

\begin{fignd}{mmt:morphism}{Morphisms}
\ianc{\elablink{\TG}{\ql}{\qS}{\qT}{\_}}
     {\omorphism{\TG}{\ql}{\qS}{\qT}}
     {\TMmor}
\\
\ianc{\elabthy{\TG}{\qT}{\_}}
     {\omorphism{\TG}{\mmtident{\qT}}{\qT}{\qT}}
     {\TMid}
\tb\tb
\ibnc{\omorphism{\TG}{\mu}{\qR}{\qS}}
     {\omorphism{\TG}{\mu'}{\qS}{\qT}}
     {\omorphism{\TG}{\ö{\mu}{\mu'}}{\qR}{\qT}}
     {\TMcomp}
\\
\ibnc{\imports{\qM}{\qS}}
     {\omorphism{\TG}{\mu}{\qS}{\qT}}
     {\omorphism{\TG}{\mu}{\qM}{\qT}}
     {\TMcontravariant}
\tb\tb
\ibnc{\omorphism{\TG}{\mu}{\qS}{\qM}}
     {\imports{\qM}{\qT}}
     {\omorphism{\TG}{\mu}{\qS}{\qT}}
     {\TMcovariant}
\\
\ianc{\def\arraystretch{1}\begin{array}{c}
            \omorphism{\TG}{\mu}{\qS}{\qT} \tb \omorphism{\TG}{\mu'}{\qS}{\qT}\\
            \otermequal{\TG}{\qT}{\ma{\spath{\qS}{\qc\,}}{\mu}}{\ma{\spath{\qS}{\qc\,}}{\mu'}}\;
            \mtext{whenever}\;\elabsym{\TG}{\qS}{\qc}{\_}{\undef}\;\;\\{}
            [\imports{\qM}{\qS} \tb \omorphequal{\TG}{\mu}{\mu'}{\qM}{\qT}]
         \end{array}}
       {\omorphequal{\TG}{\mu}{\mu'}{\qS}{\qT}}
      {\TMequal}
\end{fignd}



   \subsection{Inference Rules for Terms}\label{sec:mmt:rules:term}
      As noted above, {\mmt} relegates the judgments
 \[\otermequal[\Yps]{\TG}{T}{\omega}{\omega'} \tb\mand\tb \otermtype[\Yps]{\TG}{T}{\omega}{\omega'}\]
for typing and equality of terms to the foundation. {\mmt} only defines the judgment $\oterm[\Yps]{\TG}{\qT}{\omega}$ for \defemph{structurally well-formed} terms. Structural well-formedness guarantees in particular that only constants and variables are used that are in scope.

This judgment is axiomatized by the rules in Fig.~\ref{fig:mmt:terms}. First we define an auxiliary judgment $\ocont{\TG}{\qT}{\Yps}$ for well-formed contexts using the rules $\TOemptycon$ and $\TOvardec$. These are such that every variable may occur in the types and definitions of subsequent variables. The rules $\TOvar$, $\TOsym$, $\TOspecial$, $\TOapp$, and $\TObind$ are straightforward. $\TObind$ is such that a bound variable may occur in the type or definition of subsequent variables in the same binder.

Finally, $\TOmor$ and $\TOmeta$ formalize the cases relevant for the {\mmt} module system. $\TOmor$ moves closed terms along morphisms, and $\TOmeta$ moves terms along the meta-theory relation. Note that $\ma{\omega}{\mu}$ is well-formed independent of whether $\omega$ is filtered by $\mu$. This is important because the decision whether $\omega$ is filtered is expensive if the theory graph has not been flattened yet.

\begin{fignd}{mmt:terms}{Structurally Well-formed Terms}
\ianc{\elabthy{\TG}{\qT}{\_}}
     {\ocont{\TG}{\qT}{\cdot}}
     {\TOemptycon}
\tb\tb
\icnc{\ocont{\TG}{\qT}{\Yps}}
     {[\oterm[\Yps]{\TG}{\qT}{\tau}]}
     {[\oterm[\Yps]{\TG}{\qT}{\delta}]}
     {\oterm{\TG}{\qT}{\Yps,\yps[:\tau][=\delta]}}
     {\TOvardec}
\\
\ibnc{\ocont{\TG}{\qT}{\Yps}}
     {\ombvar{\yps}{\_}{\_} \minn \Yps}
     {\oterm[\Yps]{\TG}{\qT}{\yps}}
     {\TOvar}
\tb\tb
\ibnc{\ocont{\TG}{\qT}{\Yps}}
     {\elabsym{\TG}{\qT}{\qc}{\_}{\_}}
     {\oterm[\Yps]{\TG}{\qT}{\spath{\qT}{\qc}}}
     {\TOsym}
\\
\ianc{\ocont{\TG}{\qT}{\Yps}}
     {\oterm[\Yps]{\TG}{\qT}{\hid}}
     {\TOspecial}
%
\\
\ianc{\oterm[\Yps]{\TG}{\qT}{\omega_i}\mforall i=1,\ldots n}
     {\oterm[\Yps]{\TG}{\qT}{\oma{\omega_1,\ldots,\omega_n}}}
     {\TOapp}
\tb\tb
\ibnc{\oterm[\Yps]{\TG}{\qT}{\omega}}
     {\ocont{\TG}{\qT}{\Yps,\Yps'}}
     {\oterm[\Yps]{\TG}{\qT}{\ombind{\omega}{\Yps'}{\omega'}}}
     {\TObind}
\\
\icnc{\ocont{\TG}{\qT}{\Yps}}
     {\oterm{\TG}{\qS}{\omega}}
     {\omorphism{\TG}{\mu}{\qS}{\qT}}
     {\oterm[\Yps]{\TG}{\qT}{\ma{\omega}{\mu}}}
     {\TOmor}
\tb\tb
\ibnc{\oterm[\Yps]{\TG}{\qM}{\omega}}
     {\imports{\qM}{\qT}}
     {\oterm[\Yps]{\TG}{\qT}{\omega}}
     {\TOmeta}
\end{fignd}

It is easy to prove a subexpression property for structural well-formedness: If $\oterm[\Yps]{\TG}{\qT}{\omega}$ then all subexpressions of $\omega$ are well-formed in the respective context.


   \subsection{Module-Level Reasoning}\label{sec:mmt:modlevel}
      The extensional definition of the equality of morphisms is very inefficient because it requires the full elaboration of the domain. We encountered a similar problem in rule $\TGinstimp$. To remedy this, we introduce the following admissible rule of inference, which refines rule $\TMequal$ to avoid elaboration:

\[
\ianc{\myatoppp{
         \omorphism{\TG}{\mu}{\qS}{\qT} \tb \omorphism{\TG}{\mu'}{\qS}{\qT}
       }{
         \otermequal{\TG}{\qT}{\ma{\spath{\qS}{\qc\,}}{\mu}}{\ma{\spath{\qS}{\qc\,}}{\mu'}}\;
            \mtext{whenever}\;\symdd{\qc}{\_}{}\minn \qS
       }{
         \omorphequal{\TG}{\ö{\lpath{\qS}{\qi\,}}{\mu}}{\ö{\lpath{\qS}{\qi\,}}{\mu'}}{\qR}{\qT}\;
            \mtext{whenever}\;\impddm{\qi}{\qR}{\_}{\_}\minn \qS
       }{
        [\imports{\qM}{\qS} \tb \omorphequal{\TG}{\mu}{\mu'}{\qM}{\qT}]
       }
     }
     {\omorphequal{\TG}{\mu}{\mu'}{\qS}{\qT}}
     {\TMequal'}
\]

Both $\TMequal$ and $\TMequal'$ require $\otermequal{\TG}{\qT}{\ma{\spath{\qS}{\qc\,}}{\mu}}{\ma{\spath{\qS}{\qc\,}}{\mu'}}$ for the constants of $\qS$. But in $\TMequal$, this is required for all constants, including the ones induced by structures. $\TMequal'$, on the other hand, only requires it for the local constants of $\qS$, which can be verified without elaboration. For the induced constants of $\qS$, rule $\TMequal'$ recursively checks equality of morphisms for every structure declared in $\qS$. A variant of $\TGinstimp$ that does not require elaboration can be obtained in a similar way.

By unraveling the recursion, it is easy to see that $\TMequal'$ eventually checks the same prerequisites as $\TMequal$. Therefore, $\TMequal'$ by itself does not yield an efficiency gain. However, we can often avoid the recursive calls in $\TMequal'$ by using other, more efficient admissible rules to establish the equality of two morphisms.

These additional rules are axioms that are obtained from the invariants of {\mmt}. Firstly, we have one
equality axiom for every defined view or structure. And secondly,
Thm.~\ref{thm:mmt:strass} establishes the soundness of one equational axiom for every
structure assignment.
Recall that all nodes and edges in the theory graph have URIs, and morphisms are paths
in the theory graph, i.e., lists of URIs. Therefore, representing them and reasoning about the equality of morphisms using these equational axioms is efficient in most cases.

We call this module-level reasoning because it forgets all details about the bodies of
theories and links and only uses the theory graph. Naturally module-level reasoning about
equality of morphisms is sound but not complete. Moreover, the equational theory of paths
in the theory is not necessarily decidable. However, in our experience, module-level
reasoning succeeds in the majority of cases occurring in practice.



 \section{Formal Properties}\label{sec:mmt:metaanalysis}
   Now that we have established the grammar and well-formedness conditions for {\mmt}, we can
analyze the properties of well-formed theory graphs: In Sect.~\ref{sec:mmt:results} we
establish that normalization is well-defined and in Sect.~\ref{sec:mmt:morphism_props}
that assignment to structures can be used to establish commutativity conditions in theory
graphs. In Sect.~\ref{sec:mmt:structural_wff} we introduce the concept of structural
well-formedness, as a computationally motivated compromise between {\mmt}-well-formedness
and grammatical well-formedness. Finally, in Sect~\ref{sec:mmt:operations} we examine the
operation of flattening (i.e. copying out the modular aspects of {\mmt}) as a
semantics-giving operation of theory graphs and show that in {\mmt} it can be made
incremental, which is important for computational tractability and scalability.



  \subsection{Theory Graphs}\label{sec:mmt:results}
      To finish the formal definition of {\mmt}, we must take care of one proof obligation that we have deferred so far: the well-definedness of normalization.

\begin{lemma}\label{lem:normalize}
  Assume a regular foundation and (i) $\library{\TG}$ for a clash-free $\TG$ and (ii)
  $\oterm{\TG}{\qT}{\omega}$. Then $\rewr{\omega}$ is well-defined and does not contain
  any morphism applications.
\end{lemma}
\begin{proof}
  Inspecting the definition of $\rewr{\omega}$, we see there is exactly one case for every
  possible term $\omega$. Technically, this observation uses (i) to deduce that $\gamma$
  is clash-free so that all lookups occurring during the normalization are
  well-defined. It also uses (ii) to conclude that whenever the case
  $\rewr{\ma{\spath{D}{\qc}}{\ql}}$ occurs, $D$ is either the domain of $\ql$ or a
  possibly indirect meta-theory of it.

Furthermore, a straightforward induction shows that if the normal form is well-defined, it does not contain morphism applications.

Therefore, the only thing that must be proved is the well-foundedness of the recursive definition. Essentially, this follows because every case decreases one of the following: the size of $\TG$, the size of $\omega$, or the size of the subterms of $\omega$ to which a morphism is applied. Only some cases warrant closer attention:
\begin{itemize}
	\item $\rewr{\ma{\omega}{\ö{\mu}{\mu'}}}$ and $\rewr{\mab{\ma{\omega}{\mu}}{\ql}}$. These cases do not decrease the size of the terms involved. But it is easy to see that they recurse between themselves only finitely many times, namely until the term \[\rewr{\ma{\ma{\rewr{\ma{\omega}{\ql_1}}}{\iddots}}{\ql_n}}\]
is reached where $\ql_1,\ldots,\ql_n$ is the list of links comprising $\ö{\mu}{\mu'}$ (modulo associativity and identity morphism).
 \item $\rewr{\spath{\qT}{\qc}}$. This case may increase the size of the involved terms when a constant is replaced with its definiens. But due to the well-formedness of $\TG$ and the regularity of the foundation, the definiens must be structurally well-formed over a theory graph smaller than $\TG$. (In particular, there are no cyclic dependencies between definitions in well-formed theory graphs.)
 \item $\rewr{\ma{\spath{\qS}{\qc}\;}{\ql}}$. Similar to the previous case, this case may increase the size of the involved terms when $\ma{\spath{\qS}{\qc}\;}{\ql}$ is replaced with the assignment $\ql$ provides for $\spath{\qS}{\qc}$. The same argument applies.
\end{itemize}
\end{proof}



   \subsection{Properties of Morphisms}\label{sec:mmt:morphism_props}
     With this bureaucracy out of the way, we can prove some intended properties of
morphisms. First we show that morphisms behave as expected. In fact, the presence of
filtering makes some of these theorems quite subtle. Therefore, we use the following
definition:

\begin{definition}
A morphism $\omorphism{\TG}{\mu}{\qS}{\qT}$ is \defemph{total} if $\rewr{\ma{\spath{\qS}{\qc}\;}{\mu}}\neq\hid$ whenever $\elabsym{\TG}{\qS}{\qc}{\_}{\_}$ and if its metamorphism (if there is one) is total as well. A theory graph is total if all its links are total morphisms.
\end{definition}

Note that a morphism that filters only defined constants is still total because the normalization expands definitions. A morphism is not total if it filters undefined constants. Partial (i.e., non-total) morphisms often behave badly because they do not preserve truth: Assume a view from $\qS$ to $\qT$ that does not provide an assignment for an axiom $a$, maybe because that axiom is not provable in $\qT$ at all. Then clearly we cannot expect all theorems of $\qS$ to be translated to theorems of $\qT$. However, this property is also what makes partial morphisms interesting in practice: For example, a partial morphism can be used to represent a translation from a higher-order axiomatization of the real numbers to a first-order one: Such a translation would only translate the first-order-expressible parts, which is still useful in practice.

First, we prove the following intuitively obvious, but technically difficult lemma.

\begin{lemma}\label{lem:results:appnormal}
If $\oterm{\TG}{\qS}{\omega}$ and $\omorphism{\TG}{\mu}{\qS}{\qT}$, then
 $\rewr{\ma{\omega}{\mu}}=\rewr{\ma{\rewr{\omega}}{\mu}}$.
\end{lemma}
\begin{proof}
\newcommand{\DN}{\mathtt{Def}}
\newcommand{\IH}{\mathtt{IH}}
This is proved by a straightforward but technical induction on the structure of $\ma{\omega}{\mu}$. A notable subtlety is that the primary induction is on $\TG$ and $\mu$ using the statement for arbitrary $\omega$ as the induction hypothesis. Then the case where $\mu$ is a link uses a sub-induction on $\omega$. We give some example cases where $\DN$ refers to the definition of normalization and $\IH$ refers to the induction hypothesis:
\begin{itemize}
	\item case for a composed morphism in the induction on $\mu$:
	 \[
	 \rewr{\ma{\omega}{\ö{\mu}{\mu'}}}  \Ceq{\DN}
	 \rewr{\mab{\ma{\omega}{\mu}}{\mu'}} \Ceq{\IH\,\mu'} 
	 \rewr{\ma{\rewr{\ma{\omega}{\mu}}}{\mu'}} \Ceq{\IH\,\mu}
	 \rewr{\ma{\rewr{\ma{\rewr{\omega}}{\mu}}}{\mu'}} \Ceq{\IH\,\mu'}
	 \rewr{\mab{\ma{\rewr{\omega}}{\mu}}{\mu'}} \Ceq{\DN}
	 \rewr{\ma{\rewr{\omega}}{\ö{\mu}{\mu'}}}
	 \]
	\item case for $\ma{\omega}{\ql}$ for a single link $\ql$, proved by a sub-induction on  $\omega$:
	 \begin{itemize}
	 \item case for application:
  	\[\mathll{
	     \rewr{\ma{\oma{\omega_1,\ldots,\omega_n}}{\ql}} \Ceq{\DN}
	     \rewr{\oma{\ma{\omega_1}{\ql},\ldots,\ma{\omega_n}{\ql}}} \Ceq{\IH}
	     \rewr{\oma{\ma{\rewr{\omega_1}}{\ql},\ldots,\ma{\rewr{\omega_n}}{\ql}}} \Ceq{\DN} \nl
	     \rewr{\ma{\oma{\rewr{\omega_1},\ldots,\rewr{\omega_n}}}{\ql}} \Ceq{\DN}
	     \rewr{\ma{\rewr{\oma{\omega_1,\ldots,\omega_n}}}{\ql}}
	     
	  }\]
	  \item case for a constant $\spath{D}{\qc}$: If $\elabsym{\TG}{D}{\qc}{\_}{\undef}$, the statement is trivial because $\rewr{\spath{D}{\qc}}\Ceq{\DN}\spath{D}{\qc}$. If $\elabsym{\TG}{D}{\qc}{\_}{\delta}$ for $\delta\neq\undef$, then
	  \[\rewr{\ma{\spath{D}{\qc}}{\ql}}\Ceq{\DN} \rewr{\ma{\delta}{\ql}} \Ceq{\IH} \rewr{\ma{\rewr{\delta}}{\ql}} \Ceq{\DN} \rewr{\ma{\rewr{\spath{D}{\qc}}}{\ql}}\]
\end{itemize}
\end{itemize}
\end{proof}

Then we have the main technical results about theory graphs and morphisms.

\begin{theorem}[Morphisms]\label{thm:results:morphisms}
Assume a fixed regular foundation. Then
\begin{enumerate}
	\item For fixed $\TG$, the binary relation on morphisms induced by $\omorphequal{\TG}{\mu}{\mu'}{\qS}{\qT}$ is an equivalence relation.
	\item\label{thm:results:morphisms:comp}
	 If $\omorphequal{\TG}{\mu_1}{\mu'_1}{\qR}{\qS}$ and $\omorphequal{\TG}{\mu_2}{\mu'_2}{\qS}{\qT}$ and $\mu_2$ and $\mu'_2$ are total, then
	   \[\omorphequal{\TG}{\ö{\mu_1}{\mu_2}}{\ö{\mu'_1}{\mu'_2}}{\qR}{\qT},\]
	\item When composition is well-formed, it is associative and $\mmtident{\qT}$ is a neutral element.
	\item The identity and the composition of total morphisms are total. In particular, every well-formed total theory graph induces a category of theories and -- modulo equality -- morphisms.
	\item\label{results:enum1:appcomp}
	 If $\oterm{\TG}{\qR}{\omega}$ and $\omorphism{\TG}{\ö{\mu}{\mu'}}{\qR}{\qT}$,
	  then $\otermequal{\TG}{\qT}{\ma{\omega}{\ö{\mu}{\mu'}}}{\mab{\ma{\omega}{\mu}}{\mu'}}$.
	\item\label{results:enum1:app1}
	  If $\oterm{\TG}{\qS}{\omega}$ and $\omorphequal{\TG}{\mu}{\mu'}{\qS}{\qT}$, then
	  \[\otermequal{\TG}{\qT}{\ma{\omega}{\mu}}{\ma{\omega}{\mu'}}.\]
	\item\label{results:enum1:app}
	  If $\omorphism{\TG}{\mu}{\qS}{\qT}$, $\oterm{\TG}{\qS}{\omega}$, $\oterm{\TG}{\qS}{\omega'}$, then
	    \begin{itemize}
	      \item if $\rewr{\omega}=\rewr{\omega'}$, then $\rewr{\ma{\omega}{\mu}}=\rewr{\ma{\omega'}{\mu}}$, and
	      \item if $\otermequal{\TG}{\qS}{\omega}{\omega'}$ and $\mu$ is total, then
	      $\otermequal{\TG}{\qT}{\ma{\omega}{\mu}}{\ma{\omega'}{\mu}}$.
	    \end{itemize}
\end{enumerate}
\end{theorem}
\begin{proof}
\begin{enumerate}
	\item Reflexivity, symmetry, and transitivity follow immediately from the corresponding properties for terms using rule $\TMequal$.
	\item This is proved by induction on the number of meta-theories of $\qR$. If there is none, the result follows using rule $\TMequal$ and applying (\ref{results:enum1:appcomp}) and twice (\ref{results:enum1:app}). If $\imports{\qM}{\qR}$, the same argument applies with $\qM$ instead of $\qR$.
	\item Because of rule $\TMequal$, the equality of two morphisms is equivalent to a set of judgments of the form $\otermequal{\TG}{D}{\ma{C}{\mu}}{\ma{C}{\mu'}}$ for all constants $c$ of $\qS$ or one of its meta-theories. Because the foundation is regular, every such judgment is equivalent to $\otermequal{\TG}{\qT}{\rewr{\ma{C}{\mu}}}{\rewr{\ma{C}{\mu'}}}$.
 Then the conclusion follows from the definition of normalization.
	\item The totality properties are easy to prove. A category is obtained by taking the theories $\elabthy{\TG}{\qT}{\_}$ as the objects, and the quotient
	  \[\{\mu \;|\; \omorphism{\TG}{\mu}{\qS}{\qT}\} \;/\;\{(\mu,\mu')\;|\;\omorphequal{\TG}{\mu}{\mu'}{\qS}{\qT}\}\]
	as the set of morphisms from $\qS$ to $\qT$. Identity and composition are induced by $\mmtident{\qT}$ and $\ö{\mu}{\mu'}$. (See~\cite{categories} for the notion of a \emph{category}.)
	\item Because the foundation is regular, the conclusion is equivalent to $\otermequal{\TG}{\qT}{\rewr{\ma{\omega}{\ö{\mu}{\mu'}}}}{\rewr{\mab{\ma{\omega}{\mu}}{\mu'}}}$. And this follows directly from the definition of normalization.
	\item Using regularity, it is sufficient to show $\rewr{\ma{\omega}{\mu}}=\rewr{\ma{\omega}{\mu'}}$. Using Lem.~\ref{lem:results:appnormal}, this reduces to the case where $\omega$ is flat. For flat $\omega$, the definition of normalization shows that $\ma{\omega}{\mu}$ arises from $\omega$ by replacing all constants $C$ with $\ma{C}{\mu}$. $\omorphequal{\TG}{\mu}{\mu'}{\qS}{\qT}$ yields $\otermequal{\TG}{\qT}{\ma{C}{\mu}}{\ma{C}{\mu'}}$. Because $\omega$ is flat, the result follows from property (\ref{def:regular:congruence}) in Def.~\ref{def:foundation:regular}.
	\item The first statement follows immediately from Lem.~\ref{lem:results:appnormal}.
   Also using Lem.~\ref{lem:results:appnormal}, the conclusion of the second statement reduces to $\otermequal{\TG}{\qT}{\ma{\rewr{\omega}}{\mu}}{\ma{\rewr{\omega'}}{\mu}}$, i.e., it is sufficient to consider the case where $\omega$ and $\omega'$ are flat. And that case follows using property (\ref{def:regular:morphism}) in Def.~\ref{def:foundation:regular} and the type-preservation of well-formed total morphisms guaranteed by rule $\TGinstsymb$.
\end{enumerate}
\end{proof}

The restriction that $\mu$ must be total in part (\ref{results:enum1:app}) of Thm.~\ref{thm:results:morphisms} is necessary. To see why, assume $\omega=\oma{\pi_1,\oma{\cn{pair},a,b}}$. A foundation might define $\otermequal{\TG}{\qS}{\omega}{a}$. Now if $\mu$ filters $b$, then $\rewr{\ma{\omega}{\mu}}=\hid$ but not necessarily $\rewr{\ma{a}{\mu}}=\hid$. An even trickier example arises when $\qS$ contains an axiom $\symdd{a}{\oma{\cn{true},\oma{\cn{equal},\omega,\omega'}}}{}$ and the foundation uses $a$ to derive $\otermequal{\TG}{\qS}{\omega}{\omega'}$. If $\mu$ filters $a$, then it is possible that $\otermequal{\TG}{\qT}{\ma{\omega}{\mu}}{\ma{\omega'}{\mu}}$ does not hold even when $\mu$ filters neither $\omega$ nor $\omega'$.

\begin{wrapfigure}{r}{3.9cm}\vspace*{-1em}
\fbox{
\begin{tikzpicture}[yscale=.6,xscale=.8]
\node[thy] (R) at (0,0)
    {$\qR$};
\node[thy] (\qS) at (2,0)
    {$\qS$};
\node[thy] (T) at (2,-2)
    {$\qT$};
\draw[struct](R) -- node[above] {$\lpath{\qS}{\qh}$} (S);
\draw[morph](S) -- node[right] {$\ql:\maps{\qh}{\mu}$} (T);
\draw[morph](R) -- node[left] {$\mu$} (T);
\end{tikzpicture}
}\vspace*{-1em}
\end{wrapfigure}

The following theorem establishes a central property of {\mmt} theory graphs that plays a crucial role in adequacy proofs. In the diagram on the right, $\qS$ is a theory with a structure instantiating $\qR$, and $\ql$ is a link from $\qS$ to $\qT$ that assigns $\mu$ to $\qh$. The theorem states that the triangle commutes. This means that assignments to structures can be used to represent commutativity conditions on diagrams.

\begin{theorem}\label{thm:mmt:strass}
Assume $\library{\TG}$ relative to a fixed regular foundation. If $\omorphism{\TG}{\mu}{\qR}{\qT}$ and $\elablink{\TG}{\ql}{\qS}{\qT}{\{\sigma\}}$ such that $\sigma$ contains the assignment $\maps{\qh}{\mu}$, then \[\omorphequal{\TG}{\ö{\lpath{\qS}{\qh}\;}{\ql}}{\mu}{\qR}{\qT}.\]
\end{theorem}
\begin{proof}
By rule $\TMequal$, we have to show $\otermequal{\TG}{\qT}{\ma{\spath{\qR}{\qc\,}}{\ö{\lpath{\qS}{\qh}\;}{\ql}}}{\ma{\spath{\qR}{\qc}\,}{\mu}}$ for all $\elabsym{\TG}{\qR}{\qc}{\_}{\undef}$. Using the regularity, it is enough to show equality after normalization. A first normalization step reduces the left hand side to $\rewr{\ma{\spath{\qS}{\qh,\qc\,}}{\ql}}$. Now there are two cases differing by whether $\lapp{\qh}{\qc}$ has a definiens in $\qS$ or not.
\begin{itemize}
	\item If $\elabsym{\TG}{\qS}{\lapp{\qh}{\qc\,}}{\_}{\undef}$, then $\elabass{\TG}{\ql}{\lapp{\qh}{\qc\,}}{\ma{\spath{\qR}{\qc\,}}{\mu}}$, and the left hand side normalizes to $\rewr{\ma{\spath{\qR}{\qc\,}}{\mu}}$.
	\item If $\elabsym{\TG}{\qS}{\lapp{\qh}{\qc}}{\_}{\omd}$ for some $\omd\neq\undef$, a further normalization step reduces the left hand side to $\rewr{\ma{\delta}{\ql}}$. And	rule $\TGinstimp$ guarantees that in this case $\otermequal{\TG}{\qT}{\ma{\delta}{\ql}}{\ma{\spath{\qR}{\qc\,}}{\mu}}$.
\end{itemize}
Furthermore, we have to show that the morphisms agree on the meta-theory of $\qR$ if there is one. This is explicitly required in rule $\TGinstimp$.
\end{proof}


   \subsection{Structural Well-Formedness}\label{sec:mmt:structural_wff}
      The formal definition of structural well-formedness of theory graphs and terms was the original motivation of {\mmt}. Essentially, it means that all references to modules,
symbols, or variables exist and are in scope, and that all morphisms are well-typed. But
it does not guarantee that terms are well-typed. This yields an intermediate well-formedness level between context-free and semantic validation.

Context-free validation checks a theory graph against a context-free grammar. This is the
state of the art inXML-based languages, where the grammar is usually given as XML
schema. It is simple and widely implemented, but it is very weak and accepts many
meaningless expressions. For example, documents containing references to non-existent
knowledge item pass validation. For many knowledge management applications, this is too
weak.

Semantic validation on the other hand accepts only meaningful expressions. It checks a theory graph using a type system or an interpretation function. This is normal for formal languages such as logics and type theories. But semantic validation depends on the foundation. Therefore, it is complex, and often only one implementation is available for a specific formal language, which cannot easily be reused by other applications.

Structural well-formedness is a trade-off between these extremes. It is foun\-dation-independent and therefore easy to implement. And the added strength of full validation is not necessary as a precondition for many web scale algorithms such as browsing or versioning.

Technically, structural well-formedness can be defined using a special foundation:
\begin{definition}
The \emph{structural foundation} is the foundation where $\otermtype{\TG}{\qT}{\omega}{\omega'}$ and $\otermequal{\TG}{\qT}{\omega}{\omega'}$ always hold.
A theory graph $\TG$ is \defemph{structurally well-formed} if it is clash-free and $\library{\TG}$ holds relative to the structural foundation.
\end{definition}

Clearly, it is not a reasonable mathematical foundation, but it is useful because it is maximal or most permissive among all foundations. It is also easy to implement and can be used as a default foundation when the actual foundation is not known or an implementation for it not available.

Structural well-formedness is foundation-independent in the following sense:
\begin{theorem}
If a theory graph is well-formed relative to any foundation, then it is structurally well-formed. If $\TG$ is structurally well-formed, then $\oterm{\TG}{\qT}{\omega}$ is independent of the foundation.
\end{theorem}
\begin{proof}
The first statement holds because the use of the structural foundation simply amounts to removing the typing and equality hypotheses in the rules $\TGsymbol$ and $\TGinstsymb$. The second statement holds because the rules for the judgment $\oterm{\TG}{\qT}{\omega}$ do not refer to any other judgment.
\end{proof}

Corresponding to the notions of structural and semantic validation, we can define structural and semantic equivalence of theory graphs:

\begin{definition}
Relative to a fixed foundation, two well-formed theory graphs $\TG$ and $\TGv$ are called \defemph{structurally equivalent} if the following holds:
\begin{itemize}
	\item $\elabthy{\TG}{\qT}{\_}$ iff $\elabthy{\TGv}{\qT}{\_}$, and in that case $\qT$ has meta-theory $\qM$ in $\TG$ iff it does so in $\TGv$,
	\item $\elablink{\TG}{\ql}{\qS}{\qT}{\_}$ iff $\elablink{\TGv}{\ql}{\qS}{\qT}{\_}$,
	\item whenever $\elabthy{\TG}{\qT}{\_}$, where $\elabsym{\TG}{\qT}{\qc}{\_}{\_}$ iff $\elabsym{\TGv}{\qT}{\qc}{\_}{\_}$.
\end{itemize}
\end{definition}

The intuition behind structural equivalence is that structurally equivalent theory graphs declare the same names: they have the same theories, the same constants, and the same links. It leaves open whether a constant of name $\lapp{\qi}{\qc}$ is declared or whether a constant $\qc$ is imported via a structure $\qi$. It also leaves open whether a link is a structure or a view.

The value of structural equivalence is that it imposes no requirements on the foundation. Furthermore, structural equivalence is sufficiently strong an invariant for many applications such as indexing or cross-referencing.
This is formalized in the following next theorem. 

\begin{theorem}
Assume two structurally equivalent theory graphs $\TG$ and $\TGv$. Then for all theories $\qS,\qT$ of $\TG$:
\begin{itemize}
	\item $\oterm{\TG}{\qT}{\omega} \tb\miff \tb \oterm{\TGv}{\qT}{\omega}$,
	\item $\omorphism{\TG}{\mu}{\qS}{\qT} \tb\miff\tb \omorphism{\TGv}{\mu}{\qS}{\qT}$.
\end{itemize}
\end{theorem}
\begin{proof}
This follows by a straightforward induction on the derivations of well-formed terms and morphisms.
\end{proof}

In structurally equivalent theory graphs, the same constant might have different types. Semantic equivalence refines this:

\begin{definition}\label{def:mmt:semantequiv}
Two structurally equivalent theory graphs $\TG$ and $\TGv$ are called \defemph{semantically equivalent} if the following holds:
\begin{itemize}
	\item If $\elabthy{\TG}{\qT}{\_}$, $\elabsym{\TG}{\qT}{\qc}{\omt}{\omd}$, and $\elabsym{\TGv}{\qT}{\qc}{\omt'}{\omd'}$, then $\rewr{\omd}^{\TG}=\rewr{\omd'}^{\TGv}$ and $\rewr{\omt}^{\TG}=\rewr{\omt'}^{\TGv}$.
  \item For all $\elablink{\TG}{\ql}{\qS}{\qT}{\_}$, if $\elabass{\TG}{\ql}{\qc}{\omd}$ and $\elabass{\TGv}{\ql}{\qc}{\omd'}$, then $\rewr{\omd}^{\TG}=\rewr{\omd'}^{\TGv}$.
\end{itemize}
\end{definition}

Intuitively, if two theory graphs are semantically equivalent, then they have the same
constant declarations and the same assignments. Another way to put it, is that the theory
graphs are indiscernable in the following sense:

\begin{theorem}\label{thm:mmt:semantequiv}
Assume two semantically equivalent theory graphs $\TG$ and $\TGv$ and a regular foundation. Then for all module declarations $Mod$:
 \[\library{\TG,\;Mod} \tb\miff\tb \library{\TGv,\;Mod}.\]
\end{theorem}
\begin{proof}
First of all, due to the structural equivalence, $\TG,\;Mod$ is clash-free iff $\TGv,\;Mod$ is.
Now assume a well-formedness derivation $D$ for $\library{\TG,\;Mod}$. Let $D'$ arise from $D$ by replacing every occurrence of $\TG$ with $\TG'$, and replacing the subtree of $D$ deriving $\library{\TG}$ with some derivation of $\library{\TGv}$. We claim that every subtree of $D'$ is a well-formedness derivation for its respective root. Then in particular, $D'$ is a well-formedness derivation for $\library{\TGv,\;Mod}$. This is shown by induction on $Mod$.
All induction steps are simple because in most rules the theory graph only occurs as a fixed parameter. Those rules that ``look into'' the theory graph do so via the judgments given in Sect.~\ref{sec:mmt:syntax:lookup}, and the semantic equivalence of $\TG$ and $\TGv$ guarantees that these judgments agree up to normalization, and normalization is respected by a regular foundation.
\end{proof}

This provides systems working with {\mmt} theory graphs with an invariant for foundation-independent and semantically indiscernible transformations. Systems maintaining theory graphs can apply such transformations to increase the efficiency of storage or lookup in a way that is transparent to other applications. Moreover, it provides an easily implementable criterion to analyze the management relevance of a change.

Of course, Def.~\ref{def:mmt:semantequiv} is just a sufficient criterion for semantic
indiscernability. If a foundation adds equalities between terms, then theory graphs that
are distinguished by Def.~\ref{def:mmt:semantequiv} become equivalent with respect to that
foundation.  But the strength of Def.~\ref{def:mmt:semantequiv} and
Thm.~\ref{thm:mmt:semantequiv} is that they are foundation-independent. Therefore, it can
implemented easily and generically.

The most important examples of semantical equivalence are reordering and flattening (see Sect.~\ref{sec:mmt:operations}).

\begin{theorem}\label{thm:operations:reorder}
If $\TG$ and $\TGv$ are well-formed theory graphs that differ only in the order of modules, symbols, or assignments, then they are semantically equivalent.
\end{theorem}
\begin{proof}
Clear since the elaboration judgments are insensitive to reorderings.
\end{proof}
However, note that not all reorderings preserve the well-formedness of theory graphs as defined here -- there is a partial order on declarations that the linearization in the theory graph must respect, for example, constants must be declared before they are used. But Thm.~\ref{thm:operations:reorder} permits to generalize the definition of well-formed theory graphs as follows: A theory graph can be considered well-formed if there is some reordering for which it is well-formed. Using this relaxed definition is extremely valuable in practice because it permits applications to forget the order and thus to store theory graphs more efficiently. It is also relevant for distributed developments where keeping track of the order is often not feasible.
\ednote{FR: This is not so easy with sub-theories.}



   \subsection{Flattening}\label{sec:mmt:operations}
      The representation of theory graphs introduced in the last section is geared towards expressing mathematical knowledge in its most general form and with the least redundancy: constants can be shared by inheritance (i.e., via imports), and terms can be moved between theories via morphisms. This style of writing mathematics has been cultivated by the Bourbaki group~\cite{bourbakisets,bourbakialgebra} and lends itself well to a systematic development of theories.

However, it also has drawbacks: Items of mathematical knowledge are often not where or in the form in which we expect them, as they have been generalized to a different context. For example, a constant $c$ need not be explicitly represented in a theory $T$, if it is induced as the image of a constant $c'$ under some import into $T$.

In this section, we show that for every theory graph there is an equivalent flat one. This involves adding all induced knowledge items to every theory thus making all theories self-contained (but hugely redundant between theories). For a given {\mmt} theory graph $\TG$, we can view the flattening of $\TG$ as its semantics because flattening eliminates the specific {\mmt}-representation infrastructure of structures and morphisms.

\begin{theorem}\label{thm:mmt:flattening}
  Given a fixed regular foundation, every well-formed theory graph is semantically
  equivalent to a flat one.
\end{theorem}
\begin{proof}
  Given a $\library{\TG}$ the flat theory graph $\TGv$ is obtained as follows.
\begin{enumerate}
	\item Theories
	  \begin{itemize}
	    \item For every $\elabthy{\TG}{\qT}{\_}$, there is a theory $\qT$ in $\TGv$. It has the same meta-theory (if any) in $\TGv$ as in $\TG$.
	    \item For every $\elabsym{\TG}{\qT}{\qc}{\omt}{\omd}$, the theory $\qT$ of $\TGv$ contains a constant declaration $\symdd{\qc}{\rewr{\omt}}{\rewr{\omd}}$.
	  \end{itemize}
	\item Links with definiens: For every $\elablink{\TG}{\ql}{\qS}{\qT}{\mu}$, $\TGv$ contains a view $\vwdef{\ql}{\qS}{\qT}{\mu}$.
	\item Links with assignments:
	  \begin{itemize}
	    \item For every $\elablink{\TG}{\ql}{\qS}{\qT}{\{\_\}}$, $\TGv$ contains a view from $\qS$ to $\qT$. It has the same meta-morphism (if any) in $\TGv$ as in $\TG$.
	    \item For every $\elabass{\TG}{\ql}{\qc}{\omd}$, the view $\ql$ of $\TGv$ contains a constant assignment $\maps{\qc}{\rewr{\omd}}$.
	  \end{itemize}
\end{enumerate}
It is easy to see that these declarations can be arranged in some way that makes $\TGv$ structurally well-formed.
Furthermore, it is clear from the construction of $\TGv$ that $\TGv$ is flat and that $\TG$ and $\TGv$ are semantically equivalent. The only property that is not obvious is that $\TGv$ is well-formed. For that, we must show in particular that all assignments in all views in $\TGv$ satisfy the typing assumption of rule $\TGinstsymb$. This follows from the construction of $\TG$ and property (\ref{def:regular:normalize}) of regular foundations (which is the only property of regular foundations needed for this proof).
\end{proof}


\begin{example}[Continued from Ex.~\ref{ex:mmt:algebra}]\label{ex:mmt:flat}
  The flattening of the theory graph of our running example contains the module
  declarations in Fig~\ref{fig:moddecs}, where we omit all types for simplicity
\begin{figure}[ht]\centering\vspace*{-1em}
\fbox{$\mathll{
\thdeclm{\mpath{e}{\cn{Monoid}}}{\mpath{f}{\cn{FOL}}}{
     \symdd{\cn{comp}}{}{},\; \symdd{\cn{unit}}{}{}
} \nl
\thdeclm{\mpath{e}{\cn{CGroup}}}{\mpath{f}{\cn{FOL}}}{
     \symdd{\cnpath{mon,comp}}{}{},\; \symdd{\cnpath{mon,unit}}{}{},\; \symdd{\cn{inv}}{}{}
} \nl
\vwdeclm{\mpath{e}{\cnpath{CGroup,mon}}}{\mpath{e}{\cn{Monoid}}}{\mpath{e}{\cn{CGroup}}}
        {\mmtident{\mpath{f}{\cn{FOL}}}}{\nldecl\tb
     \maps{\cn{comp}}{\triple{e}{\cn{CGroup}}{\cnpath{mon,comp}}},\nldecl\tb
     \maps{\cn{unit}}{\triple{e}{\cn{CGroup}}{\cnpath{mon,unit}}}
} \nl
\thdeclm{\mpath{e}{\cn{Ring}}}{\mpath{f}{\cn{FOL}}}{\nldecl\tb
     \symdd{\cnpath{add,mon,comp}}{}{},\; \symdd{\cnpath{add,mon,unit}}{}{},\; \symdd{\cnpath{add,inv}}{}{},\; 
     \symdd{\cnpath{mult,comp}}{}{},\; \symdd{\cnpath{mult,unit}}{}{}
} \nl
\vwdeclm{\mpath{e}{\cnpath{Ring,add}}}{\mpath{e}{\cn{CGroup}}}{\mpath{e}{\cn{Ring}}}
        {\mmtident{\mpath{f}{\cn{FOL}}}}{\nldecl\tb
     \maps{\cnpath{mon,comp}}{\triple{e}{\cn{Ring}}{\cnpath{add,mon,comp}}},\nldecl\tb
     \maps{\cnpath{mon,unit}}{\triple{e}{\cn{Ring}}{\cnpath{add,mon,unit}}},\nldecl\tb
     \maps{\cnpath{inv}}{\triple{e}{\cn{Ring}}{\cnpath{add,inv}}}
} \nl
\vwdeclm{\mpath{e}{\cnpath{Ring,mult}}}{\mpath{e}{\cn{Monoid}}}{\mpath{e}{\cn{Ring}}}
        {\mmtident{\mpath{f}{\cn{FOL}}}}{\nldecl\tb
     \maps{\cn{comp}}{\triple{e}{\cn{Ring}}{\cnpath{mult,comp}}},\nldecl\tb
     \maps{\cn{unit}}{\triple{e}{\cn{Ring}}{\cn{mult,unit}}}
} \nl
\vwdef{\mpath{e}{\cnpath{Ring,add,mon}}}{\mpath{e}{\cn{Monoid}}}{\mpath{e}{\cn{Ring}}}
      {\ö{\mpath{e}{\cnpath{CGroup,mon}}}{\;\mpath{e}{\cnpath{Ring,add}}}}
}$}\vspace*{-.5em}
\caption{Module Declarations for the running example}\label{fig:moddecs}\vspace*{-1em}
\end{figure}
\end{example}

Two features of {\mmt} are not eliminated in the flattening: meta-theories and filtering.

Regarding meta-theories, the definitions and results in this section could be easily extended to elaborate meta-theories as well. For example, a meta-theory $\qM$ can be reduced to a structure that instantiates $\qM$ and has some reserved name. In fact, that is what we did in an earlier version of {\mmt}~\cite{rabe:thesis:08}. However, this is not desirable because both humans and machines can use meta-theories to relate {\mmt} theories to their semantics. In particular, constants of the meta-theory are often treated differently than the others; for example, their semantics might be hard-coded in an implementation.

Regarding filtering, the situation is more complicated. Imagine a constant declaration $\symdd{\qc}{\omt}{\hid}$ in the flat theory graph. This is particularly intuitive if we think of $\qc$ as a theorem stating $\omt$. Then $\symdd{\qc}{\omt}{\hid}$ means that the theorem holds but its proof is filtered because it relies on a filtered assumption.

It is now a foundational question how to handle this case. One possibility is to delete
the declaration of $\qc$. This is especially appealing from a type/proof theoretical
perspective where constant declarations are what defines the existence of objects and
their meaning. This community might argue that if the proof is filtered, then the theorem
is useless because it can never be applied or verified. Consequently, it can just as well
be removed. Another possibility is to replace $\qc$ with the declaration
$\symdd{\qc}{\omt}{}$, i.e., to turn it into an axiom. This is appealing from a set/model
theoretical perspective where constant declarations merely introduce names for objects that
exist in the models. This community might argue that it is irrelevant whether the proof is
filtered or not as long as we know that there is one.

In order to stay neutral to this foundational issues, we do not elaborate filtering. Instead, we leave all filtered declarations in the flattened signature and leave it to the foundation to decide whether they are used or not.
\medskip

The most important practical aspect of the flattening in {\mmt} is not its existence but
that it can be applied incrementally. This is significantly more difficult. Consider a theory
graph \[\TG_0,\;\thdeclm{\qT}{\qM}{\theta_0,\;\impdd{\qi}{\qS}{\sigma},\;\theta_1},\;\TG_1.\]
We would like to flatten only the structure $\lpath{\qT}{\qi}$. Then the structure can be
replaced with a translated copy of the body of $\qS$.

For example $\symdd{\qc}{\omt}{\undef}$ is translated to $\symdd{\lapp{\qi}{\qc}}{\omt'}{\undef}$, where $\omt'$ is the translation of $\omt$. In particular, in $\omt'$ all names referring to constant of $\qS$ must be prefixed with $\qi$. If $\qi$ has an assignment $\maps{\qc}{\omd'}$, then the declaration is translated to $\symdd{\lapp{\qi}{\qc}}{\omt'}{\omd'}$.

We obtain incrementality if structure declarations in $\qS$ are not flattened
recursively. This is possible in {\mmt}. For example, for a structure
$\dimpdd{\qh}{\qR}{}$ in the body of $\qS$, a structure
$\dimpdd{\lapp{\qi}{\qh}}{\qR}{\ö{\lpath{\qS}{\qh}}{\lpath{\qT}{\qi}}}$ can be added to
$\qT$ rather than adding all induced constants $\spath{\qT}{\qi,\qh,\qc}$. Individual
assignments to structures can be flattened similarly.



 \section{Specific Foundations}\label{sec:mmt:specific_foundations}
   To define a specific foundation, we need to define the judgments
$\otermequal[\Yps]{\TG}{\qT}{\omega}{\omega'}$ and $\otermtype[\Yps]{\TG}{\qT}{\omega}{\omega'}$.

For a fixed theory graph, let $<$ be the transitive closure of the relation ``$X$ has
meta-theory $Y$''. Then the foundational theories are the $<$-maximal ones; and for every
other theory $\qT$, there is a unique foundational theory $\qM$ with $\qT<\qM$, which we
call \defemph{the foundational theory of} $\qT$. A specific foundation is typically
coupled with a certain foundational theory $\qM$ and only defines
$\otermequal{\TG}{\qT}{\omega}{\omega'}$ and $\otermtype{\TG}{\qT}{\omega}{\omega'}$ for
theories $\qT<\qM$. For example, a foundation for set theory could be coupled with the
foundational theory $\cn{ZFC}$.

Foundational theories and foundations can be given for a wide variety of formal languages. As examples, we give them for two very different languages: {\openmath} and LF.

\subsection{OpenMath} 

{\openmath}~\cite{openmath} is used for the communication of set theory-based
mathematical objects over the internet. {\openmath} content dictionaries correspond to
{\mmt} theories, so that {\mmt} yields a module system for {\openmath} content
dictionaries. Pure {\openmath} is an untyped language, in which $\alpha$-conversion of bound
variables is the only non-trivial equality relation. Clearly, this foundation is very easy
to implement.

The foundational theory for {\openmath} is empty because {\openmath} does not use any
predefined constant names.
Thus the standard content dictionaries can be
introduced as {\mmt} theories with that meta-theory. We can define a
foundation for {\openmath} as follows.

Firstly, $\otermtype{\TG}{\qT}{\omega}{\omega'}$ holds iff one of the following holds:
\begin{enumerate}
	\item $\oterm{\TG}{\qT}{\omega}$ and $\omega'=\undef$,
  \item $\omega=\undef$ and $\omega'=\undef$.
\end{enumerate}

To understand why this characterizes {\openmath}, consider how it affects the rules $\TGsymbol$ and $\TGinstsymb$. According to rule $\TGsymbol$, a constant declaration $\symdd{\qc}{\omt}{\omd}$ is only well-formed if $\otermtype{\TG}{\qT}{\omd}{\omt}$. Thus, each of the above cases leads to one kind of constant declaration: The first case is used to define a constant to be equal to some term. The second case is used to declare undefined constants. All constants are untyped.

According to rule $\TGinstsymb$, an assignment $\maps{\qc}{\omd}$ must satisfy that $\omd$ is typed by the translation of the type of $\qc$. Since all constants are untyped, this is vacuous.

Secondly, $\otermequal{\TG}{\qT}{\omega}{\omega'}$ is the smallest relation on structurally well-formed terms that
\begin{itemize}
	\item is reflexive,
	\item is closed under substitution of equals,
	\item is closed under $\alpha$-renaming,
	\item respects normalization, i.e, $\otermequal{\TG}{\qT}{\omega}{\rewr{\omega}}$.
\end{itemize}

\begin{theorem}
The foundation for {\openmath} is regular.
\end{theorem}
\begin{proof}
All properties can be verified directly.
\end{proof}

Foundations for other untyped languages such as set theories can be defined similarly. The main difference is that significantly more complicated definitions of the (undecidable) equality judgment must be employed.

\subsection{The Edinburgh Logical Framework (LF)}

LF~\cite{lf} is a logical framework based on dependent type theory. Being a logical framework, it represents both logics and theories as LF signatures. {\mmt} subsumes this approach by also representing LF as a (foundational) {\mmt} theory. As for {\openmath}, {\mmt} yields a module system for LF.

The foundational theory for LF is given by:
\[\thdecl{\cn{LF}}{\symdd{\cn{type}}{}{},\;\symdd{\cn{kind}}{}{},\;\symdd{\cn{lambda}}{}{},\;
                  \symdd{\cn{Pi}}{}{}}.\]
$\cn{type}$ is the kind of all types. $\cn{kind}$ is the universe of kinds; it does not occur in concrete syntax for LF, but is needed as the {\mmt} type of all well-formed LF kinds. $\cn{Pi}$ is the dependent type constructor, and $\cn{lambda}$ its introductory form. The application of {\mmt} can be used as the eliminatory form of $\cn{Pi}$.

If $\qT=\cn{LF}$, only the typing judgment $\otermtype{\TG}{\cn{LF}}{\undef}{\undef}$ holds. This is needed to make the untyped constants in the theory $\cn{LF}$ well-formed (see rule $\TGinstsymb$). $\otermequal{\TG}{\cn{LF}}{\omega}{\omega'}$ holds iff $\rewr{\omega}=\rewr{\omega'}$.

Otherwise, typing and equality are defined according to the LF type theory:
\begin{itemize}
  \item For constants, $\otermtype{\TG}{\qT}{\spath{D}{\qc}}{\omt}$ holds if $\qT<\cn{LF}$ and $\elabsym{\TG}{D}{\qc}{\omt}{\_}$.
  \item For other terms, $\otermtype{\TG}{\qT}{\omega}{\omega'}$ holds if $\rewr{\omega}$ is a well-formed LF-term of type $\rewr{\omega'}$ or a well-formed LF-type family of kind $\rewr{\omega'}$. The details are as in~\cite{lf} except that the rule for constants is not needed.
  \item $\otermtype{\TG}{\qT}{\undef}{\omega}$ holds if $\rewr{\omega}$ is a well-formed LF-type or a well-formed LF-kind. This permits declarations of typed or kinded constants.
\end{itemize}

Similarly, the judgment $\otermequal{\TG}{\qT}{\omega}{\omega'}$ is defined by the rules given in the properties (\ref{def:regular:normalize}) and (\ref{def:regular:congruence}) of Def.~\ref{def:foundation:regular} and the equality rules for LF given in~\cite{lf}.

\begin{theorem}
The foundation for LF is regular.
\end{theorem}
\begin{proof}
The properties (\ref{def:regular:normalize}) and (\ref{def:regular:congruence}) are built into the definition.
Property (\ref{def:regular:morphism}) follows from the results in~\cite{lfencodings} after observing that every type-preserving mapping from $\qS$ to $\qT$ yields an LF signature morphisms from $\ov{\qS}$ to $\ov{\qT}$. Here $\ov{\qS}$ denotes the union of the bodies of all theories $D$ with $\qT\leq D<\cn{LF}$.
\end{proof}

Regular foundations for any pure type system and for other type theories can be given in
the same way.



 \section{Web-Scalability}\label{sec:mmtweb}
   Because {\mmt} documents are transparent to the semantics, they have been deliberately
ignored so far. But documents play a central for web-scalability because they permit the
packaging and distribution of theory graphs. We will discuss them in
Sect.~\ref{sec:mmtweb:documents}.  The basis for web-scalability is web
standards-compliance, and we introduce the XML-based concrete, external syntax for {\mmt}
theory graphs in Sect.~\ref{sec:mmtweb:xmlencoding} and a URI-based concrete syntax for
identifiers in Sect.~\ref{sec:mmtweb:URIencoding}. We will also define relative URI
references that are indispensable for scalability.  Finally, we describe in
Sect.~\ref{sec:mmtweb:api} the decomposition of {\mmt} documents into sequences of atomic
declarations, their incremental validation, and a basic query language for atomic document
fragments.



  \subsection{Documents and Libraries}\label{sec:mmtweb:documents}
      Recall that our syntax uses two-partite module identifiers $\mpath{g}{\qI}$. $g$ is a URI that identifies a package, called \defemph{document} in {\mmt}.
We use the syntax \[Doc\bnfas \docdecl{g}{\TG}\] to declare a document $g$ containing the theory graph $\TG$.
A document is called \defemph{primary} if all modules declared within $\TG$ have module identifiers of the form $\mpath{g}{\qI}$. Non-primary documents arise when documents are aggregated dynamically using fragments from different documents, i.e., as the result of a search query; we call those \defemph{virtual} documents in \cite{KRZ:mmttnt:10}.

Within {\mmt} documents, we define two relaxations of the {\mmt} syntax that are important
for scalability and that can be easily elaborated into the official syntax: relative
identifiers and remote references. \defemph{Relative identifiers} and their resolution
into the official identifiers are defined in Sect.~\ref{sec:mmtweb:URIencoding}. We speak
of {\defemph{remote references}} if a document refers to a module that is declared in
some other document. Technically, according to the rules of {\mmt}, such a non-self-contained theory graph would be invalid. Therefore, we make documents with remote references self-contained by adding all referenced remote modules in some valid order at the beginning. This is always possible if there is no cyclic dependency between documents.

The semantics of remote references is well-defined because {\mmt} identifiers are URIs and thus globally unique. However, they are not necessarily URLs and thus do not necessarily indicate physical locations from which the remote module could be retrieved. Therefore, we make use of a \defemph{catalog} that translates {\mmt} URIs into URLs, which give the physical locations. This way applications are free to retrieve content from a variety of backends, such as file systems, databases, or local working copies, in a way that is transparent to the {\mmt} semantics.
\medskip

We call a collection of documents together with a catalog an {\mmt} \defemph{library}. A library's document collection can be anything from a self-contained document to (the {\mmt}-relevant subset of) the whole internet. The central component is the catalog that defines the meaning of identifiers in terms of physical locations. Adding a document to a library may include the upload of a physical document, but may also simply consist in adding some catalog entries.

Well-formedness of libraries is checked incrementally by checking individual documents when they are added. A document 
$\docdecl{g}{\TG}$ is \defemph{well-formed} relative to a library $L$ if the following hold:
\begin{itemize}
\item If a module identifier declared in $\TG$ already exists in $L$, then the two modules must be identical.
\item $\TG_L$ is a well-formed theory graph where $\TG_L$ arises by prepending all remotely referenced modules according to their resolution in $L$.
\end{itemize}

It is easy to prove that if we only ever add well-formed documents to an initially empty library, all modules in the library can be arranged into a single well-formed theory graph. This can be realized, for example, by implementing $L$ as a database that rejects the commit of ill-formed content (see Sect.~\ref{sec:mmtweb:implementation}). Thus, libraries provide a safe and scalable way of building large theory graphs.



  \subsection{XML-based Concrete Syntax}\label{sec:mmtweb:xmlencoding}
      For the XML syntax, we build on the {\omdoc} format~\cite{omdoc}, which already integrates
some of the primitive notions of {\mmt} including the {\mathml} 3 syntax for terms
(interpreted as {\openmath} objects). In fact, the {\mmt} data model and the XML syntax
presented here will form the kernel of the upcoming version of the {\omdoc}
format.

The XML grammar mostly follows the abstract grammar of {\mmt}. Theory graphs are
{\element{omdoc}} elements with {\element{theory}} and {\element{view}} elements as
children. The children of {\element{theory}} elements are {\element{constant}} and
{\element{structure}} elements. And the children of {\element{view}} and structure
elements are {\element{conass}} and {\element{strass}} elements. Both terms and morphisms
are represented as strict content {\mathml} expressions.
\def\meta#1{\langle\kern-.2em\langle\hbox{\sl #1}\rangle\kern-.2em\rangle}
\begin{figure}
\begin{center}
\begin{tabular}{|l|l|}\hline
$\enc{\spath{\qT}{\qc}}$ &
$\enctrip{\spath{\qT}{\qc}}$
\\\hline
$\enc{\yps}$ &
\begin{lstlisting}
<m:ci>$\yps$</m:ci>
\end{lstlisting}
\\\hline
 $\enc{\hid}$ &
\begin{lstlisting}
$\omdocoms{filtered}$
\end{lstlisting}
\\\hline
 $\enc{\ma{\omega}{\mu}}$ &
\begin{lstlisting}
<m:apply>
  $\omdocoms{morphism-application}$
  $\enc{\omega}$
  $\enc{\mu}$
</m:apply>
\end{lstlisting}
\\\hline
 $\enc{\oma{\omega_1,\ldots,\omega_n}}$ &
\begin{lstlisting}
<m:apply>$\enc{\omega_1}\;\ldots\;\enc{\omega_n}$</m:apply>
\end{lstlisting}
\\\hline
$\enc{\ombind{\omega_1}{\Yps}{\omega_2}}$ &
\begin{lstlisting}
<m:bind>
  $\enc{\omega_1}$
  $\enc{\Yps}$
  $\enc{\omega_2}$
</m:bind>
\end{lstlisting}
\\\hline
$\enc{\cdot}$ &
(empty sequence)
\\\hline
$\enc{\Yps,\yps[:\tau][=\delta]}$ &
\begin{lstlisting}
$\enc{\Yps}$
<m:bvar>
  <m:semantics>
    <m:ci name="$\yps$"/>
    [<m:annotation-xml base="$\meta{MMTURI}$"
                       cd="mmt" name="type">
      $\enc{\tau}$
     </m:annotation-xml>]
    [<m:annotation-xml base="$\meta{MMTURI}$"
                       cd="mmt" name="value">
      $\enc{\delta}$
     </m:annotation-xml>]
  </m:semantics>
</m:bvar>
\end{lstlisting}
\\\hline
%
\multicolumn{2}{|l|}{$\protect\meta{MMTURI}$ = \texttt{http://cds.omdoc.org/omdoc/mmt.omdoc}}\\\hline
\end{tabular}
\caption{XML Encoding of Terms}\label{fig:mmtweb:xmlencoding:terms}
\end{center}
\end{figure}

\begin{variant}{\modexp}
\begin{figure}
\begin{center}
\begin{tabular}{|l|l||l|l|}\hline
$\enc{\ql}$ & 
\begin{lstlisting}
$\encuri{\ql}$
\end{lstlisting}
&
$\enc{\qT}$ & 
\begin{lstlisting}
<m:csymbol $\encuri{\qT}$/>
\end{lstlisting}
\\\hline
$\enc{\mmtident{\qT}}$ &
\begin{lstlisting}
<m:apply>
  $\omdocoms{identity}$
  <m:csymbol $\enc{\qT}$/>
</m:apply>
\end{lstlisting}
&
$\enc{\öö{\mu_1}{\ldots}{\mu_n}}$ &
\begin{lstlisting}
<m:apply>
  $\omdocoms{composition}$
  $\enc{\mu_1}\ldots\enc{\mu_n}$
</m:apply>
\end{lstlisting}
\\\hline
$\enc{\poiim{\mu_1}{\mu_2}}$ &
\begin{lstlisting}
<m:apply>
  $\omdocoms{morphism-union}$
  $\enc{\mu_1}$
  $\enc{\mu_2}$
</m:apply>
\end{lstlisting}
&
$\enc{\poim{\mu}{\mu_1}{\mu_2}}$ &
\begin{lstlisting}
<m:apply>
  $\omdocoms{morphism-push-along-with}$
  $\enc{\mu_1}$
  $\enc{\mu}$
  $\enc{\mu_2}$
</m:apply>
\end{lstlisting}
\\\hline
$\enc{\poiw{\mu}{\Theta}}$ &
\begin{lstlisting}
<m:apply>
  $\omdocoms{morphism-raise-to}$
  $\enc{\mu}$
  $\enc{\Theta}$
</m:apply>
\end{lstlisting}
&
$\enc{\poii{\Theta_1}{\Theta_2}}$ &
\begin{lstlisting}
<m:apply>
  $\omdocoms{theory-union}$
  $\enc{\Theta_1}$
  $\enc{\Theta_2}$
</m:apply>
\end{lstlisting}
\\\hline
$\enc{\poi{\mu}{\Theta_1}{\Theta_2}}$ &
\begin{lstlisting}
<m:apply>
  $\omdocoms{theory-push-along-with}$
  $\enc{\Theta_1}$
  $\enc{\mu}$
  $\enc{\Theta_2}$
</m:apply>
\end{lstlisting}
&&\\\hline
\end{tabular}
\caption{XML Encoding of Morphism and Theory Expressions}\label{fig:mmtweb:xmlencoding:modexp}
\end{center}
\end{figure}
\end{variant}

\begin{variant}{\not\modexp}
\begin{figure}
\begin{center}
\begin{tabular}{|l|l|}
\hline
$\enc{\ql}$ & 
\begin{lstlisting}
$\enctrip{\ql}$
\end{lstlisting}
\\\hline
$\enc{\mmtident{\qT}}$ &
\begin{lstlisting}
<m:apply>
  $\omdocoms{identity}$
  $\enctrip{\qT}$
</m:apply>
\end{lstlisting}
\\\hline
$\enc{\öö{\mu_1}{\ldots}{\mu_n}}$ &
\begin{lstlisting}
<m:apply>
  $\omdocoms{composition}$
  $\enc{\mu_1}\ldots\enc{\mu_n}$
</m:apply>
\end{lstlisting}
\\\hline
\end{tabular}
\caption{XML Encoding of Morphisms}\label{fig:mmtweb:xmlencoding:modexp}
\end{center}
\end{figure}
\end{variant}

The definition of the XML encoding $\enc\cdot$ is given in
Fig.~\ref{fig:mmtweb:xmlencoding:terms}, \ref{fig:mmtweb:xmlencoding:modexp}, and \ref{fig:mmtweb:xmlencoding:structural}. The
encoding of identifiers is given in Sect.~\ref{sec:mmtweb:URIencoding}. In the definition
of the encoding, we assume that the following namespace bindings are in effect:
\begin{lstlisting}
xmlns="http://www.omdoc.org/ns/omdoc"
xmlns:m="http://www.w3.org/1998/Math/MathML"
\end{lstlisting}
Moreover, we assume a special {\openmath} content dictionary with \texttt{cdbase} \url{http://cds.omdoc.org/omdoc/mmt.omdoc} and name \texttt{mmt} declaring the following symbols:
\begin{center}
\begin{tabular}{|l|l|l|}
\hline
Name  &  Intuition & Role \\
\hline
\texttt{type} & $:$ & attribution \\
\texttt{value} & $=$ & attribution \\
\texttt{filtered} & $\hid$ & constant \\
\texttt{identity} & $\mmtident{\qT}$ & application\\
\texttt{composition} & $\ö{\mu}{\mu}$ & application \\
\texttt{morphism-application} & $\ma{\omega}{\mu}$ & application
\iv{\modexp}{%
\\ \texttt{theory-union} & $\poii{\Theta}{\Theta}$ & application
\\ \texttt{morphism-union} & $\poiim{\mu}{\mu}$ & application
\\ \texttt{theory-push-along-with} & $\poi{\mu}{\Theta}{\Theta}$ & application
\\ \texttt{morphism-push-along-with} & $\poim{\mu}{\mu}{\mu}$ & application
\\ \texttt{morphism-raise-to} & $\poiw{\mu}{\Theta}$ & application
}
\\ \hline
\end{tabular}
\end{center}

These symbols are used to encode the {\mmt} primitives discussed in Sect.~\ref{sec:mmt:syntax}.
We write $\omdocoms{n}$ for the element
\begin{lstlisting}
<m:csymbol base="http://cds.omdoc.org/omdoc/mmt.omdoc" module="mmt" name="$n$"/>
\end{lstlisting}

The encoding of the structural levels in Fig.~\ref{fig:mmtweb:xmlencoding:structural} is
straightforward.

The encoding of terms in Fig.~\ref{fig:mmtweb:xmlencoding:terms} is similar to the
encoding of {\openmath} objects in strict content MathML \cite{mathml3}. It differs in some minor respects:
\begin{itemize}
\item We use a \lstinline{base} attribute to give the document URI (interpreted as the ``content dictionary base'' in {\mathml} 3) of {\element{csymbol}s} and annotations. This is necessary because {\mathml} 3 does not provide a way to ascribe different CD bases to individual symbols except when format-specific mechanisms are defined by the format in which {\mathml} is embedded. For {\mmt}, this mechanism is given by the \lstinline{base} attribute and Def.~\ref{def:base-URI}.
\item The {\element{csymbol}} element is used to refer to both symbols and modules. This is only necessary when encoding modular {\mmt} theory graphs.
\item Symbol and module names permit a larger set of characters. In particular, the forward slash character that we use for constructing theory paths is not allowed in names, which are restricted to NCNames (see \cite{xml}). We do not normalize these away here and assume an omitted encoding step that eliminates the offending characters.
\item We do not enclose {\openmath} objects in {\element{math}} elements. This is redundant due to the use of XML namespaces.
\end{itemize}
Alternatively, we could use the XML encoding of {\openmath} objects defined by the {\openmath} 2 standard~\cite{openmath}. 

\begin{figure}
\begin{center}
\begin{tabular}{|l|l|l|}\hline
Document & $\enc{\docdecl{g}{Mod_1,\ldots,Mod_n}}$ &
\begin{lstlisting}
<omdoc base="$g$">
  $\enc{Mod_1}\ldots\enc{Mod_n}$
</omdoc>
\end{lstlisting}
\\\hline
Theory & $\enc{\thdeclm{\qT}{[\qM]}{S_1,\ldots,S_n}}$ &
\begin{lstlisting}
<theory name="$\qT$" 
       [meta="$\encuri{\qM}$"]>
  $\enc{S_1}\ldots\enc{S_n}$
</theory>
\end{lstlisting}
\iv{\modexp}{
\\
\cline{2-3}& $\enc{\thdef{\qT}{\Theta}}$ & 
\begin{lstlisting}
<theory name="$\qT$">
  <definition>$\enc{\Theta}$</definition>
</theory>
\end{lstlisting}
}
\\\hline

View & $\enc{\vwdeclm{\ql}{\qS}{\qT}{[\mu]}{\sigma}}$ &
\begin{lstlisting}
<view name="$\ql$" from="$\encuri{\qS}$" 
      to="$\encuri{\qT}$">
  [<include>$\enc{\mu}$</include>]
  $\enc{\sigma}$
</view>
\end{lstlisting}
\\\cline{2-3}
 & $\enc{\vwdef{\ql}{\qS}{\qT}{\mu}}$ & 
\begin{lstlisting}
<view name="$\ql$" from="$\encuri{\qS}$" 
      to="$\encuri{\qT}$">
  <definition>$\enc{\mu}$</definition>
</view>
\end{lstlisting}
\\\hline
Constant
 & $\enc{\qc[:\tau][=\delta]}$ &
\begin{lstlisting}
<constant name="$\qc$">
  [<type>$\enc{\tau}$</type>]
  [<definition>$\enc{\delta}$</definition>]
</constant>
\end{lstlisting}
\\\hline
Structure & $\enc{\impddm{\qi}{\qS}{[\mu]}{\sigma}}$ & 
\begin{lstlisting}
<structure name="$\qi$" 
           from="$\encuri{\qS}$">
  [<include>$\enc{\mu}$</include>]
  $\enc{\sigma}$
</structure>
\end{lstlisting}
\\\cline{2-3}
 & $\enc{\dimpdd{\qi}{\qS}{\mu}}$ & 
\begin{lstlisting}
<structure name="$\qi$" 
           from="$\encuri{\qS}$">
  <definition>$\enc{\mu}$</definition>
</structure>
\end{lstlisting}
\\\hline
Assignment & $\enc{Ass_1,\ldots, Ass_n}$ &
\begin{lstlisting}
$\enc{Ass_1}\ldots\enc{Ass_n}$
\end{lstlisting}
\\\cline{2-3}
& $\enc{\maps{\qc}{\omega}}$ &
\begin{lstlisting}
<conass name="$\enc{\qc}$">
  $\enc{\omega}$
</conass>
\end{lstlisting}
\\\cline{2-3}
& $\enc{\maps{\qi}{\mu}}$ &
\begin{lstlisting}
<strass name="$\enc{\qi}$">
  $\enc{\mu}$
</strass>
\end{lstlisting}
\\\hline
\end{tabular}
\caption{XML Encoding of Structural Levels}\label{fig:mmtweb:xmlencoding:structural}
\end{center}
\end{figure}



  \subsection{URI-based Addressing}\label{sec:mmtweb:URIencoding}
       As defined in the {\mmt} grammar, an \defemph{absolute identifier} of an {\mmt} knowledge item is a document URI $G$, a module identifier $\mpath{G}{M}$, or a symbol identifier $\triple{G}{M}{S}$. It is convenient to unify these
three cases by assuming $M=\epsilon$ and/or $S=\epsilon$ if the respective component is
not present. Then absolute references are always triples $(G,M,S)$.

Similarly, a \defemph{relative identifier} is a triple $(g,m,s)$. $g$ is a relative document reference, i.e., a URI reference as defined in RFC 3986~\cite{uri} but without query or fragment. Note that this includes the case $g=\epsilon$.
$m$ and $s$ are usually of the form $I$, i.e., slash-separated (possibly empty) sequences of non-empty names.
For completeness, we mention that {\mmt} also permits $m$ and $s$ to be relative: If $g=\epsilon$, $m$ may be of the form $/I$, which is a module reference that is interpreted relative to the current module; and if $g=m=\epsilon$, $s$ may also be of the form $/I$, which is a symbol reference that is interpreted relative to the current symbol.

\begin{figure}
\begin{center}
\begin{tabular}{|l|l|l|}\hline
triple & $\enctrip{\triple{g}{\qI}{\qI'}}$ &
\begin{lstlisting}
<m:csymbol base="$g$" cd="$\qI$">$\qI'$</csymbol>
\end{lstlisting}
\\\cline{2-3}
& $\enctrip{\mpath{g}{\qI}}$ &
\begin{lstlisting}
<m:csymbol base="$g$" cd="$\qI$"/>
\end{lstlisting}
\\\hline
URI & $\encuri{\triple{g}{\qI}{\qI'}}$ &
$g?\qI?\qI'$
\\\cline{2-3}
& $\encuri{\mpath{g}{\qI}}$ &
$g?\qI$
\\\hline
\end{tabular}
\caption{XML Encoding of Identifiers}\label{fig:mmtweb:names}
\end{center}
\end{figure}

Since absolute and relative identifiers are both triples, they can be encoded in the same
way. There are two different ways to encode {\mmt} identifiers, which are given in Fig.~\ref{fig:mmtweb:names}. When identifiers occur as XML elements, we use $\enctrip{-}$ to obtain the triple of document, module, and symbol name. If they occur as attribute values, we use $\encuri{-}$ to obtain a string.

This triple-based addressing model takes up an idea (called ``reference by context'') from
{\omdoc} 1.1 that was dropped in {\omdoc} 1.2 because its
semantics could not be rigorously defined without the {\mmt} concepts.  In particular
triples $(g,\qT,\qc)$ correspond to the $(cdbase,cd,name)$ triples of the {\openmath}
standard~\cite{openmath}.

When names occur in attribute values, we encode identifiers as URI strings using $\globalpathsep$ as a separating character. In this encoding, concrete and abstract syntax are identical. Absolute and relative identifiers are encoded as URIs and URI references, respectively. We adopt the convention that trailing but not leading $\globalpathsep$ characters can be dropped. For example, we encode
\begin{itemize}
\item $(g,m,\epsilon)$ as $\mpath{g}{m}$,
\item $(\epsilon,m,s)$ as $\triple{}{m}{s}$,
\item $(\epsilon,\epsilon,s)$ as $\triple{}{}{s}$,
\end{itemize}
This encoding can be parsed back uniquely into triples.
\medskip

\begin{definition}[Relative URI Resolution]\label{def:mmt:uriresolution}
The \defemph{resolution} of relative identifier $R=(g,m,s)$ is defined relative to an absolute identifier $B=(G,M,S)$, which serves as the base of the resolution. The result is the following absolute identifier:
\[\resolve{B}{R}:=\cas{
  (G+g, m, s)
     \mifc g\neq\epsilon \\
  (G, M+m, s)
      \mifc g=\epsilon, m\neq\epsilon \\
  (G,M,S+s)
      \mifc g=m=\epsilon, s\neq\epsilon \\
  (G,M,S)
      \mifc g=m=s=\epsilon
}\]
where $G+g$ denotes the resolution of the URI reference $g$ relative to the URI $G$ as
defined in RFC 3986~\cite{uri}. Furthermore, $M+m$ resolves $m$ relative to
$M$: If $m=/m'$, then $M+m$ arises by appending $m'$ to $M$; otherwise $M+m=m$. $S+s$ is
defined accordingly.
\end{definition}

The above definition yields the ill-formed result $(G,\epsilon,s)$ when resolving a symbol
level reference $R=(\epsilon,\epsilon,s)$ against a document level base
$B=(G,\epsilon,\epsilon)$. We forbid that pathological case, which would correspond to a
symbol being declared outside a theory.
\medskip

To resolve relative identifiers within a document, we need the following:
\begin{definition}[Base URI]\label{def:base-URI}
  Let $\TG$ a theory graph, then we define a {\defemph{base URI}} for {\mmt} expressions
  occurring in $\TG$:
  \begin{enumerate}
  \item The base of a module declaration or a remote reference to a module is the URI of
    the containing document.
  \item The base of symbol declaration is the URI of the containing theory.
  \item The base of an assignment to a symbol is the URI of the codomain theory.
  \item If $\mu$ is a morphism with domain $\qS$, then the bases of $\omega$ in
    $\ma{\omega}{\mu}$, and $\mu'$ in $\ö{\mu'}{\mu}$, are $\qS$.
  \item In all other cases, the base of an expression is the base of the parent node in
    the syntax tree.
  \end{enumerate}
  Furthermore, in the XML encoding of documents, authors may override the base reference
  by using an attribute {\snippet{base}}.  This attribute may be present on any XML element
  occurring in the encoding, and all relative {\mmt} URIs are interpreted relative to the closest enclosing {\snippet{base}} attribute. Thus, {\snippet{base}} is similar to the {\snippet{xml:base}} attribute except that its value is an {\mmt} URI and that relative identifiers are resolved according to Def.~\ref{def:mmt:uriresolution}.
  Note that the value of {\snippet{base}} may itself be relative -- this implies that there is no semantic difference between an empty and an omitted {\snippet{base}} attribute.
\end{definition}

\medskip

URIs are the main data structure needed for cross-application scalability, and our
experience shows that they must be implemented by almost every peripheral system, even
those that do not implement {\mmt} itself. Already at this point, we had to implement them
in SML~\cite{RS:twelfmod:09}, Javascript~\cite{GLR:jobad:09},
XQuery~\cite{ZKR:tntbase:10}, Haskell (for Hets, \cite{hets}), and Bean Shell (for a jEdit
plugin) -- in addition to the Scala-based reference API presented in
Sect.~\ref{sec:reference-implementation}.

This was only possible because {\mmt}-URIs constitute a well-balanced trade-off between
mathematical rigor, feasibility, and URI-compatibility: In particular, due to the use of
the two separators $/$ and $?$ (rather than only one), they can be parsed locally, i.e.,
without access to or understanding of the surrounding {\mmt} document.



  \subsection{An API for Knowledge Management}\label{sec:mmtweb:api}
      We use the concepts introduced above to specify an API for {\mmt} documents. It is
designed around {\mmt} library operations that add or retrieve atomic URI-identified
knowledge items. It is easy to implement and can be quickly integrated with different
front and back ends ranging from HTTP servers to interactive editors.

\paragraph*{Adding Knowledge Items}
{\mmt} fragments are added during the validation algorithm. In general, we distinguish three levels of \defemph{validation} with varying strictness strictness. Plain XML validation is quick but cannot guarantee {\mmt}-well-formedness. The latter is guaranteed by structural validation, which implements the inference system given in this paper. Structural validation uses a default foundation, in which typing and equality of terms is always true. Foundation-relative validation refines structural validation by additionally checking typing and equality constraints by using a plugin for specific foundations.

Structural validation of {\mmt} theory graphs can be implemented by decomposing the theory graph into a sequence of atomic declarations that are validated and added incrementally. Except for structures, every symbol or assignment is an atomic declaration. Declarations of documents, theories, views, and structures, are atomic if the body is empty. For example, a view is decomposed into the declaration of an empty view and one declaration for each assignment.

The {\mmt} inference system is designed such that structural validation is possible. In particular, later atomic declarations can never invalidate earlier ones.
\medskip

\paragraph*{Retrieving Knowledge Items}
To retrieve knowledge items from a library, we use \defemph{atomic queries} given in Fig.~\ref{fig:mmt:xmlget}. These take an {\mmt} URI and return the {\mmt} declaration identified by the elaboration judgments.

Atomic queries permit not only the retrieval of all declarations of the original documents in the library, but also of all induced declarations. Note that there are two ways to combine a structure URI $\mpath{g}{\qT,\qi}$ and a constant $\qc$: The query $\triple{g}{\qT,\qi}{\qc}$ retrieves an assignment provided by $\qi$ for $\qc$ (defaulting
to $\maps{\qc}{\triple{g}{\qT}{\qi,\qc}}$ if there is none); and the query
$\triple{g}{\qT}{\qi,\qc}$ retrieves the induced constant declaration of $\qT$.

\begin{figure}[hbt]
\begin{center}
\begin{tabular}{|l|ll|l|}
\hline
URI & Definedness condition & Result \\ \hline
$g$ & $\docdecl{g}{\TG}$ in $L$ & $\docdecl{g}{\TG}$ \\ \hdashline
$\qT$ & $\elabthy{\TG_L}{\qT}{\theta}$,\tb [$\imports{\qM}{\qT}$] & $\thdeclm{\qT}{[M]}{\theta}$ \\
$\ql$ & $\elablink{\TG_L}{\ql}{\qS}{\qT}{\{\sigma\}}$, [$\imports{\mu}{\ql}$] & $\vwdeclm{\ql}{\qS}{\qT}{[\mu]}{\sigma}$ \\
$\ql$ & $\elablink{\TG_L}{\ql}{\qS}{\qT}{\mu}$ & $\vwdef{\ql}{\qS}{\qT}{\mu}$ \\
\hdashline
$\mpath{\qT}{\qc}$ & $\elabsym{\TG_L}{\qT}{\qc}{\omt}{\omd}$ & $\symdd{\qc}{\omt}{\omd}$ \\
$\mpath{\ql}{\qc}$ & $\elabass{\TG_L}{\ql}{\qc}{\omd}$ & $\maps{\qc}{\omd}$ \\
\hline
\end{tabular}
\caption{Atomic Queries}\label{fig:mmt:xmlget}
\end{center}
\end{figure}

\begin{example}[Continued from Ex.~\ref{ex:mmt:lookup}]
Examples for atomic queries were already indicated in Ex.~\ref{ex:mmt:lookup}.
\begin{center}\footnotesize
\begin{tabular}{|l|l|p{4cm}|}\hline
  atomic query  & returns:& comment\\\hline
  $\triple{e}{\cnpath{CGroup,mon}}{\cn{comp}}$ &
  $\maps{\cnpath{mon,comp}}{\triple{e}{\cn{CGroup}}{\cnpath{mon,comp}}}$ &
  the default assignment for lack of an explicit
  assignment\\\hline
  $\triple{e}{\cn{CGroup}}{\cnpath{mon,comp}}$ &
  $\symdd{\cnpath{mon,comp}}{\ma{\tau}{\mpath{e}{\cnpath{CGroup,mon}}}}{\undef}$ &
  where $\tau$ is as in Ex.~\ref{ex:mmt:lookup} \\\hline 
\end{tabular}
\end{center}
\end{example}

Atomic queries are relatively easy to implement and provide a sufficient interface for higher knowledge management layers to implement many additional services.
For example, we can use them to implement \defemph{local validation}. Given a library $L$ and an implementation of atomic queries, we can validate documents and document fragments relative to $L$ without having to read all of $L$. Instead, the respective atomic query is sent to $L$ whenever a reference to an unknown knowledge item is encountered.

Moreover, if we are interested in structural validation only, it is sufficient to know only the type of a query result (i.e., theory, view, constant, structure, assignment to constant, assignment to structure). This information can be precomputed and cached by the library.

Atomic queries also yield an easy implementation of \defemph{flattening} because they already return all declarations of induced constants and assignments that occur in the flattened theory graph. Therefore, to implement flattening, we only have to know the URIs of the induced declarations. Again this information can be cached by the library, and applications can aggregate the flattened theory graph without having to implement {\mmt}.

\paragraph*{Foundations as Plugins}
An implementation of {\mmt} should provide a plugin interface for foundations. Every
plugin must identify a theory $\qM$, and implement functions that decide or (attempt to
prove) instances of the typing and equality judgments for any theory
$\qT<\qM$. Foundations should be regular, and due to Lem.~\ref{def:foundation:normalize},
they only need to consider flat instances of these judgments. Moreover, due to
Thm.~\ref{thm:mmt:flattening}, they can assume that $\qT$ is flat. Thus, existing
implementations of formal systems for $\qM$ can easily be reused to obtain plugins for the
corresponding foundational theory.



 \section{Implementations}\label{sec:mmtweb:implementation}
     The design of the {\mmt} language has been driven in a tight feedback loop between
  theoretical analysis of knowledge structures and practical implementation efforts. In
  particular, we have evaluated the space of possible module systems along the
  classifications developed in Sect.~\ref{sec:mmt:aspects} in terms of expressivity,
  computational tractability, and scalability in a variety of case studies. The most
  relevant one for {\mmt} is the logic atlas in the LATIN
  project~\cite{project:latin,CHKMR:latinabs:11}, where we are attempting a modular development of logics
  and inference systems currently used in mathematical/logic-based software systems with a
  focus on concept sharing and trans-logic interoperability. These efforts led to three
  implementations, which we will present here. All of them are open source, and can be
  obtained from the authors.

\subsection{The {\mmt} Reference Implementation}\label{sec:reference-implementation}

The {\mmt} implementation~\cite{project:mmt} provides a Scala-based \cite{scala} (and thus fully Java-compatible) open-source implementation for the API from Sect.~\ref{sec:mmtweb:api}.
The core of the implementation acts as a library with atomic add and retrieve methods. XML documents are decomposed, validated, and added incrementally, and retrieval of document fragments is implemented via atomic queries.

The \defemph{validation} algorithm provides a plugin interface for foundations. Every plugin must identify a foundational theory $\qM$, and implement functions that decide or (attempt to prove) instances of the typing and equality judgments for any theory $\qT<\qM$.

Foundations should be regular, and due to Lem.~\ref{def:foundation:normalize}, they only need to consider flat instances of these judgments. Moreover, due to Thm.~\ref{thm:mmt:flattening}, they can assume that $\qT$ is flat. Thus, existing algorithms and implementations for the flat case can be reused, and the {\mmt} reference implementation adds a module system to them. Currently, one such plugin exists for the foundation for LF from Sect.~\ref{sec:mmt:specific_foundations}.

As a by-product of validation, a \defemph{relational representation} of the validated document is generated, which corresponds to an ABox in the {\mmt} ontology. The individuals of this ontology are the valid {\mmt} URIs, and the relations between them include for example ``Constant $c_1$ occurs in the type of constant $c_2$.'' or ``View $v$ has domain $\qS$.''. This information is cached, and the implementation includes a simple relational query language. The combination of atomic queries and relational queries is a simple but powerful interface to {\mmt} libraries.

The library component can be combined with various back and front ends. The back ends implement the \defemph{catalog} that translates {\mmt} URIs into physical locations. The current implementation includes back ends that retrieve documents from remote {\mmt} libraries via HTTP or from local working copies of repositories via file system access. This catalog is fully transparent to the library component.

The front ends provide users and systems access to the library. The current implementation includes a \defemph{shell} and a \defemph{web server} front end. The shell is scriptable and can be used to explicitly retrieve, validate, and query {\mmt} documents. The web server is implemented using the Lift web framework for Scala \cite{lift}. The interaction with the web server proceeds like with the shell except that input and output are passed via HTTP. In particular, the web server can easily be run as a local proxy that provides {\mmt} functionality and file system abstraction to local applications. An example instance of the web server is serving the content of the {\tntbase} repository of the LATIN project \cite{project:latin}.

In fact, the {\mmt} language is significantly larger than presented here. Going beyond the
scope of this paper, it also provides an {\omdoc}-style \defemph{notation language} with a simple declarative syntax
to define renderings of {\mmt} content in arbitrary human- or machine-oriented formats; see~\cite{KMR:presentation:08} for an overview of a precursor of the {\mmt} notation system. Notations can be
grouped into styles, which are themselves subject to the {\mmt} module system. The web
server mentioned above can serve documents as XHTML with presentation {\mathml} and
integrates the JOBAD technology for \defemph{interactive browsing} we presented
in~\cite{GLR:jobad:09}. Another use of notations is as a fast way of translating {\mmt}
into system's concrete input syntax so that {\mmt} can serve as an interchange language.

\subsection{TNTbase -- a Scalable {\mmt}-Compliant Database}\label{sec:tntbase}

The {\tntbase} system~\cite{tntbase} is an open-source versioned XML \defemph{database} developed at
Jacobs University. It was obtained by integrating Berkeley DB XML~\cite{berkeleydbxml}
into the Subversion Server~\cite{svn}, is intended as a basis for collaborative
editing and sharing XML-based documents, and integrates versioning and access of document
fragments. We have extended {\tntbase} with an {\mmt} \defemph{plugin} that makes it {\mmt}-aware~\cite{KRZ:mmttnt:10,ZKR:tntbase:10}.

The most important aspect of this plugin is validation-upon-commit. Using the \texttt{tntbase:validate} property, folders and files can be configured to require validation. Thus, users can choose between no, XML-based, structural, or foundational validation of {\mmt} files. Since the commit of ill-formed files can be rejected, {\tntbase} can guarantee that it only contains well-formed documents. Thus, other systems can use {\mmt}-enriched {\tntbase} for the long term storage of their system libraries and can trust in the correctness of documents retrieved from the database.

Moreover, the plugin computes the
relational representation of a committed well-formed document, which {\tntbase} stores as an XML file along with every {\mmt} file. {\tntbase} exposes the relational representation of {\mmt} documents via an XQuery
interface, and we have implemented a variety of custom queries in an XQuery module that is
integrated into {\tntbase}. {\tntbase} indexes the files containing the relational
representation so that such queries scale very well. For example, our XQuery module
includes a function that computes the transitive closure of the structural dependency
relation between {\mmt} modules and dynamically generates a self-contained {\mmt} document
that includes all dependencies of a given module. Even for small libraries and even if
{\tntbase} runs on a remote server, this query outperforms the straightforward implementation
based on local files.

Moreover, using the virtual documents of {\tntbase}, such generated documents are editable;
see~\cite{tntbase_virtual} for details. {\tntbase} keeps track of how a document was
aggregated and propagates the necessary patches when a changed version of the virtual
document is committed.

\subsection{Twelf -- an {\mmt}-Compliant Logical Framework}\label{sec:twelf}

The {\mmt} implementation from Sect.~\ref{sec:reference-implementation} starts with a generic {\mmt} implementation and adds a plugin for a specific formal language $F$. Alternatively, an invasive implementation is possible, which starts with an implementation of $F$ and adds the {\mmt} module system to it. Such implementations are restricted to theory graphs with a single foundational theory for $F$, but can reuse special features for $F$ such as user interfaces and type inference. We have implemented this for LF as well~\cite{RS:twelfmod:09} using the Twelf implementation \cite{twelf} of LF.

The effect of adding {\mmt} to Twelf is that Twelf becomes a tool for authoring theory graphs with LF as the single foundational theory. A major advantage of this approach is that authors can benefit from the advanced Twelf features, in particular infix parsing, type reconstruction, and implicit arguments. This implementation was used successfully to generate large case studies of {\mmt} theory graphs in~\cite{DHS:algebra:09}, \cite{HR:folsound:10}, and~\cite{IR:foundations:10}.

Twelf also supports several advanced language features that are part of {\mmt} but were
not mentioned in this paper. In particular, this includes nested theories and unnamed
imports between theories and links. Furthermore, fixity and precedence declarations of
Twelf are preserved as {\mmt} notations that are used when rendering the {\mmt} theory
graph.

Twelf can produce {\mmt} documents in XML syntax from its input that are guaranteed to be well-formed. In \cite{CHKMRS:lfhets:11}, we showed how logics written in Twelf can be exported in {\mmt} concrete syntax and imported into and used in the Hets system \cite{hets}.



\section{Related Work}\label{sec:mmt:related}
 \ednote{check name clash resolution in Isabelle}

\Defprop[p = presentation, c = content]{form}{formality}
\defprop[p = physical, l = logical, s = single package]{pack}{packages}
\Defprop[o = open, c = closed]{pimp}{package imports}
\defprop[i = interspersed, s = separated, a = axiom-inheritance]{nimp}{named inheritance}
\defprop[i/f = interfaced/free, e/i = explicit/implicit, t/p = total/partial]{nins}{\quad instantiation}
\defprop{nren}{\quad renaming}
\defprop[s = simple, c = complex, f = filtering]{nhid}{\quad hiding}
\defprop{uimp}{unnamed inheritance$^4$}
\defprop[i = identify, d = distinguish, a = identify iff instantiations agree, e = error]{udia}{\quad diamond semantics}
\defprop[i = overload/identify, q = qualified names, s = shadowing, e = error]{unc}{\quad name clash resolution}
\defprop{uins}{\quad instantiation$^5$}
\defprop{uren}{\quad renaming}
\Defprop{uhid}{\quad hiding$^6$}
\defprop{real}{realizations as objects}
\defprop{gro}{grounded realizations}
\defprop{view}{views/functors}
\defprop{ho}{higher-order}
\Defprop[explicit syntax for the translation along views/functors: y = syntactic, e = semantic]{trans}{translations}
\defprop[m = model theory, e = elaboration]{sem}{semantics}
\defprop[+ = internalized, $\ast$ = additionally an internalized module system as in last column]{int}{internalized}
\defprop{lind}{logic-independent}
\Defprop{find}{foundation-independent}
\defprop{uri}{URIs as identifiers}
\defprop{xml}{XML syntax}

\defsystem[id=om,form=pc,pack=l,pimp=o,find=+,uri=+,xml=+]{OpenMath (CD)}
\defsystem[id=omdoc,form=pc,pack=l,pimp=o,uimp=i,udia=i,unc=q,uins=fep,uhid=f,view=+,find=+,uri=+,xml=+]{OMDoc (theory)}
\defsystem[id=obj,form=c,pack=s,nimp=s,nins=iit,uimp=s,udia=i,unc=i,view=+,sem=m,lind=+]{OBJ (theory)}
\defsystem[id=asl,form=c,uimp=a,uins=fet,uhid=c,view=+,trans=y,sem=m,lind=+]{ASL (specification)}
\defsystem[id=devgraph,form=c,uimp=a,uins=fet,uhid=c,view=+,trans=y,sem=m,lind=+]{dev. graphs (node)}
\defsystem[id=casl,form=c,pack=p,pimp=c,uimp=i,udia=i,unc=i,uins=iep,uhid=s,uren=+,view=+,sem=m,lind=+]{CASL (specification)}
\defsystem[id=imps,form=c,pack=s,uimp=a,view=+,sem=e]{IMPS (theory)}
\defsystem[id=pvs,form=c,pack=p,pimp=c,uimp=i,udia=a,unc=q,uins=iit,uhid=s,sem=m]{PVS (theory)}
\defsystem[id=isa,form=c,pack=s,pimp=c,nimp=s,nins=fep,uimp=s,udia=a,unc=e,uins=fep,gro=+,view=+,sem=e,lind=+]{Isabelle (locale)}
\defsystem[id=nuprl,form=c,int=$\ast$]{Nuprl}
\defsystem[id=coq,form=c,pack=p,pimp=o,uimp=i,udia=e,unc=e,uins=fep,uhid=s,nimp=i,nins=fep,nhid=s,real=+,gro=+,view=+,trans=e,sem=e,int=$\ast$]{Coq (module type)}
\defsystem[id=agda,form=c,pack=p,pimp=c,nimp=i,nins=iit,nren=+,nhid=s,sem=e,int=$\ast$]{Agda (module)}
\defsystem[id=sml,form=c,pack=s,uimp=i,udia=d,unc=s,uins=fep,uhid=s,nimp=i,nins=fep,nhid=s,real=+,gro=+,view=+,trans=e,sem=e]{SML (signature)}
\defsystem[id=java,form=c,pack=l,pimp=o,nimp=i,nins=iit,nhid=s,uimp=s,udia=i,unc=i,uins=iit,uhid=s,real=+,gro=+,view=+,ho=+,trans=e,sem=e,int=+,uri=+]{Java/Scala (class)}
\defsystem[id=mmt,form=c,pack=l,pimp=o,nimp=i,nins=fep,nhid=f,real=+,gro=+,view=+,trans=ye,sem=e,lind=+,find=+,uri=+,xml=+]{MMT (theory)}
\defsystem[id=twelf,form=c,pack=l,pimp=o,nimp=i,nins=fep,nren=+,uimp=i,udia=i,unc=q,uren=+,real=+,gro=+,view=+,trans=ye,sem=e,lind=+,uri=+]{Twelf (signature)}
\defsystem[id=internal,nimp=i,nins=iit,real=+,gro=+,view=+,ho=+,trans=e]{Internalized}

In this section, we survey the state of the art in module systems for formal languages using the terminology developed in Sect.~\ref{sec:mmt:aspects} and relate {\mmt} to them. Fig.~\ref{fig:reltable} gives an overview of the discussed systems.

\begin{figure}[ht!]
\centering\setlength\tabcolsep{.85mm}
{\footnotesize\reltable}
\caption{Features of Module Systems}\label{fig:reltable}
\end{figure}

\paragraph*{Mathematical Language}
Even though mathematical knowledge can vary greatly in its presentation as
well as its level of formality and rigor, there is a level of deep semantic structure that
is common to all forms of mathematics. This large-scale structure of mathematical
knowledge is much less apparent than that of formulas and is usually implicit in informal
representations.  Experienced mathematicians are nonetheless aware of it, and use it for
navigating in and communicating mathematical knowledge.

Much of this structure can be found in networks of theories such as those in a monograph
``Introduction to Group Theory'' or a chapter in a textbook. The relations among such
theories are described in the text, sometimes supported by mathematical statements called
``representation theorems''. We can observe that mathematical texts can only be understood
with respect to a particular mathematical context given by a theory which the reader can
usually infer from the document, e.g., from the title or the specialization of the
author. The intuitive notion of meta-theory is well-established in mathematics, but again
it is mainly used informally. Formal definitions are found in the area of logic where a
logic is used as the meta-language of a logical theory.

Mathematical theories have been studied by mathematicians and logicians in the search of a
rigorous foundation for mathematical practice. They have usually been formalized as
collections of symbol declarations and axioms. Mathematical reasoning often involves
several related mathematical theories, and it is desirable to exploit these relationships
by moving theorems between theories. The first systematic, large-scale applications of
this technique in mathematics are found in the works by
Bourbaki~\cite{bourbakisets,bourbakialgebra}, which tried to prove every theorem in the
theory with the smallest possible set of axioms.

This technique was formalized in~\cite{littletheories}, which introduced the \emph{little theories} approach. Theories are studied as formal objects. And structural relationships between them are represented as theory morphisms, which serve as conduits for passing information (e.g., definitions and theorems) between theories (see~\cite{farmerintertheory}).

\paragraph*{Web Scale Languages}
The challenge in putting mathematics on the World Wide Web is to capture both notation and
meaning in a way that documents can utilize the human-oriented notational forms of
mathematics and provide machine-supported interactions at the same time. The W3C
recommendation for mathematics on the web is the {\mathml} language~\cite{mathml3}. It
provides two sublanguages: \defemph{presentation {\mathml}} permits the specification of
notations for mathematical formulas, and \defemph{content {\mathml}} is geared towards
specifying the meaning in a machine-processable way. The latter is structurally equivalent
to {\openmath}. In particular, both formats represent the structure of mathematical
formulas as {\openmath} objects, i.e. tree-like expressions built up from constants,
variables, and primitive data types via function applications and bindings.

{\mmt} constants correspond to symbols in {\mathml} and {\openmath} and {\mmt} theories to
{\defemph{content dictionaries}} (CDs). CDs are machine-readable and web-accessible
documents that provide a very simple way to declare mathematical objects for the
communication over the WWW and attach meaning to them. Meaning can be expressed in the
form of axioms or types given as {\openmath} objects representing logical formulas or in
the form informal mathematical text.

{\openmath} provides a certain communication safety over traditional mathematics: It can
no longer be the case that the author writes $\mathbb{N}$ for the set of natural numbers
with $0$, and the reader understands the set of natural number without $0$, as the two
notions of ``natural numbers'' --- even though presented identically --- are represented
by different symbols (probably from different CDs). Thus, the service offered by the
{\openmath/\mathml} approach is one of disambiguation as a base for further machine
support.

In {\mmt} terms, the productions for constants, variables, application, and binding correspond closely to {\openmath}. {\mmt} adds morphism application and the special term $\hid$, and we omit the primitive data types. We use typed and defined variables in analogy to {\mmt} constant declarations and do not use the attributions of {\openmath}.

OpenMath CDs enable formula disambiguation and web scale communication, but the lack of
machine-understandable intra-CD knowledge structure and inter-CD relations preclude
higher-level machine support. Therefore, {\omdoc}~\cite{omdoc} represents mathematical
knowledge at the levels of objects, statements, theories, and documents: {\openmath} and
content {\mathml} are subsumed to represent objects. Statements are symbols, axioms,
definitions, theorems, proofs and occur as declarations within theories. Moreover,
theories may declare unnamed, interspersed, free instantiations, and structured theory morphisms can be declared as in development graphs. Documents provide a basic content-oriented infrastructure for communication and archival.

Syntactically, {\omdoc} and {\openmath} are distinguished from purely formal representation languages by the fact that all formal mathematical elements of the language can be augmented or replaced by natural language text fragments. Semantically, they differ because they do not supplement the formal syntax with a formal semantics.

Some implementations of purely formal representation languages have made use of XML,
{\openmath}/{\mathml}, or {\omdoc} as primary or secondary representation formats. For
example, Mizar~\cite{mizar} uses XML as the primary internal format, and
Matita~\cite{matita} uses content and presentation {\mathml};
Coq~\cite{calcconstructions,coq} provides an OMDoc export, and Isabelle~\cite{isabelle} a
partial XML export. Web-scale languages can in principle serve as standardized interchange
formats between such systems. Some examples of interoperability mediated by {\omdoc} and
{\openmath} are \cite{CHKMRS:lfhets:11,CO:CommProofIntMathDoc00,HorRoz:ossp09}. But
applications have so far been limited due to the lack of an interchange format with a
standardized semantics.

{\mmt} provides such a semantics. It keeps {\omdoc}'s leveled representation but restricts
attention to a subset for which a formal semantics can be developed. Syntactically, the
main addition of {\mmt} is the use of named imports and of theory morphisms as objects.

{\openmath} and {\omdoc} use URIs~\cite{uri} to identify symbol by triples of symbol name,
CD id, and CD base. The CD base is a URI acting as a namespace identifier, which
corresponds to the triples in {\mmt} identifiers. But the formation of {\openmath} URIs is
only straightforward via the one-CD-one-file restriction imposed by {\openmath}, which is
too restrictive in general. {\mmt} is designed such that all knowledge items have
canonical URIs. Moreover, the formation of symbol URIs in {\openmath} and {\omdoc} uses
the \emph{fragment} components of URIs. Therefore, fragment access does not scale well
because clients have to download a complete document and then execute the fragment access
locally. {\mmt} avoids this by using the \emph{query} component of the URI.

\paragraph*{Algebraic Specification Languages}
In algebraic specification, theories are used to specify the behavior of programs and software
components, and realizations (theory morphisms in {\mmt}) are used to enable reuse of
components (structures in {\mmt}) and to formalize refinements of specifications (views in
{\mmt}).

In this setting, implementations can be regarded as refinements into executable
specifications, which we have called \emph{grounded realizations}. This approach naturally
leads to a regime of specification and implementation co-development, where initial,
declarative specifications are refined to take operational issues into
account. Implementations are adapted to changing specifications, and verification
conditions and their proofs have to be adapted as programming errors are found and
fixed. This has been studied extensively, and a number of systems have been developed. We
will discuss OBJ~\cite{obj}, ASL~\cite{asl,specificationarbitrary},
CASL~\cite{caslmanual,hets}, and development graphs~\cite{devgraphs,devgraphshiding} as
representative examples.

\defemph{OBJ} refers to a family of languages based on variants of sorted first-order
logic. It was originally developed in the 1970s based on the Clear programming language
and pioneered many ideas of modular specifications, in particular the use of initial model
semantics~\cite{initialsemantics}. The most important variant is OBJ3; Maude~\cite{maude}
is a closely related system based on rewriting logic. OBJ is a single-package system. Theories and views are similar to {\mmt}. OBJ permits unnamed imports without instantiation and with identify-semantics, and named imports with interfaced, implicit, and total instantiations. All imports are separated. Named imports can be instantiated with views, but more complex realizations cannot be formed.
\medskip

\defemph{ASL} is a generic module system over an arbitrary
institution~\cite{institutions} with a model theoretical semantics. Similar to
institutions, the focus is on abstract modeling rather than concrete syntax. Modules are
called ``specifications'' and are formed using the operations of union (which corresponds
to concatenation of theory bodies in {\mmt}), imports, and complex hiding (which was
introduced by ASL). Imports between specifications are unnamed and do not use
instantiations but only axiom-inheritance and renaming. Unnamed views are used to express
refinement theorems. We gave a representation of ASL in {\mmt} in \cite{CHKMR:hiding:11},
which uses an extension of {\mmt} to accommodate hiding.
\medskip

The \defemph{development graph} language is an extension of ASL specifically designed for
the management of change. The central data structure are theory graphs of theories and two
kinds of links, which correspond to the ones in {\mmt}. (Global) ``definitional links''
are unnamed imports like in ASL and provide axiom-inheritance; (global) ``theorem links''
are partial views where the missing instantiations are treated as proof obligations that
are to be discharged by theorem proving systems. ASL style hiding is supported by
\emph{hiding links}. The Maya system~\cite{maya} implements development graphs for
first-order logic. Like the {\mmt} implementation and contrary to most other systems
discussed here, Maya does not flatten the specification while reading it in. Thus, the
modular information, in particular the theory graph, is available in the internal data
structures. This is much more robust against changes in the underlying modules and
provides a good basis for theorem reuse and management of change.

The development graph calculus uses local links. From the {\mmt} perspective, a local link is a link which filters all but the local constants of its domain. A global theorem theorem link can be decomposed into a set of commuting local theorem links. By finding these local theorem links individually and reusing them where possible, development graphs can avoid redundancy and move theorems between theories. From the {\mmt} perspective, a decomposed global theorem link is simply a set of total views without deep assignments, i.e., views where all structures are mapped to morphisms. Thus, {\mmt} provides not only a representation format for development graphs and decomposed theorem links, but also for intermediate development graphs in which theorem links have been partially decomposed or where local theorem links are postulated but have not been found yet.

Our rules for the module-level reasoning about morphisms are very similar to such decompositions: A judgment about all (possibly imported) constants in $\qS$ is decomposed into separate judgments about the local constants and the structures declared of $\qS$.
\medskip

The common algebraic specification language (\defemph{CASL}) was initiated in 1994 in an attempt to unify and standardize existing specification languages. As such, it was strongly influenced by other languages such as OBJ and ASL. The CASL logics are centered around partial subsorted first-order logic, and specific logics are obtained by specializing (e.g., total functions, no subsorting) or extending (e.g., modal logic or higher-order logic).
CASL uses closed physical packages based on files and called ``libraries''. The modules are called ``specifications'', the imports are unnamed and interspersed, permit renaming, and use the identify-semantics. The overload/identify-semantics is used to handle import name clashes. Instantiations are interfaced, explicit, and total, and map constants to constants. In parametric specifications, special separated imports are used that can be instantiated with views. CASL offers simple hiding.

In HetCASL~\cite{habilmossa} and the Hets system~\cite{hets}, CASL is extended to heterogeneous specifications using  different logics and logic morphism the same specification). Imports and views may go across logics if logic morphisms are attached. This is a very similar to the use of meta-theories and meta-morphisms in {\mmt}. Contrary to {\mmt}, the logics and logic morphisms are implemented in the underlying programming language and not declared within the formal language itself. Hets implements the development graph calculus for heterogeneous specifications.

\paragraph*{Type Theories}
Type theories and related formal languages utilize strong logical systems to express both mathematical statements and proofs as mathematical objects. Some systems like AutoMath~\cite{automath}, Isabelle~\cite{isabelle}, or Twelf~\cite{twelf} even allow the specification of the logical language itself, in which the reasoning takes place. Semi-automated theorem proving systems have been used to formalize substantial parts of mathematics and mechanically verify many theorems in the respective areas.

These systems usually come with a module system that manages and structures the body of knowledge formalized in the system and a library containing a large set of modules. We will consider the module systems of IMPS~\cite{imps}, PVS~\cite{pvs,pvssemantics}, Isabelle~\cite{isabelle}, Coq~\cite{calcconstructions,coq}, Agda~\cite{agda}, and Nuprl~\cite{nuprl}.
We have already discussed the module system of Twelf, which was designed based on {\mmt}, in Sect.~\ref{sec:twelf}.

\defemph{IMPS} was the first theorem proving system that systematically exploited the
``little theories approach'' of separating theories into small modules and moving theorems
along theory morphisms. It was initiated in 1990 and is built around a custom variant of
higher-order logic. It is a single-package system, the imports are unnamed and separated without
instantiations; there is no renaming. Modules can be related via views, which map symbols to symbols.  \medskip

\defemph{PVS} is an interactive theorem prover for a variant of
classical higher-order logic with a rich undecidable type system. The PVS packages are
called ``libraries'' and are physical packages based on directories. Unnamed,
interspersed imports have interfaced, total, and implicit instantiations, which map symbols
to terms. Unnamed imports of the same module are identified if the instantiations agree. There is no renaming, and the import name clash situation is handled using the overload/identify semantics. Simple hiding is supported by export declarations that
determine which names become available upon import. \medskip

\defemph{Isabelle} is an interactive theorem prover based on
simple type theory~\cite{churchtypes} with a structured high-level proof language. Its
packages are called ``theories'' and are identified physically based on files, packaging
is closed. Isabelle provides two generic module systems.

Originally, only axiomatic type classes were used as modules. They permit only inheritance
via unnamed, separated imports without instantiations. Type class ascriptions to type
variables and overloading resolution are used to access the symbols of a type class. Later
locales were introduced as modules in~\cite{isabelle_locales} and gradually extended. In
the current release, locales offer unnamed, separated imports with free instantiations;
renaming is possible. Type classes are recovered as a special case.

Realizations are treated differently depending on whether they are grounded or not and
whether the domain is a type class or a locale: Theory morphism between locales are called
``sublocale'' and ``subclass declarations'', and grounded realizations are called
``interpretation'' for locales and ``instantiation'' for type classes.

Isabelle assigns the semantics of a modular theory by elaboration. Locales are
internalized by locale predicates that abstract over all symbols and assumptions of the
locale; every theorem proved in the locale is relativized by the locale predicate and
exported to the toplevel. Thus, instantiation is reduced to $\beta$-reduction.  \medskip
	
\defemph{Nuprl} is an interactive theorem prover based on a rich undecidable type
theory. It does not provide an explicit module system. However, its type theory is so expressive that it can in principle be used to define an internalized module system as shown in \cite{nuprl_classes}. Then modules, grounded realizations, and higher-order functors can be defined using Nuprl types, terms, and function terms, respectively. Named and unnamed imports are defined using intersection and dependent sum types. But Nuprl does not provide specific module system-like syntax for these notions.
\medskip

\defemph{Coq} is an interactive theorem prover based on the calculus of
constructions~\cite{calcconstructions}. Physical open packages are called ``libraries'' and correspond to directories and files.

The Coq module system is modeled after the SML module system (see below). SML signatures, structures, and functors correspond to Coq module types, modules without parameters, and modules with parameters, respectively. Contrary to SML, no shadowing is used, and errors are signaled instead. In addition, Coq can be used with an internalized higher-order module system using record types. As for Nuprl, this yields modules, grounded realizations, and higher-order functors. Both module systems are used independently. The standard library mainly uses the former. The latter is used systematically in \cite{coq_ssreflect}.
\medskip

\defemph{Agda} is a functional programming language based on Martin-L\"of's dependent type theory \cite{martinlof}. It uses dependent record types to internalize certain theories. In addition, the notion of ``modules'' combines aspects of what we call packages and modules. These modules are physical closed packages based on files and are used mainly for namespace management. Named interspersed imports between modules are possible using nested module declarations where the inner one is defined in terms of a parametric module. These imports carry interfaced, implicit, and total instantiations that map symbols to term. Named imports may not occur as parameters so that this does not yield a notion of functors.

\paragraph*{Programming Languages}
Programming languages differ from the languages mentioned above in that they focus on aspects of execution including input/output and state. But if we ignore those aspects, we find the same module system patterns as in the other languages. We discuss the functional language SML~\cite{sml} and the object-oriented language Java~\cite{java} as examples.
\medskip

\defemph{SML} uses a single-package system that permits the modular design of specifications (called ``signatures'') and realizations (called ``functors'', and if grounded ``structures'').

The specification level module system has signatures as modules. Imports are interspersed and can be named (called ``structure declarations'') or unnamed (called ``inclusions''). Both kinds of imports carry free, explicit, and partial instantiations that map symbols to symbols or structures to realizations. If unnamed imports lead to a diamond situation or a name clash, the later declarations always shadow the previous ones. Views are restricted to inclusion morphisms between signatures (called ``structural subtyping''); these views are implicit and inferred by implementations.

Realizations can themselves be given modularly. A functor is a realization of a signature that is parametric in symbols or structure declarations. Imports between realizations are possible by declaring a structure and defining it to be equal to the result of a functor application. Consequently, these imports are named and interspersed, and the instantiations are interfaced, explicit, and total, map symbols to symbols and structures to realizations. Structures are typed structurally by signatures, which permits simple hiding.

From an {\mmt} perspective, SML signatures, structures, and functors can be unified conceptually. Signatures correspond to {\mmt} theories in which no constant has a definition; structures to {\mmt} theories in which all constants have definitions; and functors to {\mmt} theories where only a few declarations at the beginning (the interface of the functor) have no definition. Both the structural subtyping relation between signatures and the typing relation between structures and signatures correspond to an inclusion view between the respective {\mmt} theories.
\medskip

\defemph{Java} uses open packages with optional imports. Package names are the authority components of URIs~\cite{uri}. Packages are provided in \verb|jar| archive files, and implementations provide a catalog to locate packages that is based on the \verb|classpath|. Java packages are very close to {\mmt} documents. Similar to {\mmt}, Java identifiers are logical and formed from the three hierarchical components package URI, class name, and field name. However, Java uses ``.'' as a separator character both between and within these components and resolves ambiguities dynamically; {\mmt} uses ``?'' and ``/'' so that {\mmt} URIs can be understood statically.

Java modules are called ``classes''. There are two kinds of imports. Firstly, unnamed, separated imports without renaming are called ``class inheritance''; a class may only inherit from one other class though. Secondly, named, interspersed imports are called ``object instantiation'', and the resulting structures ``objects''. Instantiations are interfaced, implicit, and total, but a class may provide multiple interfaces (called ``constructors''), which map symbols to expressions or objects to objects. As constructors may execute code, the expressions passed to the constructor do not have to correspond to symbols or objects declared in the class. Views are restricted to inclusion morphism out of special modules (called ``interfaces'').
Simple hiding is realized via private declarations.

Java internalizes its module system, and functors are subsumed by the concept of methods.

Scala \cite{scala} is a higher-order extension of Java that retains all features listed above for Java. Moreover, Scala permits multiple unnamed imports into the same class by using ``traits''.


\section{Conclusion and Future Work}\label{sec:mmt:conc}
  Formal knowledge is at the core of mathematics, logic, and computer science, and we are
seeing a trend towards employing computational systems like (semi-)automated theorem
provers, model checkers, computer algebra systems, constraint solvers, or concept
classifiers to deal with it. It is a characteristic feature of these systems that they
either have mathematical knowledge implicitly encoded in their critical algorithms or
(increasingly) manipulate explicit representations of this knowledge, often in the form of
logical formulas. Unfortunately, these systems have differing domains of applications,
foundational assumptions, and input languages, which makes them non-interoperable and
difficult to compare and relate in practice. Moreover, the quantity of mathematical
knowledge is growing faster than our ability to formalize and organize it, aggravating the
problem that mathematical software systems cannot easily share knowledge representations.

In this work, we contributed to the solution of this problem by providing a scalable
representation language for mathematical knowledge. We have focused on the modular
organization of formal, explicitly represented mathematical knowledge. We have developed a
classification of modular knowledge representation languages and evaluated the space of
possible module systems in terms of expressivity, computational tractability, and
scalability. We have distilled our findings into one particularly well-behaved system --
{\mmt} -- discussed its properties, and described a set of loosely coupled
implementations.

\subsection{The {\mmt} Language}\label{sec:concl:mmt}

{\mmt} is a foundationally unconstrained module system that serves as a web-scalable
interface layer between computational systems working with formally represented
knowledge. 

{\mmt} integrates successful features of existing paradigms
\begin{itemize}
\item reuse along theory morphisms from the ``little theories'' approach,
\item the theory graph abstraction from algebraic specification languages,
\item categories of theories and logics from model theoretical logical frameworks,
\item the logics-as-theories representation from proof theoretical logical frameworks,
\item declarations of constants and named realizations from type theory,
\item the Curry-Howard correspondence from type/proof theory,
\item URIs as logical namespace identifiers from {\openmath}/{\omdoc} and Java,
\item standardized XML-based concrete syntax from web-oriented representation languages,
\end{itemize}
and makes them available in a single, coherent representational system for the first time.

The combination of these features is reduced to a small set of carefully chosen, orthogonal primitives in order to obtain a simple and extensible language design. In fact, some of the primitives combine so many intuitions that it was rather difficult to name them.
\ednote{MK/FR: discuss naming; documents/packages to namespaces; structure
  to imports; maybe rename view to morphism; link to (atomic) morphism; morphism to
  morphism (expression)?}

{\mmt} contributes three new features:
\begin{list}{}{\setlength{\leftmargin}{0.5em}\setlength{\itemindent}{2.5em}}
  \item[\bf Canonical identifiers]
    By making morphisms named objects, {\mmt} can provide globally unique, web-scalable identifiers for all knowledge items. Even in the presence of modularity and reuse, all induced knowledge items become addressable via URIs. Moreover, identifiers are invariant under {\mmt} operations such as flattening.
\item[\bf Meta-theories]
  The logical foundations of domain representations of mathematical knowledge can be represented as modules themselves and can be structured and interlinked   via meta-morphisms. Thus, the different foundations of systems can be related and the systems made interoperable. The explicit representation of epistemic foundations also benefits systems whose mathematical knowledge is only implicitly embedded into the algorithms: The explicit representation can serve as a documentation of the system interface as well as a basis for verification or testing attempts.
\item[\bf Foundation-independence]
  The design, implementation, and maintenance of large scale logical knowledge management services will realistically only pay off if the same framework can be reused for different foundations of mathematics. Therefore, {\mmt} does not commit to a particular foundation and provides an interface layer between the logical-mathematical core of a mathematical foundation and knowledge management services. Thus, the latter can respect the semantics of the former without knowing or implementing the foundation.
\end{list}

{\mmt} is \defemph{web-scalable} in the sense that it supports the distribution of resources (theories, proofs, etc.) over the internet thus permitting their collaborative development and application. We can encapsulate {\mmt}-based or {\mmt}-aware systems as web-services and use {\mmt} as a universal interface language.
At the same time {\mmt} is \defemph{fully formal} in the sense that its semantics is specified rigorously in a self-contained formal system, namely using the type-theoretical style of judgments and inference rules. Such a level of formality is rare among module systems, SML being one of the few examples.

We contend that the dream of formalizing large parts of mathematics to make them machine-understandable can only be reached based on a system with both these features. However, in practice, they are often in conflict, and their combination makes {\mmt} unique. In particular, it is easy to write large scale implementations in {\mmt}, and it is easy to verify and trust them.

\subsection{Beyond {\mmt}}\label{sec:concl:beyond}

We have designed {\mmt} as the simplest possible language that combines foundation-independence, modularity, web-scalability, and formality. Future work can now build on {\mmt} and add individual orthogonal language features -- in each case preserving these four qualities. In particular, for each feature, we have to define grammar and inference rules, the induced knowledge items and their URIs, and their behavior under theory morphisms. In fact, we have already developed some of these features but excluded them in this paper to focus on a minimal core language.

In the following we list some language features that we will carefully add to {\mmt} in the future:
\begin{list}{}{\setlength{\leftmargin}{0.5em}\setlength{\itemindent}{2.5em}}
\item[\bf Unnamed Imports]
 In addition to the described named imports with distin\-guish-semantics, {\mmt} is designed to provide also unnamed imports with identify-semantics. They are already part of the {\mmt} API, and the main reason to omit them here was to simplify the presentation of the formal semantics of {\mmt}.
\item[\bf Cyclic Imports]
 Inspecting the flattening theorem reveals that cyclic imports are not as harmful as one might think: Cyclic imports can be elaborated easily if we permit theories with infinitely many constant declarations. In particular, cyclic imports will permit elegant representations of languages with an infinite hierarchy of universes or with an infinite hierarchy of reflection.
\item[\bf Nested Theories]
  Nested theories will provide a scalable mechanism for representing hierarchic scopes and visibility. Many language features naturally suggest such a nesting of scopes such as mutual recursion, local functions, record types, or proofs with local definitions.
  
  Intuitively, if $\qS$ is a subtheory of $\qT$, the declarations of $\qT$ occurring before $\qS$ are implicitly imported into $\qS$ via an unnamed import, and the declarations of $\qT$ succeeding $\qS$ can refer to $\qS$, e.g., by importing it. The main difficulty here is to add nested theories in a way that preserves the order-invariance of declarations.
\item[\bf (Co-)Inductive Data Types]
  Using some of the above features, it is possible to give foundation-independent definitions of inductive and coinductive data types. An inductive data type over $T$ is declared as a theory $I$ with a distinguished type $t$ over $I$: The values of the induced type are defined using the closed terms $\omega$ such that $\otermtype{\TG}{I}{\omega}{t}$. Functions from this type to some type $u$ over $T$ can be defined by induction, which amounts to giving a theory morphism from $I$ to $T$ that maps $t$ to $u$.
  
  A coinductive data type over $T$ is declared as a theory $C$ with a distinguished partial morphism $m$ from $C$ to $T$. The values of the induced type are defined using the valid morphisms $\omorphism{\TG}{\mu}{C}{T}$ that agree with $m$. Thus, definitions by coinduction are reduced to theory morphisms. In particular, $C$ specializes to a record type if it does not contain cyclic imports.
  
  Coinductive types can be used to reflect the {\mmt}-concept of realizations into individual foundations. For example, consider Ex.~\ref{ex:mmt:FOL} with $C=\cn{Monoid}$, $T=\cn{ZFC}$, and $m=\cn{FOLSem}$. Then the values of the coinductive type over $T$ given by $C$ and $m$ are the models of $C$.
\begin{variant}{\not\modexp}
\item[\bf Theory Expressions]
  Some module systems, e.g., CASL or Isabelle, provide complex theory expressions. For example, $\qS\cup\qT$ can denote the union of the theories $\qS$ and $\qT$. Other examples are the translation of a theory along a morphism, the extension of a theory with some declarations, or the pushout of certain morphisms. Similarly, we can add further productions for morphism expression, e.g., for the mediating morphism out of a pushout.
  
  The main difficulty here is that these complex theories and consequently their declarations do not have canonical identifiers. Indeed, most systems handle theory expressions by decomposing them internally and generating fresh internal names for the involved subexpressions.
  Similarly, all of these constructions can be expressed in {\mmt} already by introducing auxiliary theories as we showed in \cite{CHKMR:hiding:11}.
  But certain theory expressions -- most importantly unions and pushouts along unnamed imports -- can be added to {\mmt} in a way that preserves canonical identifiers without using generated names.
\end{variant}
\item[\bf Conservative Extensions]
  A common practice is to give a theory $\qS$ with undefined constants -- the primitive concepts -- and then another theory $\qT$ that imports $\qS$ and adds with defined constants -- the derived concepts. This is particularly important when the declarations of $\qS$ represent axioms and those of $\qT$ theorems. In that case, it is desirable to make this kind of conservativity of $\qT$ explicit in order to exploit it later. For example, if $\qT$ is conservative over $\qS$, then a theory importing $\qS$ should implicitly also gain access to $\qT$.
\item[\bf Hiding and Filtering]
  In \cite{KRS:integration:10}, we showed how a slight extension of the semantics of filtering yields a substantial increase in expressivity. In particular, it becomes possible to safely relax the strictness of filtering. The key idea is that foundations do not only say ``yes'' when confirming a typing or equality relation but also return a list of dependencies, which {\mmt} maintains and uses to propagate filtering. We use a syntactically similar but semantically different extension of {\mmt} in \cite{CHKMR:hiding:11} to extend {\mmt} with model theoretical hiding. We expect that further research will permit the unification of these two features.
\item[\bf Sorting]
  The components of a constant declaration -- type and definiens -- correspond to the base judgments provided by the foundations -- typing and equality. In particular, {\mmt} uses the constant declarations to provide the axioms of the inference systems used in specific foundations.
  It is natural but not necessary to consider exactly typing and equality.
  For example, we can extend {\mmt} with constant declarations $c <: \tau$ that declare $c$ as a \emph{sort} refining $\tau$. Examples are subtypes (refining  types), type classes (refining the kind of types), and set theoretical classes (refining the universe of sets). This extension would go together with a subsorting judgment $\otermsort{\TG}{\qT}{\omega}{\omega'}$ in the foundation.
\item[\bf Logical Relations]
  The notions of theory and theory morphisms between theories can be extended with logical relations between theory morphisms. {\mmt} logical relations will be purely syntactical notions that correspond to the well-known semantic ones. A preliminary account was given in \cite{kristina:ml_fol}. They will permit natural representations of relations between realizations -- such as model morphisms -- as well as of extensional equality relations.
\item[\bf Computation]
  {\mmt} is currently restricted to declarative languages thus excluding the important role of computation, e.g., in computer algebra systems, decision procedures, and programs extracted from proofs. Generating code from appropriate {\mmt} theories is relatively simple. But we also want to permit literal code snippets in the definiens of a constant. This will provide a formal interface between a formal semantics and scalable implementations.
\item[\bf Aliases]
  {\mmt} avoids the introduction of new names for symbols; instead, canonical qualified identifiers are formed. But this often leads to long unfriendly identifiers. Aliases for individual identifiers or identifier prefixes are a simple syntactic device for providing human-friendly names, e.g., by declaring the aliases $+$ and $\ast$ for $\cnpath{add,mon,comp}$ and $\cnpath{mult,comp}$ in the theory $\cn{Ring}$. Moreover, such names can be used to make the modular structure of a theory transparent. This is already part of our implementation.
\item[\bf Declaration Patterns and Functors]
 A common feature of declarative languages is that the declarations in a theory $T$ with meta-theory $M$ must follow one out of several patterns. For example, if $M$ is first-order logic, then $T$ should contain only function symbol, predicate symbol, and axiom declarations. We can capture this foundation-independently in {\mmt} by declaring such patterns in $M$ and then pattern-checking the declarations in $T$ against them.
 
Patterns also permit adding a notion of functors to {\mmt} whose input is an arbitrary well-patterned theory $T$ with meta-theory $M$. The output is a theory defined by induction on the list of declarations in $T$. This permits concise representations of functors between categories of theories, e.g., the functor that takes a sorted first-order theory and returns its translation to unsorted first-order logic by relativization of quantifiers.
This can be extended to functors between categories of diagrams.
\item[\bf Minimal Foundations]
Not all language features can be defined foundation-independently. Consider Mizar-style \cite{mizar} implicit definitions of the form
 \[\mathtt{func}\;c\;\mathtt{means}\;F(c);\; \mathtt{correctness}\;P;\]
where $P$ is a proof of $\exists^!x.F(x)$ and $c$ is defined as that unique value. Such a definition is meaningful iff the foundational theory can express the quantifier $\exists^!$ of unique existence. Moreover, in that case it can be elaborated into the two declarations $\symdd{c}{}{}$ and $\symdd{c\_def}{F(c)}{}$ (which is in fact what Mizar and most other systems are doing).

In the spirit of little foundations, we will add such pragmatic language features to {\mmt} together with the minimal foundations needed to define their semantics. If an individual foundational theory $M$ imports one of these distinguished minimal foundations, the corresponding pragmatic feature becomes available in theories with meta-theory $M$.

Further pragmatic declarations include, for example, function declarations (possible if $M$ can express $\lambda$-abstraction) and constants with multiple types (possible if $M$ can express intersection types). The above-mentioned features of sorting and (co-)inductive data types as well as the Curry-Howard representation of axioms, theorems, and proof rules can become special cases of pragmatic features as well. We can even generalize the notion of foundations and then recover the type and definiens of a constant as pragmatic features that are possible if $M$ can express typing and equality.

\item[\bf Narrative and Informal Representations]
One motivation of {\mmt} has been to give a formal semantics to {\omdoc} 1.2, and the present work does this for the {\omdoc} fragment concerned with formal theory development. It omits narrative aspects (e.g., document structuring, notations, examples, citations) as well as informal and semi-formal representations. We will extend {\mmt} towards all of {\omdoc}, and this effort will culminate in the {\omdoc} 2 language. As a first step, we have included sectioning and notations in the {\mmt} API. Many other features of {\omdoc} 1.2 will be recovered as pragmatic features in the above sense.
\end{list}

\subsection{Applying {\mmt}}\label{sec:concl:apply}

The development of {\mmt} and its implementations has been driven by our ongoing and intended applications.
Most importantly, we have evaluated {\mmt} on the logic atlas built in the LATIN project as described in Sect.~\ref{sec:mmtweb:implementation}. Here, {\mmt} is applied in two ways.

Firstly, {\mmt} provides the ontology used to organize the highly interlinked theories in the logic graph. In particular, the {\mmt} principles of meta-theories and foundation-independence provide a clean separation of concerns between the logical framework (LF in the case of LATIN), the logics, and the domain theories written in these logics.

Secondly, {\mmt} serves as the scalable interface language between the various {\mmt}-aware software systems used in LATIN. Twelf \cite{twelf} is used to write logics, {\tntbase} \cite{tntbase} for persistent storage, the {\mmt} API for presentation and indexing, JOBAD \cite{GLR:jobad:09} for interactive browsing, and Hets \cite{hets} for institution-based cross-logic proof management, and we are currently adding sTeXIDE \cite{stexide} for semantic authoring support. {\mmt} is crucial to communicate the content and its semantics between both the heterogeneous platforms and the respective developers. In particular, the canonical {\mmt} identifiers have proved pivotal for the integration of software systems.
\medskip

Building on the LATIN atlas, we are creating an ``Open Archive of FlexiForms'' (OAFF). It will store flexiformal (i.e., represented at flexible degrees of formality) representations of mathematical knowledge and supply them with {\mmt}-base knowledge management services.
OAFF will contain the domain theories and libraries written in the logics that are part of the LATIN atlas.
Using {\mmt}, it becomes possible to represent libraries developed in different foundational systems in one uniform formalism. Since {\mmt} can also represent relations between the underlying foundational system, this provides a base for practical reliable system integration.
For example, we are currently importing the libraries of TPTP \cite{tptp} and Mizar \cite{mizar} into OAFF. Other systems like Coq \cite{coq}, Isabelle \cite{isabelle}, or PVS \cite{pvs} already have XML or {\omdoc} 1.2 exports that can be updated to export {\mmt}.
Variables: 


\newcommand{\etalchar}[1]{$^{#1}$}
\providecommand\seen{seen } \providecommand\webpageat{web page at }
  \providecommand\homepageat{home page at }
  \providecommand\projectpageat{project page at }
  \providecommand\systempageat{system home page at }
  \providecommand\svnrepoat{Subversion repository at }
  \providecommand\January{January} \providecommand\February{February}
  \providecommand\March{March} \providecommand\April{April}
  \providecommand\May{May} \providecommand\June{June}
  \providecommand\July{July} \providecommand\August{August}
  \providecommand\September{September} \providecommand\October{October}
  \providecommand\November{November} \providecommand\December{December}
  \providecommand\AUSTRALIA{Australia} \providecommand\ROMANIA{Romania}
  \providecommand\MEXICO{Mexico} \providecommand\ITALY{Italy}
  \providecommand\USA{USA} \providecommand\IRELAND{Ireland}
  \providecommand\HUNGARY{Hungary} \providecommand\JAPAN{Japan}
  \providecommand\CANADA{Canada} \providecommand\SPAIN{Spain}
  \providecommand\NETHERLANDS{Netherlands} \providecommand\UK{UK}
  \providecommand\SWEDEN{Sweden} \providecommand\GERMANY{Germany}
  \providecommand\openmath{OpenMath} \providecommand\fc{forthcoming}
  \providecommand\PROC{Proceedings} \providecommand\omdoc{OMDoc}
  \providecommand\activemath{ActiveMath} \hyphenation{Wiki-Sym}

\end{document}